# Energy Levels, Transition Probabilities and Electron-Impact Excitations Of Ge-Like Pr, Nd, Pm, Sm and Eu ions

O. Nagy and Fatma El_Sayed[*]

Physics Department, Faculty of Science, Kafrelsheikh University, Egypt

**Abstract**

Energies, wavelengths, transition probabilities, and oscillator strengths have been calculated for the $4s^2 4p^2 - 4s4p^3$, $4s^2 4p^2 - 4s^2 4p4d$ and $4s4p^3 - 4p^4$ allowed transitions in heavy Ge-like ions with Z=59-63. The fully relativistic Multiconfiguration Dirac-Fock (MCDF) method taking into account both the correlations within the $n=4$ complex and the quantum electrodynamic (QED) effects have been used in the calculations. MCDFGME code is used to calculate electron impact excitation cross sections for the $4s^2 4p^2 - 4s4p^3$, and $4s^2 4p^2 - 4s^2 4p4d$ transitions with plane-wave Born approximation. The results of $Pr\ XXVIII$, $Nd\ XXIX$, $Pm\ XXX$, $Sm\ XXXI$ and $Eu\ XXXII$ are compared with HFR method results.

## Contents

## 1. Introduction

Atomic collision processes are important in many fundamental and applied areas of current research [1]. The modeling and diagnostics of astrophysical and laboratory plasmas depend on accurate atomic data for the collision strengths and cross sections of the various collision processes that might occur in them. The data required involves mainly electron-ion collisions, and includes excitation, recombination (radiative and dielectronic), and ionization processes. The most important indirect ionization process is excitation-autoionization.

---
[*] fatmamahrous@sci.kfs.edu.eg

Electron impact excitation cross section of highly ionized ions are needed in developing lasers in extreme ultraviolet (EUV) and soft X-ray regimes, in astrophysics and in the study of inertial confinement fusion and other laboratory-produced plasmas.

Recently, extreme ultraviolet (EUV) light sources for microlithography are receiving much attention as an application of laser-produced high-Z plasmas where the 13.5 nm light source produced in tin plasmas was thought to be attractive due to its compactness and high emissivity [2]. Theoretical analysis of the opacity and the emissivity in both EUV light sources and X-ray lasers is important to design and optimize the laser produced high-Z plasmas. Therefore, the needs for atomic data for those high-Z elements are urgent. The most successful method to produce saturated X-ray laser output is a scheme where the collisional excitation is one of the dominant atomic processes in these plasmas.

Accurate electron-impact excitation cross sections are needed for proper interpretation of the spectral emission of highly charged ions and utilizing spectral measurements for plasma diagnostics [3]. The determination of the ionization balance for a given plasma temperature is especially sensitive to the values of the excitation cross sections used to correlate the intensity of a given line to the abundance of the corresponding ion species.

Knowing the correct charge state distribution is critical for understanding radiation levels, energy deposition and energy balance of high temperature plasmas.

In this study, energy levels, radiative rates and electron impact excitation for fine-structure levels of **$Pr^{27+}$, $Nd^{28+}$, $Pm^{29+}$, $Sm^{30+}$** and **$Eu^{31+}$** ions are calculated using mcdfgme [4] code (**M**ulti-**C**onfiguration **D**irac-**F**ock and **G**eneral **M**atrix **E**lements).

**2. Calculations**

The wavelengths, transition probabilities and oscillator strengths for the $4s^2 4p^2 - 4s 4p^3$, $4s^2 4p^2 - 4s^2 4p 4d$ and $4s 4p^3 - 4p^4$ allowed electric dipole transitions were calculated for ions belonging to Ge- isoelectronic sequence. These calculations are performed using the fully relativistic MCDF approach with the *mcdfgme* program.

The orthogonality of the wavefunctions was consistently included in the differential equations by using off-diagonal Lagrange multipliers. Most of the odd and even configurations within the $n=4$ Layzer complex are included in the calculations. These configurations are $4s^2 4p^2$, $4s^2 4d^2$, $4s^2 4p 4f$, $4s^2 4f^2$, $4s 4p^2 4d$, $4s 4p 4d 4f$, $4p^4$, $4p^2 4d^2$, $4p^2 4f^2$ and $4s^2 4p 4d$, $4s^2 4d 4f$, $4s 4p^3$, $4s 4p^2 4f$, $4s 4p 4d^2$, $4s 4p 4f^2$, $4s 4d^2 4f$, $4p^3 4d$, $4p^2 4d 4f$, $4p 4d^3$ for the even and odd parities, respectively. There are some configurations like $4s 4d^3$, $4d^2 4f^2$, $4d^4$, $4p^3 4f$, $4f^4$, $4p 4f^3$, $4d^3 4f$, $4d 4f^3$, and $4s 4f^3$ have not been included in calculations because they are located high in energy. The high energy difference between configurations gives normally small contribution. The participations from the configurations with n=5 are so weak that to be cancelled, about 0.03%, thus these interactions are not included in our calculations.

The MCDF calculations were completed with the inclusion of the relativistic two-body Breit interaction and the quantum electrodynamic corrections (QED) which arise due to self-energy and vacuum polarization.

The HFR code [5, 6, 7] used in an ab-initio way for $Pr\ XXVIII$, $Nd\ XXIX$, $Pm\ XXX$, $Sm\ XXXI$ and $Eu\ XXXII$ ions, because of the lack of experimental data of these ions.

## 3. Results and Discussion

### 3.1 Energy Levels

The energy level values obtained using the MCDF and HFR method for the $4s^2 4p^2$, $4s 4p^3$, $4s^2 4p 4d$ and $4p^4$ configurations in heavy Ge-like ions with $(59 \leq Z \leq 63)$ are presented in tables (1- 5). The main components of the computed eigenvectors in both the *jj*-coupling and the *LS*-coupling schemes are also given in these tables. In the notations, the superscript "*" corresponds to the case $j^* = l - 1/2$ while no superscript is given for $j = l + 1/2$.

Tables 1-5 show that the agreement between MCDF and HFR energy levels is within 1% in Ge-like ions.

For heavy atoms and highly charged ions, the *jj*-coupling is realized because the electrostatic interaction is directly proportional with Z value while the spin-orbit interaction increases as $Z^4$ [8].

### 3.2 Wavelengths, Transition Probabilities and Oscillator Strengths

The wavelengths, transition probabilities and oscillator strengths for the $4s^2 4p^2 - 4s 4p^3$, $4s^2 4p^2 - 4s^2 4p 4d$ and $4s 4p^3 - 4p^4$ allowed transitions were calculated using the fully relativistic MCDF method are reported in tables (6-10). The calculated MCDF transition probabilities are presented in both gauges- length and – velocity, while the oscillator strengths are only shown in length gauge. The agreement between the length and velocity forms of the oscillator strengths is within 10-20% for all the strong transitions (i.e. for $f \geq 0.10$) in Ge-like ions are presented. Exceptions to this are some few transitions; where the agreement is within $\approx 40\%$. Not only is this overall good agreement highly satisfactory, but it also gives a clear indication of the accuracy of the results. We present in table (11) a comparison between the MCDF, and HFR transition probabilities (in sec$^{-1}$) and oscillator strengths.

### 3.3 Collision Strengths and Cross Sections for Electron-Impact Excitation

Excitation rate coefficients (in cm$^3$ sec$^{-1}$) for a transition from $i \rightarrow j$ are calculated using the following formula [9]

$$C(T) = \frac{8.629 \times 10^{-6}}{g_i \sqrt{kT_e}} \exp\left(-\Delta E_{ij}/kT_e\right) \gamma(T) \qquad (1)$$

where γ is the effective collision strength, $T_e$ is the electron temperature in eV, $g_i$ is the statistical weight of the level *i*, $\Delta E_{ij}$ is the excitation energy, and $k$ is the Boltzmann constant. Using a Maxwellian velocity distribution, the effective collision strengths can be defined as a function of electron temperature.

$$\gamma(i \rightarrow j) = \int_0^\infty \Omega_{i \rightarrow j} \exp\left(-E/kT_e\right) d\left(E/kT_e\right) \qquad (2)$$

where *E* is the scattered electron energy, and $\Omega_{i \rightarrow j}$ is the collision strength. The unit of temperature is Kelvin.

The cross section $\sigma(i,j)$ for the transition $i \rightarrow j$ is expressed in terms of the collision strength $\Omega(i,j)$ as follows:

$$\sigma(i,j) = \frac{\Omega(i,j)}{\omega_i k_i^2}\left(\pi a_o^2\right) \qquad (3)$$

where the subscripts $i$ and $j$ refer to the initial and final states, $a_o$ is the Bohr radius ($\pi a_o^2 = 8.797 \times 10^{-17}\,cm^2$), $\omega_i = (2S_i+1)(2L_i+1)$ and $k_i^2$ is the energy of the incident electron in Rydbergs (1 Ryd = 13.6057 eV).

The standard form of a Plane-Wave Born excitation cross section is

$$\sigma_{PWB} = \frac{4\pi a_0^2 R}{T}\,\Omega_{PWB}(T) \qquad (4)$$

Where $\Omega_{PWB}$ is the collision strength and $R$ is the Rydberg energy.

The electron impact collision strengths for the $4s^2 4p^2 - 4s\,4p^3$ and $4s^2 4p^2 - 4s^2 4p\,4d$ allowed transitions have been calculated at nine scattered electron energies ranging from 200 to 10000 eV. Tables 12-16 list the collision strengths at different scattered electron energies.

Figures 1-5 show the behavior of cross section (m²) with the incident energy (eV) for $4s^2 4p^{*2}(J=0) - 4s^2 4p^* 4d^*(J=1)$, $4s^2 4p^* 4p(J=1) - 4s^2 4p\,4d^*(J=0)$, $4s^2 4p^2(J=0) - 4s\,4p^3(J=1)$, $4s^2 4p^* 4p(J=2) - 4s^2 4p\,4d^*(J=3)$ and $4s^2 4p^2(J=2) - 4s^2 4p\,4d(J=3)$ transitions in Ge-like ions $59 \leq Z \leq 63$. It is shown that the cross section shapes at the threshold and starts with a finite value for the excitation of ions.

## 4. Conclusions

In conclusion, we report energy levels, radiative rates and collision strengths for allowed transitions with the fully relativistic mutliconfiguration Dirac-Fock method. The accuracy of the collision strengths presented in this paper is difficult to assess due to the paucity of data available for comparison. However, since the present calculation (a) uses extensive configuration-interaction wavefunctions, (b) includes correlation terms in the total wavefunction, and (c) include the relativistic two-body Breit interaction and the quantum electrodynamic corrections (QED). It is expected that the collision strengths are accurate to approximately 20-30%. The cross section data should be a useful reference in the development of X-ray lasers and other related fields involving highly stripped ions.

## Acknowledgment

The authors wish to thank Dr. J. P. Desclaux and Dr. P. Indelicato for allowing us to use the mcdfgme code.

## References

[1] R. E. H. Clark and D. H. Reiter, "Nuclear Fusion Research, understanding Plasma-Surface Interactions",(Springer-Verlag Berlin, Heidelberg, 2005).

[2] S. Fujioka, H. Nishimura, K. Nishihara, A. Sasaki, A. Sunahara, T. Okuno, N. Ueda, T. Ando, Y. Tao, Y. Shimada, K. Hashimoto, M. Yamaura, K. Shigemori, M. Nakai, K. Nagai, T. Norimatsu, T. Nishikawa, N. Miyanaga, Y. Izawa, and K. Mima, Phys. Rev. Lett **95**, 235004 (2005).

[3] M. J. May, P. Beiersdorfer, N. Jordan, J. H. Scofield, K. J. Reed, S. B. Hansen, K. B. Fournier, M. F. Gu, G. V. Brown, F. S. Porter, R. Kelley, C. A. Kilbourne and K. R. Boyce, Nuclear Instruments and Methods in Physics Research B **235**, 231-234 (2005).

[4] J. P. Desclaux and P. Indelicato , *MCDFGME*, http://dirac.spectro.jussieu.fr/mcdf.

[5] R. D. Cowan, "The Theory of Atomic Structure and Spectra", (Berkeley, University of California Press, 1981).

[6] Htttp://das101.isan.troitsk.ru/.

[7] R. D. Cowan, J. Opt. Soc. Am **58,** 808 (1968).

[8] H. F. Beyer and V. P. Shevelko, "Introduction to the Physics of Highly Charged Ions", (IOP Publishing Ltd. 2003).

[9] Y. Hahn, A. K. Pradhan, H. Tawara, and H. L. Zhang, "Photon and Electron Interactions with Atoms, Molecules and Ions", (Springer-Verlag Berlin Heidelberg, Germany, 2001).

Table (1): **MCDF** and **HFR** energy levels (in cm$^{-1}$) of **4s$^2$4p$^2$**, **4s4p$^3$**, **4s$^2$4p4d**, and **4p$^4$** configurations for **Pr XXVIII** ion.

| | JJ-coupling | J | LS | MCDF (cm$^{-1}$) | HFR (cm$^{-1}$) |
|---|---|---|---|---|---|
| 1 | 4s$^2$ 4p$^{*\,2}$ | 0 | $^3P_0^e$ | 0 | 0 |
| 2 | 4s$^2$ 4p$^*$ 4p | 1 | $^3P_1^e$ | 200201 | 202316 |
| 3 | 4s$^2$ 4p$^*$ 4p | 2 | $^1D_2^e$ | 224962 | 224270 |
| 4 | 4s$^2$ 4p$^2$ | 2 | $^3P_2^e$ | 443435 | 443008 |
| 5 | 4s$^2$ 4p$^2$ | 0 | $^1S_0^e$ | 491235 | 491593 |
| 6 | 4s 4p$^{*\,2}$ 4p | 2 | $^5S_2^o$ | 681165 | 691953 |
| 7 | 4s 4p$^{*\,2}$ 4p | 1 | $^3D_1^o$ | 778476 | 779413 |
| 8 | 4s 4p$^2$ 4p$^*$ | 2 | $^3D_2^o$ | 880789 | 888636 |
| 9 | 4s 4p$^2$ 4p$^*$ | 3 | $^3D_3^o$ | 935926 | 936713 |
| 10 | 4s 4p$^2$ 4p$^*$ | 0 | $^3P_0^o$ | 972431 | 971481 |
| 11 | 4s 4p$^2$ 4p$^*$ | 1 | $^3P_1^o$ | 996114 | 997903 |
| 12 | 4s 4p$^2$ 4p$^*$ | 2 | $^1D_2^o$ | 1019895 | 1020722 |
| 13 | 4s 4p$^2$ 4p$^*$ | 1 | $^3S_1^o$ | 1029589 | 1030987 |
| 14 | 4s$^2$ 4p$^*$ 4d$^*$ | 2 | $^3F_2^o$ | 1031939 | 1038144 |
| 15 | 4s$^2$ 4p$^*$ 4d$^*$ | 1 | $^3D_1^o$ | 1111392 | 1113271 |
| 16 | 4s$^2$ 4p$^*$ 4d | 3 | $^3F_3^o$ | 1121390 | 1123447 |
| 17 | 4s$^2$ 4p$^*$ 4d | 2 | $^3D_2^o$ | 1125434 | 1125949 |
| 18 | 4s 4p$^3$ | 2 | $^3P_2^o$ | 1192041 | 1191338 |
| 19 | 4s 4p$^3$ | 1 | $^1P_1^o$ | 1270737 | 1271832 |
| 20 | 4s$^2$ 4p 4d$^*$ | 2 | $^1D_2^o$ | 1292717 | 1295382 |
| 21 | 4s$^2$ 4p 4d | 4 | $^3F_4^o$ | 1301159 | 1308274 |
| 22 | 4s$^2$ 4p 4d$^*$ | 0 | $^3P_0^o$ | 1312852 | 1313313 |
| 23 | 4s$^2$ 4p 4d$^*$ | 1 | $^3P_1^o$ | 1316876 | 1319065 |
| 24 | 4s$^2$ 4p 4d$^*$ | 3 | $^3D_3^o$ | 1322933 | 1321375 |
| 25 | 4s$^2$ 4p 4d | 2 | $^3P_2^o$ | 1347102 | 1347380 |
| 26 | 4s$^2$ 4p 4d | 3 | $^1F_3^o$ | 1411411 | 1411173 |
| 27 | 4s$^2$ 4p 4d | 1 | $^1P_1^o$ | 1417679 | 1417722 |
| 28 | 4p$^{*\,2}$ 4p$^2$ | 2 | $^3P_2^e$ | 1537430 | 1547016 |
| 29 | 4p$^{*\,2}$ 4p$^2$ | 0 | $^1S_0^e$ | 1584081 | 1590566 |
| 30 | 4p$^*$ 4p$^3$ | 1 | $^3P_1^e$ | 1750619 | 1763967 |
| 31 | 4p$^*$ 4p$^3$ | 2 | $^1D_2^e$ | 1780873 | 1788252 |
| 32 | 4p$^4$ | 0 | $^3P_0^e$ | 2014312 | 2024246 |

Table (2): **MCDF** and **HFR** energy levels (in cm$^{-1}$) of **4s$^2$4p$^2$**, **4s4p$^3$**, **4s$^2$4p4d**, and **4p$^4$** configurations for **Nd XXIX** ion.

| | JJ-coupling | J | LS | MCDF (cm$^{-1}$) | HFR (cm$^{-1}$) |
|---|---|---|---|---|---|
| 1 | 4s$^2$ 4p$^{*\,2}$ | 0 | $^3P_0^e$ | 0 | 0 |
| 2 | 4s$^2$ 4p$^*$ 4p | 1 | $^3P_1^e$ | 221615 | 224505 |
| 3 | 4s$^2$ 4p$^*$ 4p | 2 | $^1D_2^e$ | 247172 | 247124 |
| 4 | 4s$^2$ 4p$^2$ | 2 | $^3P_2^e$ | 487606 | 488526 |
| 5 | 4s$^2$ 4p$^2$ | 0 | $^1S_0^e$ | 536537 | 538250 |
| 6 | 4s 4p$^{*\,2}$ 4p | 2 | $^5S_2^o$ | 721822 | 732828 |
| 7 | 4s 4p$^{*\,2}$ 4p | 1 | $^3D_1^o$ | 821639 | 822168 |
| 8 | 4s 4p$^2$ 4p$^*$ | 2 | $^3D_2^o$ | 939867 | 949269 |
| 9 | 4s 4p$^2$ 4p$^*$ | 3 | $^3D_3^o$ | 997814 | 999419 |
| 10 | 4s 4p$^2$ 4p$^*$ | 0 | $^3P_0^o$ | 1035369 | 1035099 |
| 11 | 4s 4p$^2$ 4p$^*$ | 1 | $^3P_1^o$ | 1060931 | 1063093 |
| 12 | 4s$^2$ 4p$^*$ 4d$^*$ | 2 | $^3F_2^o$ | 1080696 | 1087372 |
| 13 | 4s 4p$^2$ 4p$^*$ | 2 | $^1D_2^o$ | 1087278 | 1086259 |
| 14 | 4s 4p$^2$ 4p$^*$ | 1 | $^3S_1^o$ | 1092968 | 1095006 |
| 15 | 4s$^2$ 4p$^*$ 4d$^*$ | 1 | $^3D_1^o$ | 1162888 | 1164692 |
| 16 | 4s$^2$ 4p$^*$ 4d | 3 | $^3F_3^o$ | 1177511 | 1179638 |
| 17 | 4s$^2$ 4p$^*$ 4d | 2 | $^3D_2^o$ | 1180541 | 1181319 |
| 18 | 4s 4p$^3$ | 2 | $^3P_2^o$ | 1275469 | 1276306 |
| 19 | 4s 4p$^3$ | 1 | $^1P_1^o$ | 1357027 | 1359381 |
| 20 | 4s$^2$ 4p 4d$^*$ | 2 | $^1D_2^o$ | 1365276 | 1368743 |
| 21 | 4s$^2$ 4p 4d | 4 | $^3F_4^o$ | 1377169 | 1385402 |
| 22 | 4s$^2$ 4p 4d$^*$ | 0 | $^3P_0^o$ | 1385658 | 1386850 |
| 23 | 4s$^2$ 4p 4d$^*$ | 1 | $^3P_1^o$ | 1389923 | 1392882 |
| 24 | 4s$^2$ 4p 4d$^*$ | 3 | $^3D_3^o$ | 1396066 | 1395207 |
| 25 | 4s$^2$ 4p 4d | 2 | $^3P_2^o$ | 1423644 | 1425000 |
| 26 | 4s$^2$ 4p 4d | 3 | $^1F_3^o$ | 1489096 | 1489851 |
| 27 | 4s$^2$ 4p 4d | 1 | $^1P_1^o$ | 1496468 | 1497342 |
| 28 | 4p$^{*\,2}$ 4p$^2$ | 2 | $^3P_2^e$ | 1620136 | 1629921 |
| 29 | 4p$^{*\,2}$ 4p$^2$ | 0 | $^1S_0^e$ | 1668309 | 1674826 |
| 30 | 4p$^*$ 4p$^3$ | 1 | $^3P_1^e$ | 1855288 | 1869582 |
| 31 | 4p$^*$ 4p$^3$ | 2 | $^1D_2^e$ | 1886205 | 1894431 |
| 32 | 4p$^4$ | 0 | $^3P_0^e$ | 2142074 | 2153497 |

Table (3): **MCDF** and **HFR** energy levels (in cm$^{-1}$) of **4s$^2$4p$^2$**, **4s4p$^3$**, **4s$^2$4p4d**, and **4p$^4$** configurations for **Pm XXX** ion.

|    | JJ-coupling | J | LS | MCDF (cm$^{-1}$) | HFR (cm$^{-1}$) |
|----|---|---|---|---|---|
| 1  | 4s$^2$ 4p$^{*\,2}$ | 0 | $^3$P$_0^e$ | 0 | 0 |
| 2  | 4s$^2$ 4p$^*$ 4p | 1 | $^3$P$_1^e$ | 249494 | 248488 |
| 3  | 4s$^2$ 4p$^*$ 4p | 2 | $^1$D$_2^e$ | 275841 | 271768 |
| 4  | 4s$^2$ 4p$^2$ | 2 | $^3$P$_2^e$ | 539901 | 537630 |
| 5  | 4s$^2$ 4p$^2$ | 0 | $^1$S$_0^e$ | 589970 | 588496 |
| 6  | 4s 4p$^{*\,2}$ 4p | 2 | $^5$S$_2^o$ | 769570 | 775523 |
| 7  | 4s 4p$^{*\,2}$ 4p | 1 | $^3$D$_1^o$ | 865227 | 866784 |
| 8  | 4s 4p$^2$ 4p$^*$ | 2 | $^3$D$_2^o$ | 1007307 | 1013638 |
| 9  | 4s 4p$^2$ 4p$^*$ | 3 | $^3$D$_3^o$ | 1067977 | 1065797 |
| 10 | 4s 4p$^2$ 4p$^*$ | 0 | $^3$P$_0^o$ | 1106584 | 1102389 |
| 11 | 4s 4p$^2$ 4p$^*$ | 1 | $^3$P$_1^o$ | 1132493 | 1131951 |
| 12 | 4s$^2$ 4p$^*$ 4d$^*$ | 2 | $^3$F$_2^o$ | 1134604 | 1138348 |
| 13 | 4s 4p$^2$ 4p$^*$ | 2 | $^1$D$_2^o$ | 1158199 | 1155412 |
| 14 | 4s 4p$^2$ 4p$^*$ | 1 | $^3$S$_1^o$ | 1164650 | 1162721 |
| 15 | 4s$^2$ 4p$^*$ 4d$^*$ | 1 | $^3$D$_1^o$ | 1217771 | 1217851 |
| 16 | 4s$^2$ 4p$^*$ 4d | 2 | $^3$D$_2^o$ | 1238708 | 1238807 |
| 17 | 4s$^2$ 4p$^*$ 4d | 3 | $^3$F$_3^o$ | 1238924 | 1237952 |
| 18 | 4s 4p$^3$ | 2 | $^3$P$_2^o$ | 1368887 | 1366796 |
| 19 | 4s$^2$ 4p 4d$^*$ | 2 | $^1$D$_2^o$ | 1445988 | 1445639 |
| 20 | 4s 4p$^3$ | 1 | $^1$P$_1^o$ | 1453266 | 1452404 |
| 21 | 4s$^2$ 4p 4d$^*$ | 0 | $^3$P$_0^o$ | 1461020 | 1463923 |
| 22 | 4s$^2$ 4p 4d | 4 | $^3$F$_4^o$ | 1461690 | 1466448 |
| 23 | 4s$^2$ 4p 4d$^*$ | 1 | $^3$P$_1^o$ | 1470934 | 1470238 |
| 24 | 4s$^2$ 4p 4d$^*$ | 3 | $^3$D$_3^o$ | 1470993 | 1472577 |
| 25 | 4s$^2$ 4p 4d | 2 | $^3$P$_2^o$ | 1508700 | 1506544 |
| 26 | 4s$^2$ 4p 4d | 3 | $^1$F$_3^o$ | 1575274 | 1572440 |
| 27 | 4s$^2$ 4p 4d | 1 | $^1$P$_1^o$ | 1583765 | 1580874 |
| 28 | 4p$^{*\,2}$ 4p$^2$ | 2 | $^3$P$_2^e$ | 1711287 | 1716606 |
| 29 | 4p$^{*\,2}$ 4p$^2$ | 0 | $^1$S$_0^e$ | 1760973 | 1762856 |
| 30 | 4p$^*$ 4p$^3$ | 1 | $^3$P$_1^e$ | 1970057 | 1980763 |
| 31 | 4p$^*$ 4p$^3$ | 2 | $^1$D$_2^e$ | 2001639 | 2006177 |
| 32 | 4p$^4$ | 0 | $^3$P$_0^e$ | 2281604 | 2290109 |

Table (4): **MCDF** and **HFR** energy levels (in cm$^{-1}$) of **4s$^2$4p$^2$**, **4s4p$^3$**, **4s$^2$4p4d**, and **4p$^4$** configurations for **Sm XXXI** ion.

|    | JJ-coupling | J | LS | MCDF (cm$^{-1}$) | HFR (cm$^{-1}$) |
|----|---|---|---|---|---|
| 1  | 4s$^2$ 4p$^{*\,2}$ | 0 | $^3$P$_0^e$ | 0 | 0 |
| 2  | 4s$^2$ 4p$^*$ 4p | 1 | $^3$P$_1^e$ | 274985 | 274369 |
| 3  | 4s$^2$ 4p$^*$ 4p | 2 | $^1$D$_2^e$ | 300258 | 298308 |
| 4  | 4s$^2$ 4p$^2$ | 2 | $^3$P$_2^e$ | 591556 | 590529 |
| 5  | 4s$^2$ 4p$^2$ | 0 | $^1$S$_0^e$ | 642768 | 642538 |
| 6  | 4s 4p$^{*\,2}$ 4p | 2 | $^5$S$_2^o$ | 815001 | 820156 |
| 7  | 4s 4p$^{*\,2}$ 4p | 1 | $^3$D$_1^o$ | 912576 | 913375 |
| 8  | 4s 4p$^2$ 4p$^*$ | 2 | $^3$D$_2^o$ | 1074337 | 1081955 |
| 9  | 4s 4p$^2$ 4p$^*$ | 3 | $^3$D$_3^o$ | 1129853 | 1136067 |
| 10 | 4s 4p$^2$ 4p$^*$ | 0 | $^3$P$_0^o$ | 1177304 | 1173571 |
| 11 | 4s$^2$ 4p$^*$ 4d$^*$ | 2 | $^3$F$_2^o$ | 1187477 | 1191061 |
| 12 | 4s 4p$^2$ 4p$^*$ | 1 | $^3$P$_1^o$ | 1204821 | 1204693 |
| 13 | 4s 4p$^2$ 4p$^*$ | 2 | $^1$D$_2^o$ | 1230972 | 1228406 |
| 14 | 4s 4p$^2$ 4p$^*$ | 1 | $^3$S$_1^o$ | 1235868 | 1234354 |
| 15 | 4s$^2$ 4p$^*$ 4d$^*$ | 1 | $^3$D$_1^o$ | 1273388 | 1272739 |
| 16 | 4s$^2$ 4p$^*$ 4d | 3 | $^3$F$_3^o$ | 1299914 | 1298395 |
| 17 | 4s$^2$ 4p$^*$ 4d | 2 | $^3$D$_2^o$ | 1304232 | 1298424 |
| 18 | 4s 4p$^3$ | 2 | $^3$P$_2^o$ | 1463609 | 1463125 |
| 19 | 4s$^2$ 4p 4d$^*$ | 2 | $^1$D$_2^o$ | 1506402 | 1526162 |
| 20 | 4s$^2$ 4p 4d$^*$ | 0 | $^3$P$_0^o$ | 1542543 | 1544624 |
| 21 | 4s$^2$ 4p 4d | 4 | $^3$F$_4^o$ | 1545983 | 1551524 |
| 22 | 4s 4p$^3$ | 1 | $^1$P$_1^o$ | 1550772 | 1551224 |
| 23 | 4s$^2$ 4p 4d$^*$ | 1 | $^3$P$_1^o$ | 1551523 | 1551226 |
| 24 | 4s$^2$ 4p 4d$^*$ | 3 | $^3$D$_3^o$ | 1551599 | 1553577 |
| 25 | 4s$^2$ 4p 4d | 2 | $^3$P$_2^o$ | 1578874 | 1592126 |
| 26 | 4s$^2$ 4p 4d | 3 | $^1$F$_3^o$ | 1662685 | 1659057 |
| 27 | 4s$^2$ 4p 4d | 1 | $^1$P$_1^o$ | 1671326 | 1668433 |
| 28 | 4p$^{*\,2}$ 4p$^2$ | 2 | $^3$P$_2^e$ | 1802109 | 1807270 |
| 29 | 4p$^{*\,2}$ 4p$^2$ | 0 | $^1$S$_0^e$ | 1853301 | 1854856 |
| 30 | 4p$^*$ 4p$^3$ | 1 | $^3$P$_1^e$ | 2086241 | 2097815 |
| 31 | 4p$^*$ 4p$^3$ | 2 | $^1$D$_2^e$ | 2118487 | 2123796 |
| 32 | 4p$^4$ | 0 | $^3$P$_0^e$ | 2424302 | 2434492 |

Table (5): **MCDF** and **HFR** energy levels (in cm$^{-1}$) of **4s$^2$4p$^2$**, **4s4p$^3$**, **4s$^2$4p4d**, and **4p$^4$** configurations for **Eu XXXII** ion.

| | JJ-coupling | J | LS | MCDF (cm$^{-1}$) | HFR (cm$^{-1}$) |
|---|---|---|---|---|---|
| 1  | 4s$^2$ 4p$^{*\,2}$ | 0 | $^3P_0^e$ | 0 | 0 |
| 2  | 4s$^2$ 4p$^*$ 4p | 1 | $^3P_1^e$ | 301522 | 302192 |
| 3  | 4s$^2$ 4p$^*$ 4p | 2 | $^1D_2^e$ | 327601 | 326786 |
| 4  | 4s$^2$ 4p$^2$ | 2 | $^3P_2^e$ | 646099 | 647308 |
| 5  | 4s$^2$ 4p$^2$ | 0 | $^1S_0^e$ | 698460 | 700465 |
| 6  | 4s 4p$^{*\,2}$ 4p | 2 | $^5S_2^o$ | 848063 | 866736 |
| 7  | 4s 4p$^{*\,2}$ 4p | 1 | $^3D_1^o$ | 959882 | 961948 |
| 8  | 4s 4p$^2$ 4p$^*$ | 2 | $^3D_2^o$ | 1144514 | 1154272 |
| 9  | 4s 4p$^2$ 4p$^*$ | 3 | $^3D_3^o$ | 1202520 | 1210284 |
| 10 | 4s$^2$ 4p$^*$ 4d$^*$ | 2 | $^3F_2^o$ | 1241426 | 1245740 |
| 11 | 4s 4p$^2$ 4p$^*$ | 0 | $^3P_0^o$ | 1252957 | 1248700 |
| 12 | 4s 4p$^2$ 4p$^*$ | 1 | $^3P_1^o$ | 1280227 | 1281373 |
| 13 | 4s 4p$^2$ 4p$^*$ | 2 | $^1D_2^o$ | 1306780 | 1305303 |
| 14 | 4s 4p$^2$ 4p$^*$ | 1 | $^3S_1^o$ | 1310344 | 1309965 |
| 15 | 4s$^2$ 4p$^*$ 4d$^*$ | 1 | $^3D_1^o$ | 1330091 | 1329585 |
| 16 | 4s$^2$ 4p$^*$ 4d | 2 | $^3D_2^o$ | 1352210 | 1360416 |
| 17 | 4s$^2$ 4p$^*$ 4d | 3 | $^3F_3^o$ | 1362383 | 1361208 |
| 18 | 4s 4p$^3$ | 2 | $^3P_2^o$ | 1563290 | 1565394 |
| 19 | 4s$^2$ 4p 4d$^*$ | 2 | $^1D_2^o$ | 1591064 | 1610584 |
| 20 | 4s$^2$ 4p 4d$^*$ | 0 | $^3P_0^o$ | 1625781 | 1629229 |
| 21 | 4s$^2$ 4p 4d | 4 | $^3F_4^o$ | 1632600 | 1640921 |
| 22 | 4s$^2$ 4p 4d$^*$ | 1 | $^3P_1^o$ | 1635036 | 1636119 |
| 23 | 4s$^2$ 4p 4d$^*$ | 3 | $^3D_3^o$ | 1635131 | 1638483 |
| 24 | 4s 4p$^3$ | 1 | $^1P_1^o$ | 1653206 | 1655947 |
| 25 | 4s$^2$ 4p 4d | 2 | $^3P_2^o$ | 1666883 | 1682038 |
| 26 | 4s$^2$ 4p 4d | 3 | $^1F_3^o$ | 1751952 | 1750000 |
| 27 | 4s$^2$ 4p 4d | 1 | $^1P_1^o$ | 1761715 | 1760312 |
| 28 | 4p$^{*2}$ 4p$^2$ | 2 | $^3P_2^e$ | 1896199 | 1901961 |
| 29 | 4p$^{*2}$ 4p$^2$ | 0 | $^1S_0^e$ | 1948890 | 1950876 |
| 30 | 4p$^*$ 4p$^3$ | 1 | $^3P_1^e$ | 2207530 | 2220831 |
| 31 | 4p$^*$ 4p$^3$ | 2 | $^1D_2^e$ | 2240442 | 2247383 |
| 32 | 4p$^4$ | 0 | $^3P_0^e$ | 2573953 | 2586788 |

Table (6): **MCDF** wavelengths (in Å), transition probabilities (in sec$^{-1}$), and oscillator strengths for lines of **Pr XXVIII** ion.

| Lower | Upper | λ (Å) | $A_L$ (sec$^{-1}$) | $A_V$ (sec$^{-1}$) | $f_L$ | Lower | Upper | λ (Å) | $A_L$ (sec$^{-1}$) | $A_V$ (sec$^{-1}$) | $f_L$ |
|---|---|---|---|---|---|---|---|---|---|---|---|
| 4s$^2$ 4p$^2$ $^3P_0^e$ | 4s 4p$^3$ $^3D_1^o$ | 129.180 | 1.187E+11 | 1.117E+11 | 9.247E-01 | 4s$^2$ 4p$^2$ $^3P_2^e$ | 4s$^2$ 4p 4d $^3P_1^o$ | 114.522 | 3.191E+10 | 3.387E+10 | 3.765E-02 |
| | 4s 4p$^3$ $^3P_1^o$ | 100.829 | 6.743E+08 | 1.879E+08 | 3.083E-03 | | 4s$^2$ 4p 4d $^3P_2^o$ | 110.680 | 1.399E+11 | 1.534E+11 | 2.569E-01 |
| | 4s 4p$^3$ $^3S_1^o$ | 97.544 | 1.977E+10 | 1.119E+10 | 8.459E-02 | | 4s$^2$ 4p 4d $^3F_2^o$ | 169.997 | 1.749E+08 | 6.057E+08 | 7.576E-04 |
| | 4s 4p$^3$ $^1P_1^o$ | 78.967 | 1.293E+07 | 1.876E+05 | 3.627E-05 | | 4s$^2$ 4p 4d $^3F_3^o$ | 147.540 | 2.717E+08 | 6.083E+08 | 1.242E-03 |
| | 4s$^2$ 4p 4d $^3D_1^o$ | 89.995 | 3.698E+11 | 3.747E+11 | 1.347E+00 | | 4s$^2$ 4p 4d $^1F_3^o$ | 103.329 | 3.689E+11 | 3.194E+11 | 8.266E-01 |
| | 4s$^2$ 4p 4d $^3P_1^o$ | 75.947 | 2.096E+08 | 2.143E+07 | 5.438E-04 | | 4s$^2$ 4p 4d $^1D_2^o$ | 117.779 | 7.518E+08 | 2.974E+09 | 1.564E-03 |
| | 4s$^2$ 4p 4d $^1P_1^o$ | 70.542 | 8.506E+08 | 2.328E+09 | 1.904E-03 | | 4s$^2$ 4p 4d $^1P_1^o$ | 102.661 | 9.460E+09 | 1.404E+10 | 8.969E-03 |
| 4s$^2$ 4p$^2$ $^3P_1^e$ | 4s 4p$^3$ $^3D_1^o$ | 174.258 | 7.611E+08 | 2.899E+07 | 3.465E-03 | 4s$^2$ 4p$^2$ $^1S_0^e$ | 4s 4p$^3$ $^1P_1^o$ | 129.021 | 8.805E+10 | 7.376E+10 | 6.592E-01 |
| | 4s 4p$^3$ $^3D_2^o$ | 147.883 | 4.383E+10 | 4.374E+10 | 2.395E-01 | | 4s 4p$^3$ $^3S_1^o$ | 187.303 | 9.511E+09 | 9.816E+09 | 1.501E-01 |
| | 4s 4p$^3$ $^3P_0^o$ | 130.234 | 9.799E+10 | 9.081E+10 | 8.305E-02 | | 4s 4p$^3$ $^3P_1^o$ | 199.804 | 3.576E+09 | 3.221E+09 | 6.420E-02 |
| | 4s 4p$^3$ $^3P_1^o$ | 126.338 | 1.215E+11 | 1.052E+11 | 2.907E-01 | | 4s 4p$^3$ $^3D_1^o$ | 353.576 | 1.962E+07 | 3.173E+08 | 1.103E-03 |
| | 4s 4p$^3$ $^3P_2^o$ | 101.272 | 4.787E+08 | 2.157E+08 | 1.227E-03 | | 4s$^2$ 4p 4d $^1P_1^o$ | 107.956 | 2.032E+11 | 2.388E+11 | 1.065E+00 |
| | 4s 4p$^3$ $^3S_1^o$ | 121.222 | 6.672E+10 | 4.831E+10 | 1.470E-01 | | 4s$^2$ 4p 4d $^3P_1^o$ | 121.152 | 1.448E+09 | 1.757E+09 | 9.557E-03 |
| | 4s 4p$^3$ $^5S_2^o$ | 209.834 | 6.435E+09 | 6.122E+09 | 7.079E-05 | | 4s$^2$ 4p 4d $^3D_1^o$ | 161.319 | 2.499E+07 | 7.027E+06 | 2.925E-04 |
| | 4s 4p$^3$ $^1D_2^o$ | 122.660 | 7.127E+09 | 7.979E+09 | 2.679E-02 | 4s 4p$^3$ $^1D_2^o$ | 4p$^4$ $^1D_2^e$ | 132.170 | 9.695E+10 | 1.536E+11 | 2.539E-01 |
| | 4s 4p$^3$ $^1P_1^o$ | 93.799 | 2.056E+10 | 1.044E+10 | 2.712E-02 | | 4p$^4$ $^3P_1^e$ | 137.675 | 4.097E+10 | 4.928E+10 | 6.985E-02 |
| | 4s$^2$ 4p 4d $^3D_1^o$ | 109.778 | 3.094E+10 | 2.686E+10 | 5.590E-02 | | 4p$^4$ $^3P_2^e$ | 194.885 | 1.305E+10 | 2.210E+10 | 7.432E-02 |
| | 4s$^2$ 4p 4d $^3D_2^o$ | 108.098 | 1.514E+11 | 1.489E+11 | 4.421E-01 | 4s 4p$^3$ $^1P_1^o$ | 4p$^4$ $^1D_2^e$ | 197.722 | 1.974E+10 | 3.171E+10 | 1.928E-01 |
| | 4s$^2$ 4p 4d $^3P_0^o$ | 89.894 | 2.690E+11 | 2.904E+11 | 1.086E-01 | | 4p$^4$ $^1S_0^e$ | 323.695 | 4.700E+07 | 1.374E+09 | 2.461E-04 |
| | 4s$^2$ 4p 4d $^3P_1^o$ | 89.570 | 2.224E+11 | 2.236E+11 | 2.675E-01 | | 4p$^4$ $^3P_0^e$ | 135.276 | 2.718E+11 | 4.217E+11 | 2.485E-01 |
| | 4s$^2$ 4p 4d $^3P_2^o$ | 87.202 | 3.005E+10 | 2.387E+10 | 5.709E-02 | | 4p$^4$ $^3P_1^e$ | 210.302 | 5.944E+09 | 1.286E+10 | 3.941E-02 |
| | 4s$^2$ 4p 4d $^3F_2^o$ | 120.264 | 1.568E+09 | 1.310E+09 | 5.666E-03 | | 4p$^4$ $^3P_2^e$ | 381.273 | 1.598E+07 | 6.688E+08 | 5.805E-04 |
| | 4s$^2$ 4p 4d $^1D_2^o$ | 91.549 | 1.631E+11 | 1.535E+11 | 3.415E-01 | 4s 4p$^3$ $^3D_1^o$ | 4p$^4$ $^1D_2^e$ | 100.200 | 1.712E+07 | 1.660E+07 | 4.296E-05 |
| | 4s$^2$ 4p 4d $^1P_1^o$ | 82.146 | 7.497E+09 | 6.295E+09 | 7.585E-03 | | 4p$^4$ $^1S_0^e$ | 124.816 | 1.296E+11 | 1.268E+11 | 1.009E-01 |
| 4s$^2$ 4p$^2$ $^1D_2^e$ | 4s 4p$^3$ $^1P_1^o$ | 96.028 | 2.926E+04 | 1.500E+07 | 2.427E-08 | | 4p$^4$ $^3P_0^e$ | 81.203 | 3.511E+08 | 5.566E+07 | 1.157E-04 |
| | 4s 4p$^3$ $^1D_2^o$ | 126.498 | 1.979E+11 | 1.664E+11 | 4.748E-01 | | 4p$^4$ $^3P_1^e$ | 103.332 | 2.698E+10 | 1.659E+10 | 4.320E-02 |
| | 4s 4p$^3$ $^3D_1^o$ | 182.109 | 9.727E+09 | 4.736E+09 | 2.902E-02 | | 4p$^4$ $^3P_2^e$ | 132.534 | 5.844E+06 | 6.301E+10 | 2.565E-01 |
| | 4s 4p$^3$ $^3D_2^o$ | 153.500 | 1.215E+09 | 5.468E+08 | 4.290E-03 | 4s 4p$^3$ $^3D_2^o$ | 4p$^4$ $^1D_2^e$ | 111.650 | 1.338E+09 | 6.009E+08 | 2.501E-03 |
| | 4s 4p$^3$ $^3D_3^o$ | 141.523 | 3.228E+10 | 3.280E+10 | 1.357E-01 | | 4p$^4$ $^3P_1^e$ | 115.553 | 5.625E+10 | 4.577E+10 | 6.756E-02 |
| | 4s 4p$^3$ $^3P_1^o$ | 130.414 | 1.119E+10 | 1.172E+10 | 1.712E-02 | | 4p$^4$ $^3P_2^e$ | 153.332 | 2.906E+10 | 2.646E+10 | 1.024E-01 |
| | 4s 4p$^3$ $^3P_2^o$ | 103.874 | 1.633E+10 | 1.035E+10 | 2.642E-02 | 4s 4p$^3$ $^3D_3^o$ | 4p$^4$ $^1D_2^e$ | 118.973 | 4.609E+10 | 3.859E+10 | 6.986E-02 |
| | 4s 4p$^3$ $^5S_2^o$ | 221.324 | 3.306E+09 | 2.284E+09 | 2.428E-02 | | 4p$^4$ $^3P_2^e$ | 167.491 | 4.637E+10 | 3.637E+10 | 1.393E-01 |
| | 4s 4p$^3$ $^3S_1^o$ | 124.970 | 1.894E+11 | 1.388E+11 | 2.660E-01 | 4s 4p$^3$ $^3P_0^o$ | 4p$^4$ $^3P_1^e$ | 129.238 | 3.151E+10 | 3.285E+10 | 2.367E-01 |
| | 4s$^2$ 4p 4d $^3D_1^o$ | 112.843 | 3.452E+09 | 5.380E+09 | 3.954E-03 | 4s 4p$^3$ $^3P_1^o$ | 4p$^4$ $^1D_2^e$ | 128.149 | 3.369E+10 | 3.796E+10 | 1.382E-01 |
| | 4s$^2$ 4p 4d $^3D_2^o$ | 111.068 | 2.521E+10 | 3.485E+10 | 4.663E-02 | | 4p$^4$ $^1S_0^e$ | 171.376 | 5.194E+10 | 4.175E+10 | 7.624E-02 |
| | 4s$^2$ 4p 4d $^3D_3^o$ | 91.092 | 3.611E+11 | 3.180E+11 | 6.289E-01 | | 4p$^4$ $^3P_0^e$ | 98.638 | 4.358E+10 | 2.710E+10 | 2.119E-02 |
| | 4s$^2$ 4p 4d $^3P_1^o$ | 91.600 | 5.090E+10 | 5.438E+10 | 3.842E-02 | | 4p$^4$ $^3P_1^e$ | 133.318 | 9.088E+09 | 1.086E+10 | 2.422E-02 |
| | 4s$^2$ 4p 4d $^3P_2^o$ | 89.125 | 5.117E+10 | 4.489E+10 | 6.094E-02 | | 4p$^4$ $^3P_2^e$ | 186.269 | 6.086E+09 | 6.639E+09 | 5.276E-02 |
| | 4s$^2$ 4p 4d $^3F_2^o$ | 123.952 | 1.487E+10 | 1.885E+10 | 3.426E-02 | 4s 4p$^3$ $^3P_2^o$ | 4p$^4$ $^1D_2^e$ | 171.109 | 4.407E+10 | 7.022E+10 | 1.934E-01 |
| | 4s$^2$ 4p 4d $^3F_3^o$ | 111.570 | 8.025E+10 | 6.713E+10 | 2.097E-01 | | 4p$^4$ $^3P_1^e$ | 180.450 | 2.050E+10 | 2.502E+10 | 6.003E-02 |
| | 4s$^2$ 4p 4d $^1D_2^o$ | 93.671 | 8.578E+10 | 9.836E+10 | 1.128E-01 | | 4p$^4$ $^3P_2^e$ | 293.304 | 1.014E+08 | 1.324E+09 | 1.308E-03 |
| | 4s$^2$ 4p 4d $^1P_1^o$ | 83.850 | 1.667E+09 | 3.337E+09 | 1.054E-03 | 4s 4p$^3$ $^3S_1^o$ | 4p$^4$ $^1D_2^e$ | 133.880 | 4.574E+09 | 2.359E+09 | 2.048E-02 |
| | 4s$^2$ 4p 4d $^1F_3^o$ | 84.296 | 1.197E+11 | 9.138E+10 | 1.786E-01 | | 4p$^4$ $^1S_0^e$ | 181.782 | 5.211E+10 | 1.011E+11 | 8.605E-02 |
| 4s$^2$ 4p$^2$ $^3P_2^e$ | 4s 4p$^3$ $^3D_1^o$ | 302.477 | 1.871E+08 | 1.910E+08 | 1.540E-03 | | 4p$^4$ $^3P_0^e$ | 101.998 | 5.534E+08 | 2.656E+09 | 2.877E-04 |
| | 4s 4p$^3$ $^3D_2^o$ | 230.974 | 1.095E+09 | 9.252E+08 | 8.756E-03 | | 4p$^4$ $^3P_1^e$ | 139.532 | 9.398E+10 | 1.401E+11 | 2.743E-01 |
| | 4s 4p$^3$ $^3D_3^o$ | 204.885 | 1.414E+10 | 1.480E+10 | 1.246E-01 | | 4p$^4$ $^3P_2^e$ | 198.627 | 1.207E+10 | 2.605E+10 | 1.189E-01 |
| | 4s 4p$^3$ $^3P_1^o$ | 182.392 | 6.678E+08 | 3.643E+05 | 1.998E-03 | 4s 4p$^3$ $^5S_2^o$ | 4p$^4$ $^1D_2^e$ | 91.299 | 3.058E+09 | 1.999E+09 | 3.822E-03 |
| | 4s 4p$^3$ $^3P_2^o$ | 134.375 | 1.376E+11 | 1.182E+11 | 3.724E-01 | | 4p$^4$ $^3P_1^e$ | 93.893 | 4.614E+08 | 7.433E+08 | 3.659E-04 |
| | 4s 4p$^3$ $^3S_1^o$ | 171.918 | 9.040E+09 | 6.188E+09 | 2.403E-02 | | 4p$^4$ $^3P_2^e$ | 117.396 | 6.066E+10 | 5.749E+10 | 1.253E-01 |
| | 4s 4p$^3$ $^5S_2^o$ | 428.618 | 2.441E+07 | 3.412E+07 | 6.722E-04 | | | | | | |
| | 4s 4p$^3$ $^1D_2^o$ | 174.823 | 8.064E+09 | 4.483E+09 | 3.695E-02 | | | | | | |
| | 4s 4p$^3$ $^1P_1^o$ | 121.529 | 2.317E+11 | 1.743E+11 | 3.078E-01 | | | | | | |
| | 4s$^2$ 4p 4d $^3D_1^o$ | 149.774 | 3.035E+07 | 1.227E+07 | 6.124E-05 | | | | | | |
| | 4s$^2$ 4p 4d $^3D_2^o$ | 146.664 | 2.723E+08 | 3.819E+08 | 8.782E-04 | | | | | | |
| | 4s$^2$ 4p 4d $^3D_3^o$ | 113.730 | 2.380E+10 | 2.593E+10 | 6.462E-02 | | | | | | |

Table (7): **MCDF** wavelengths (in Å), transition probabilities (in sec$^{-1}$), and oscillator strengths for lines of **Nd XXIX** ion.

| Lower | Upper | MCDF code λ (Å) | $A_L$ (sec$^{-1}$) | $A_V$ (sec$^{-1}$) | $f_L$ | Lower | Upper | MCDF code λ (Å) | $A_L$ (sec$^{-1}$) | $A_V$ (sec$^{-1}$) | $f_L$ |
|---|---|---|---|---|---|---|---|---|---|---|---|
| $4s^2\,4p^2\ ^3P_0^e$ | $4s\,4p^3\ ^3D_1^o$ | 122.413 | 1.302E+11 | 1.257E+11 | 9.343E-01 | $4s^2\,4p^2\ ^3P_2^e$ | $4s^2\,4p\,4d\ ^3P_1^o$ | 110.860 | 3.347E+10 | 3.550E+10 | 3.700E-02 |
| | $4s\,4p^3\ ^3P_1^o$ | 94.677 | 6.522E+08 | 1.681E+08 | 2.629E-03 | | $4s^2\,4p\,4d\ ^3P_2^o$ | 106.853 | 1.485E+11 | 1.633E+11 | 2.543E-01 |
| | $4s\,4p^3\ ^3S_1^o$ | 91.896 | 2.011E+10 | 1.130E+10 | 7.638E-02 | | $4s^2\,4p\,4d\ ^3F_2^o$ | 168.686 | 1.493E+08 | 5.566E+08 | 6.370E-04 |
| | $4s\,4p^3\ ^1P_1^o$ | 73.950 | 1.037E+07 | 4.517E+04 | 2.550E-05 | | $4s^2\,4p\,4d\ ^3F_3^o$ | 144.985 | 2.721E+08 | 6.278E+08 | 1.200E-03 |
| | $4s^2\,4p\,4d\ ^3D_1^o$ | 86.011 | 4.008E+11 | 4.064E+11 | 1.333E+00 | | $4s^2\,4p\,4d\ ^1F_3^o$ | 99.872 | 3.893E+11 | 3.382E+11 | 8.151E-01 |
| | $4s^2\,4p\,4d\ ^3P_1^o$ | 71.958 | 2.015E+08 | 3.047E+07 | 4.691E-04 | | $4s^2\,4p\,4d\ ^1D_2^o$ | 113.972 | 3.390E+08 | 2.063E+09 | 6.602E-04 |
| | $4s^2\,4p\,4d\ ^1P_1^o$ | 66.830 | 8.323E+08 | 2.353E+09 | 1.672E-03 | | $4s^2\,4p\,4d\ ^1P_1^o$ | 99.139 | 1.024E+10 | 1.506E+10 | 9.051E-03 |
| $4s^2\,4p^2\ ^3P_1^e$ | $4s\,4p^3\ ^3D_1^o$ | 167.996 | 1.010E+09 | 7.329E+07 | 4.272E-03 | $4s^2\,4p^2\ ^1S_0^e$ | $4s\,4p^3\ ^1P_1^o$ | 122.595 | 9.629E+08 | 8.102E+10 | 6.509E-01 |
| | $4s\,4p^3\ ^3D_2^o$ | 140.153 | 4.721E+10 | 4.720E+10 | 2.317E-01 | | $4s\,4p^3\ ^3S_1^o$ | 181.286 | 1.008E+10 | 1.050E+10 | 1.490E-01 |
| | $4s\,4p^3\ ^3P_0^o$ | 123.608 | 1.088E+11 | 1.010E+11 | 8.304E-02 | | $4s\,4p^3\ ^3P_1^o$ | 192.437 | 4.069E+09 | 3.651E+09 | 6.777E-02 |
| | $4s\,4p^3\ ^3P_1^o$ | 119.823 | 1.362E+11 | 1.182E+11 | 2.932E-01 | | $4s\,4p^3\ ^3D_1^o$ | 356.709 | 1.566E+07 | 3.028E+08 | 8.963E-04 |
| | $4s\,4p^3\ ^3P_2^o$ | 95.322 | 4.542E+08 | 1.908E+08 | 1.031E-03 | | $4s^2\,4p\,4d\ ^1P_1^o$ | 104.191 | 2.137E+11 | 2.508E+11 | 1.043E+00 |
| | $4s\,4p^3\ ^3S_1^o$ | 115.403 | 7.131E+10 | 5.208E+10 | 1.424E-01 | | $4s^2\,4p\,4d\ ^3P_1^o$ | 117.215 | 1.717E+09 | 2.078E+09 | 1.061E-02 |
| | $4s\,4p^3\ ^5S_2^o$ | 201.836 | 7.386E+09 | 7.098E+09 | 7.518E-02 | | $4s^2\,4p\,4d\ ^3D_1^o$ | 159.726 | 2.141E+07 | 5.115E+06 | 2.457E-04 |
| | $4s\,4p^3\ ^1D_2^o$ | 116.162 | 8.404E+09 | 9.304E+09 | 2.834E-02 | $4s\,4p^3\ ^1D_2^o$ | $4p^4\ ^1D_2^e$ | 125.915 | 1.096E+11 | 1.754E+11 | 2.605E-01 |
| | $4s\,4p^3\ ^1P_1^o$ | 88.448 | 2.124E+10 | 1.069E+10 | 2.491E-02 | | $4p^4\ ^3P_1^e$ | 131.015 | 4.326E+10 | 5.262E+10 | 6.680E-02 |
| | $4s^2\,4p\,4d\ ^3D_1^o$ | 106.272 | 3.215E+10 | 2.795E+10 | 5.444E-02 | | $4p^4\ ^3P_2^e$ | 189.362 | 1.350E+10 | 2.350E+10 | 7.257E-02 |
| | $4s^2\,4p\,4d\ ^3D_2^o$ | 104.301 | 1.619E+11 | 1.593E+11 | 4.401E-01 | $4s\,4p^3\ ^1P_1^o$ | $4p^4\ ^1D_2^e$ | 190.677 | 2.052E+10 | 3.300E+10 | 1.864E-01 |
| | $4s^2\,4p\,4d\ ^3P_0^o$ | 85.927 | 2.915E+11 | 3.145E+11 | 1.076E-01 | | $4p^4\ ^1S_0^e$ | 326.238 | 3.817E+07 | 1.321E+09 | 2.030E-04 |
| | $4s^2\,4p\,4d\ ^3P_1^o$ | 85.613 | 2.414E+11 | 2.427E+11 | 2.652E-01 | | $4p^4\ ^3P_0^e$ | 128.149 | 3.061E+11 | 4.683E+11 | 2.512E-01 |
| | $4s^2\,4p\,4d\ ^3P_2^o$ | 83.204 | 2.909E+10 | 2.299E+10 | 5.032E-02 | | $4p^4\ ^3P_1^e$ | 202.622 | 6.217E+09 | 1.336E+10 | 3.827E-02 |
| | $4s^2\,4p\,4d\ ^3F_2^o$ | 116.439 | 1.675E+09 | 1.402E+09 | 5.675E-03 | | $4p^4\ ^3P_2^e$ | 387.073 | 1.223E+07 | 6.277E+08 | 4.579E-04 |
| | $4s^2\,4p\,4d\ ^1D_2^o$ | 87.457 | 1.759E+11 | 1.654E+11 | 3.362E-01 | $4s\,4p^3\ ^3D_1^o$ | $4p^4\ ^1D_2^e$ | 94.358 | 1.759E+07 | 1.787E+07 | 3.913E-05 |
| | $4s^2\,4p\,4d\ ^1P_1^o$ | 78.450 | 7.853E+09 | 6.574E+09 | 7.246E-03 | | $4p^4\ ^1S_0^e$ | 118.782 | 1.479E+11 | 1.453E+11 | 1.043E-01 |
| $4s^2\,4p^2\ ^1D_2^e$ | $4s\,4p^3\ ^1P_1^o$ | 90.492 | 2.999E+05 | 1.497E+07 | 2.209E-07 | | $4p^4\ ^3P_0^e$ | 76.006 | 3.163E+08 | 4.625E+07 | 9.130E-05 |
| | $4s\,4p^3\ ^1D_2^o$ | 119.713 | 2.229E+11 | 1.866E+11 | 4.790E-01 | | $4p^4\ ^3P_1^e$ | 97.193 | 2.762E+10 | 1.673E+10 | 3.911E-02 |
| | $4s\,4p^3\ ^3D_1^o$ | 175.525 | 1.120E+10 | 5.549E+09 | 3.105E-02 | | $4p^4\ ^3P_2^e$ | 125.992 | 6.685E+10 | 7.223E+10 | 2.651E-01 |
| | $4s\,4p^3\ ^3D_2^o$ | 145.355 | 1.288E+09 | 5.841E+08 | 4.081E-03 | $4s\,4p^3\ ^3D_2^o$ | $4p^4\ ^1D_2^e$ | 106.209 | 1.245E+09 | 5.277E+08 | 2.106E-03 |
| | $4s\,4p^3\ ^3D_3^o$ | 134.064 | 3.550E+10 | 3.608E+10 | 1.339E-01 | | $4p^4\ ^3P_1^e$ | 109.814 | 6.294E+10 | 5.158E+10 | 6.827E-02 |
| | $4s\,4p^3\ ^3P_1^o$ | 123.605 | 1.259E+10 | 1.310E+10 | 1.730E-02 | | $4p^4\ ^3P_2^e$ | 148.051 | 2.937E+10 | 2.681E+10 | 9.650E-02 |
| | $4s\,4p^3\ ^3P_2^o$ | 97.700 | 1.580E+10 | 9.920E+09 | 2.260E-02 | $4s\,4p^3\ ^3D_3^o$ | $4p^4\ ^1D_2^e$ | 113.173 | 5.171E+10 | 4.349E+10 | 7.093E-02 |
| | $4s\,4p^3\ ^5S_2^o$ | 212.804 | 3.805E+09 | 2.646E+09 | 2.584E-02 | | $4p^4\ ^3P_2^e$ | 161.941 | 4.835E+10 | 3.801E+10 | 1.358E-01 |
| | $4s\,4p^3\ ^3S_1^o$ | 118.907 | 2.052E+11 | 1.517E+11 | 2.610E-01 | $4s\,4p^3\ ^3P_0^o$ | $4p^4\ ^3P_1^e$ | 122.680 | 3.500E+10 | 3.646E+10 | 2.369E-01 |
| | $4s^2\,4p\,4d\ ^3D_1^o$ | 109.236 | 3.618E+09 | 5.644E+09 | 3.883E-03 | $4s\,4p^3\ ^3P_1^o$ | $4p^4\ ^1D_2^e$ | 121.879 | 3.909E+10 | 4.400E+10 | 1.451E-01 |
| | $4s^2\,4p\,4d\ ^3D_2^o$ | 107.154 | 2.597E+10 | 3.587E+10 | 4.470E-02 | | $4p^4\ ^1S_0^e$ | 165.957 | 5.434E+10 | 4.389E+10 | 7.479E-02 |
| | $4s^2\,4p\,4d\ ^3D_3^o$ | 87.056 | 3.906E+11 | 3.440E+11 | 6.213E-01 | | $4p^4\ ^3P_0^e$ | 92.904 | 4.531E+10 | 2.789E+10 | 1.954E-02 |
| | $4s^2\,4p\,4d\ ^3P_1^o$ | 87.527 | 5.455E+10 | 5.830E+10 | 3.759E-02 | | $4p^4\ ^3P_1^e$ | 126.651 | 1.067E+10 | 1.271E+10 | 2.566E-02 |
| | $4s^2\,4p\,4d\ ^3P_2^o$ | 85.010 | 5.009E+10 | 4.345E+10 | 5.427E-02 | | $4p^4\ ^3P_2^e$ | 180.378 | 6.274E+09 | 6.860E+09 | 5.100E-02 |
| | $4s^2\,4p\,4d\ ^3F_2^o$ | 120.007 | 1.578E+10 | 1.999E+10 | 3.406E-02 | $4s\,4p^3\ ^3P_2^o$ | $4p^4\ ^1D_2^e$ | 165.024 | 4.444E+10 | 7.028E+10 | 1.814E-01 |
| | $4s^2\,4p\,4d\ ^3F_3^o$ | 107.504 | 8.976E+10 | 7.516E+10 | 2.177E-01 | | $4p^4\ ^3P_1^e$ | 173.897 | 2.251E+10 | 2.750E+10 | 6.124E-02 |
| | $4s^2\,4p\,4d\ ^1D_2^o$ | 89.455 | 9.748E+10 | 1.113E+11 | 1.169E-01 | | $4p^4\ ^3P_2^e$ | 294.229 | 7.751E+07 | 1.200E+09 | 1.006E-03 |
| | $4s^2\,4p\,4d\ ^1P_1^o$ | 80.054 | 1.685E+09 | 3.404E+09 | 9.713E-04 | $4s\,4p^3\ ^3S_1^o$ | $4p^4\ ^1D_2^e$ | 126.819 | 5.185E+09 | 2.756E+09 | 2.084E-02 |
| | $4s^2\,4p\,4d\ ^1F_3^o$ | 80.532 | 1.209E+11 | 9.214E+10 | 1.646E-01 | | $4p^4\ ^1S_0^e$ | 175.254 | 5.444E+10 | 1.050E+11 | 8.355E-02 |
| $4s^2\,4p^2\ ^3P_2^e$ | $4s\,4p^3\ ^3D_1^o$ | 303.721 | 1.605E+08 | 1.719E+08 | 1.331E-03 | | $4p^4\ ^3P_0^e$ | 95.747 | 5.071E+08 | 2.620E+09 | 2.323E-04 |
| | $4s\,4p^3\ ^3D_2^o$ | 223.462 | 1.232E+09 | 1.043E+09 | 9.220E-03 | | $4p^4\ ^3P_1^e$ | 131.995 | 1.059E+11 | 1.558E+11 | 2.767E-01 |
| | $4s\,4p^3\ ^3D_3^o$ | 197.847 | 1.505E+10 | 1.577E+10 | 1.236E-01 | | $4p^4\ ^3P_2^e$ | 191.415 | 1.226E+10 | 2.646E+10 | 1.122E-01 |
| | $4s\,4p^3\ ^3P_1^o$ | 175.883 | 9.028E+08 | 5.117E+06 | 2.512E-03 | $4s\,4p^3\ ^5S_2^o$ | $4p^4\ ^1D_2^e$ | 86.237 | 3.365E+09 | 2.173E+09 | 3.752E-03 |
| | $4s\,4p^3\ ^3P_2^o$ | 127.701 | 1.486E+11 | 1.286E+11 | 3.634E-01 | | $4p^4\ ^3P_1^e$ | 88.599 | 3.594E+08 | 6.576E+08 | 2.538E-04 |
| | $4s\,4p^3\ ^3S_1^o$ | 166.521 | 9.670E+09 | 6.661E+09 | 2.412E-02 | | $4p^4\ ^3P_2^e$ | 111.919 | 7.156E+10 | 6.773E+10 | 1.344E-01 |
| | $4s\,4p^3\ ^5S_2^o$ | 435.830 | 1.998E+07 | 2.908E+07 | 5.689E-04 | | | | | | |
| | $4s\,4p^3\ ^1D_2^o$ | 168.107 | 9.194E+09 | 5.127E+09 | 3.895E-02 | | | | | | |
| | $4s\,4p^3\ ^1P_1^o$ | 115.660 | 2.542E+11 | 1.928E+11 | 3.059E-01 | | | | | | |
| | $4s^2\,4p\,4d\ ^3D_1^o$ | 148.152 | 2.702E+07 | 9.785E+06 | 5.334E-05 | | | | | | |
| | $4s^2\,4p\,4d\ ^3D_2^o$ | 144.349 | 2.311E+08 | 3.181E+08 | 7.220E-04 | | | | | | |
| | $4s^2\,4p\,4d\ ^3D_3^o$ | 110.106 | 2.212E+10 | 2.442E+10 | 5.629E-02 | | | | | | |

Table (8): **MCDF** wavelengths (in Å), transition probabilities (in sec$^{-1}$), and oscillator strengths for lines of **Pm XXX** ion.

| Lower | Upper | MCDF code | | | |
|---|---|---|---|---|---|
| | | λ (Å) | $A_L$ (sec$^{-1}$) | $A_V$ (sec$^{-1}$) | $f_L$ |
| $4s^2\,4p^2$ $^3P_0^e$ | $4s\,4p^3\,^3D_1^o$ | 116.266 | 1.444E+11 | 1.579E+11 | 8.778E-01 |
| | $4s\,4p^3\,^3P_1^o$ | 88.703 | 2.483E+09 | 2.990E+09 | 8.788E-03 |
| | $4s\,4p^3\,^3S_1^o$ | 86.246 | 2.055E+10 | 1.000E+09 | 6.876E-02 |
| | $4s\,4p^3\,^1P_1^o$ | 69.056 | 2.454E+08 | 1.095E+08 | 5.264E-04 |
| | $4s^2\,4p\,4d\,^3D_1^o$ | 82.137 | 4.274E+11 | 4.366E+11 | 1.297E+00 |
| | $4s^2\,4p\,4d\,^3P_1^o$ | 67.996 | 9.267E+07 | 6.380E+07 | 1.927E-04 |
| | $4s^2\,4p\,4d\,^1P_1^o$ | 63.147 | 1.260E+09 | 4.160E+08 | 2.259E-03 |
| $4s^2\,4p^2$ $^3P_1^e$ | $4s\,4p^3\,^3D_1^o$ | 163.782 | 1.382E+09 | 1.966E+08 | 5.559E-03 |
| | $4s\,4p^3\,^3D_2^o$ | 132.858 | 5.094E+10 | 5.101E+10 | 2.247E-01 |
| | $4s\,4p^3\,^3P_0^o$ | 117.377 | 1.207E+11 | 1.124E+11 | 8.309E-02 |
| | $4s\,4p^3\,^3P_1^o$ | 113.918 | 2.006E+11 | 1.727E+11 | 3.903E-01 |
| | $4s\,4p^3\,^3P_2^o$ | 89.749 | 4.305E+08 | 1.676E+08 | 8.665E-04 |
| | $4s\,4p^3\,^3S_1^o$ | 109.897 | 7.629E+10 | 5.618E+10 | 1.381E-01 |
| | $4s\,4p^3\,^5S_2^o$ | 194.195 | 8.315E+09 | 8.010E+09 | 7.835E-02 |
| | $4s\,4p^3\,^1D_2^o$ | 110.679 | 9.924E+09 | 1.091E+10 | 3.037E-02 |
| | $4s\,4p^3\,^1P_1^o$ | 83.433 | 2.191E+10 | 1.093E+10 | 2.286E-02 |
| | $4s^2\,4p\,4d\,^3D_1^o$ | 103.311 | 3.309E+10 | 2.902E+10 | 5.295E-02 |
| | $4s^2\,4p\,4d\,^3D_2^o$ | 101.108 | 1.706E+11 | 1.692E+11 | 4.357E-01 |
| | $4s^2\,4p\,4d\,^3P_0^o$ | 82.565 | 3.086E+11 | 3.362E+11 | 1.051E-01 |
| | $4s^2\,4p\,4d\,^3P_1^o$ | 81.891 | 2.618E+11 | 2.633E+11 | 2.632E-01 |
| | $4s^2\,4p\,4d\,^3P_2^o$ | 79.427 | 2.812E+10 | 2.210E+10 | 4.433E-02 |
| | $4s^2\,4p\,4d\,^3F_2^o$ | 113.020 | 1.731E+09 | 1.452E+09 | 5.525E-03 |
| | $4s^2\,4p\,4d\,^1D_2^o$ | 83.597 | 1.898E+11 | 1.784E+11 | 3.313E-01 |
| | $4s^2\,4p\,4d\,^1P_1^o$ | 74.958 | 8.195E+09 | 6.838E+09 | 6.903E-03 |
| $4s^2\,4p^2$ $^1D_2^e$ | $4s\,4p^3\,^1P_1^o$ | 85.307 | 7.848E+05 | 1.519E+07 | 5.137E-07 |
| | $4s\,4p^3\,^1D_2^o$ | 114.000 | 2.470E+11 | 2.088E+11 | 4.812E-01 |
| | $4s\,4p^3\,^3D_1^o$ | 171.162 | 1.329E+10 | 7.214E+09 | 3.503E-02 |
| | $4s\,4p^3\,^3D_2^o$ | 137.673 | 1.361E+09 | 6.203E+08 | 3.866E-03 |
| | $4s\,4p^3\,^3D_3^o$ | 127.062 | 3.907E+10 | 3.972E+10 | 1.324E-01 |
| | $4s\,4p^3\,^3P_1^o$ | 117.440 | 5.132E+08 | 1.222E+09 | 6.367E-04 |
| | $4s\,4p^3\,^3P_2^o$ | 91.920 | 1.531E+10 | 9.531E+09 | 1.940E-02 |
| | $4s\,4p^3\,^5S_2^o$ | 204.657 | 4.336E+09 | 3.022E+09 | 2.723E-02 |
| | $4s\,4p^3\,^3S_1^o$ | 113.171 | 2.226E+11 | 1.660E+11 | 2.565E-01 |
| | $4s^2\,4p\,4d\,^3D_1^o$ | 106.199 | 3.737E+09 | 5.873E+09 | 3.791E-03 |
| | $4s^2\,4p\,4d\,^3D_2^o$ | 103.873 | 2.645E+10 | 3.670E+10 | 4.279E-02 |
| | $4s^2\,4p\,4d\,^3D_3^o$ | 83.692 | 4.130E+11 | 3.688E+11 | 6.072E-01 |
| | $4s^2\,4p\,4d\,^3P_1^o$ | 83.695 | 5.849E+10 | 6.255E+10 | 3.685E-02 |
| | $4s^2\,4p\,4d\,^3P_2^o$ | 81.123 | 4.896E+10 | 4.197E+10 | 4.831E-02 |
| | $4s^2\,4p\,4d\,^3F_2^o$ | 116.485 | 1.663E+10 | 2.113E+10 | 3.384E-02 |
| | $4s^2\,4p\,4d\,^3F_3^o$ | 103.851 | 9.946E+10 | 8.363E+10 | 2.251E-01 |
| | $4s^2\,4p\,4d\,^1D_2^o$ | 85.479 | 1.101E+11 | 1.252E+11 | 1.206E-01 |
| | $4s^2\,4p\,4d\,^1P_1^o$ | 76.467 | 1.698E+09 | 3.464E+09 | 8.932E-04 |
| | $4s^2\,4p\,4d\,^1F_3^o$ | 76.969 | 1.218E+11 | 9.259E+10 | 1.514E-01 |
| $4s^2\,4p^2$ $^3P_2^e$ | $4s\,4p^3\,^3D_1^o$ | 312.345 | 1.078E+08 | 1.083E+08 | 9.462E-04 |
| | $4s\,4p^3\,^3D_2^o$ | 216.320 | 1.378E+09 | 1.170E+09 | 9.665E-03 |
| | $4s\,4p^3\,^3D_3^o$ | 191.228 | 1.597E+10 | 1.675E+10 | 1.226E-01 |
| | $4s\,4p^3\,^3P_1^o$ | 170.236 | 3.544E+09 | 9.249E+08 | 9.239E-03 |
| | $4s\,4p^3\,^3P_2^o$ | 121.387 | 1.609E+11 | 1.400E+11 | 3.554E-01 |
| | $4s\,4p^3\,^3S_1^o$ | 161.411 | 1.030E+10 | 7.139E+09 | 2.414E-02 |
| | $4s\,4p^3\,^5S_2^o$ | 445.357 | 1.659E+07 | 2.586E+07 | 4.932E-04 |
| | $4s\,4p^3\,^1D_2^o$ | 163.104 | 1.052E+10 | 6.127E+09 | 4.195E-02 |
| | $4s\,4p^3\,^1P_1^o$ | 110.113 | 2.790E+11 | 2.133E+11 | 3.042E-01 |
| | $4s^2\,4p\,4d\,^3D_1^o$ | 147.591 | 2.309E+07 | 7.330E+06 | 4.524E-05 |
| | $4s^2\,4p\,4d\,^3D_2^o$ | 143.136 | 2.134E+08 | 2.986E+08 | 6.556E-04 |
| | $4s^2\,4p\,4d\,^3D_3^o$ | 107.437 | 1.982E+10 | 2.251E+10 | 4.803E-02 |
| $4s^2\,4p^2$ $^3P_2^e$ | $4s^2\,4p\,4d\,^3P_1^o$ | 107.442 | 3.498E+10 | 3.708E+10 | 3.632E-02 |
| | $4s^2\,4p\,4d\,^3P_2^o$ | 103.241 | 1.572E+11 | 1.733E+11 | 2.512E-01 |
| | $4s^2\,4p\,4d\,^3F_2^o$ | 168.238 | 1.205E+08 | 4.958E+08 | 5.114E-04 |
| | $4s^2\,4p\,4d\,^3F_3^o$ | 143.095 | 2.658E+08 | 6.382E+08 | 1.143E-03 |
| | $4s^2\,4p\,4d\,^1F_3^o$ | 96.605 | 4.105E+11 | 3.577E+11 | 8.041E-01 |
| | $4s^2\,4p\,4d\,^1D_2^o$ | 110.399 | 8.789E+07 | 1.314E+09 | 1.606E-04 |
| | $4s^2\,4p\,4d\,^1P_1^o$ | 95.816 | 1.107E+10 | 1.614E+10 | 9.138E-03 |
| $4s^2\,4p^2$ $^1S_0^e$ | $4s\,4p^3\,^1P_1^o$ | 116.535 | 1.053E+11 | 8.904E+10 | 6.434E-01 |
| | $4s\,4p^3\,^3S_1^o$ | 175.595 | 1.066E+10 | 1.120E+10 | 1.478E-01 |
| | $4s\,4p^3\,^3P_1^o$ | 186.090 | 2.598E+09 | 2.243E+09 | 4.046E-02 |
| | $4s\,4p^3\,^3D_1^o$ | 370.214 | 8.818E+06 | 2.585E+08 | 5.436E-04 |
| | $4s^2\,4p\,4d\,^1P_1^o$ | 100.642 | 2.245E+11 | 2.633E+11 | 1.023E+00 |
| | $4s^2\,4p\,4d\,^3P_1^o$ | 113.548 | 2.025E+09 | 2.446E+09 | 1.174E-02 |
| | $4s^2\,4p\,4d\,^3D_1^o$ | 159.362 | 1.856E+07 | 4.164E+06 | 2.119E-04 |
| $4s\,4p^3$ $^1D_2^o$ | $4p^4\,^1D_2^e$ | 119.287 | 1.290E+11 | 1.999E+11 | 2.753E-01 |
| | $4p^4\,^3P_1^e$ | 123.956 | 4.625E+10 | 5.559E+10 | 6.393E-02 |
| | $4p^4\,^3P_2^e$ | 182.499 | 1.429E+10 | 2.477E+10 | 7.137E-02 |
| $4s\,4p^3$ $^1P_1^o$ | $4p^4\,^1D_2^e$ | 184.070 | 2.129E+10 | 3.429E+10 | 1.802E-01 |
| | $4p^4\,^1S_0^e$ | 330.481 | 3.033E+07 | 1.262E+09 | 1.655E-04 |
| | $4p^4\,^3P_0^e$ | 121.471 | 3.444E+11 | 5.198E+11 | 2.539E-01 |
| | $4p^4\,^3P_1^e$ | 195.431 | 6.487E+09 | 1.386E+10 | 3.714E-02 |
| | $4p^4\,^3P_2^e$ | 395.412 | 9.142E+06 | 5.861E+08 | 3.571E-04 |
| $4s\,4p^3$ $^3D_1^o$ | $4p^4\,^1D_2^e$ | 88.397 | 2.870E+07 | 9.240E+06 | 5.604E-05 |
| | $4p^4\,^1S_0^e$ | 112.286 | 1.406E+11 | 2.300E+11 | 8.861E-02 |
| | $4p^4\,^3P_0^e$ | 70.860 | 1.995E+08 | 1.997E+08 | 5.006E-05 |
| | $4p^4\,^3P_1^e$ | 90.935 | 2.315E+10 | 1.686E+10 | 2.870E-02 |
| | $4p^4\,^3P_2^e$ | 118.921 | 7.328E+10 | 9.549E+10 | 2.590E-01 |
| $4s\,4p^3$ $^3D_2^o$ | $4p^4\,^1D_2^e$ | 101.097 | 1.151E+09 | 4.569E+08 | 1.764E-03 |
| | $4p^4\,^3P_1^e$ | 104.431 | 7.040E+10 | 5.808E+10 | 6.906E-02 |
| | $4p^4\,^3P_2^e$ | 143.107 | 2.963E+10 | 2.712E+10 | 9.098E-02 |
| $4s\,4p^3$ $^3D_3^o$ | $4p^4\,^1D_2^e$ | 107.702 | 5.793E+10 | 4.893E+10 | 7.196E-02 |
| | $4p^4\,^3P_2^e$ | 156.711 | 5.036E+10 | 3.968E+10 | 1.324E-01 |
| $4s\,4p^3\,^3P_0^o$ | $4p^4\,^3P_1^e$ | 116.510 | 3.888E+10 | 4.045E+10 | 2.374E-01 |
| $4s\,4p^3$ $^3P_1^o$ | $4p^4\,^1D_2^e$ | 115.740 | 3.813E+10 | 5.142E+10 | 1.276E-01 |
| | $4p^4\,^1S_0^e$ | 160.430 | 2.644E+10 | 3.136E+10 | 3.401E-02 |
| | $4p^4\,^3P_0^e$ | 87.414 | 4.077E+10 | 2.457E+10 | 1.557E-02 |
| | $4p^4\,^3P_1^e$ | 120.131 | 3.091E+10 | 5.563E+10 | 6.688E-02 |
| | $4p^4\,^3P_2^e$ | 174.327 | 1.062E+10 | 1.827E+10 | 8.061E-02 |
| $4s\,4p^3$ $^3P_2^o$ | $4p^4\,^1D_2^e$ | 159.335 | 4.484E+10 | 7.041E+10 | 1.706E-01 |
| | $4p^4\,^3P_1^e$ | 167.777 | 2.454E+10 | 2.998E+10 | 6.215E-02 |
| | $4p^4\,^3P_2^e$ | 296.524 | 5.838E+07 | 1.086E+09 | 7.695E-04 |
| $4s\,4p^3$ $^3S_1^o$ | $4p^4\,^1D_2^e$ | 120.209 | 5.870E+09 | 3.211E+09 | 2.119E-02 |
| | $4p^4\,^1S_0^e$ | 169.146 | 5.675E+10 | 1.089E+10 | 8.114E-02 |
| | $4p^4\,^3P_0^e$ | 89.939 | 4.617E+08 | 2.585E+09 | 1.867E-04 |
| | $4p^4\,^3P_1^e$ | 124.952 | 1.193E+11 | 1.731E+11 | 2.791E-01 |
| | $4p^4\,^3P_2^e$ | 184.666 | 1.244E+10 | 2.685E+10 | 1.060E-01 |
| $4s\,4p^3$ $^5S_2^o$ | $4p^4\,^1D_2^e$ | 81.507 | 3.688E+09 | 2.353E+09 | 3.673E-03 |
| | $4p^4\,^3P_1^e$ | 83.661 | 3.158E+06 | 6.232E+08 | 1.988E-04 |
| | $4p^4\,^3P_2^e$ | 106.779 | 8.302E+10 | 7.874E+10 | 1.419E-01 |

Table (9): **MCDF** wavelengths (in Å), transition probabilities (in sec$^{-1}$), and oscillator strengths for lines of **Sm XXXI** ion.

| Lower | Upper | MCDF code | | | |
|---|---|---|---|---|---|
| | | λ (Å) | $A_L$ (sec$^{-1}$) | $A_V$ (sec$^{-1}$) | $f_L$ |
| $4s^2\ 4p^2$ $^3P_0^e$ | $4s\ 4p^3\ ^3D_1^o$ | 110.249 | 1.624E+11 | 1.764E+11 | 8.878E-01 |
| | $4s\ 4p^3\ ^3P_1^o$ | 83.385 | 2.515E+09 | 2.987E+09 | 7.864E-03 |
| | $4s\ 4p^3\ ^3S_1^o$ | 81.283 | 2.122E+10 | 1.008E+10 | 6.306E-02 |
| | $4s\ 4p^3\ ^1P_1^o$ | 64.718 | 2.014E+08 | 8.872E+07 | 3.794E-04 |
| | $4s^2\ 4p\ 4d\ ^3D_1^o$ | 78.551 | 4.645E+11 | 4.726E+11 | 1.289E+00 |
| | $4s^2\ 4p\ 4d\ ^3P_1^o$ | 64.466 | 1.217E+08 | 5.634E+07 | 2.275E-04 |
| | $4s^2\ 4p\ 4d\ ^1P_1^o$ | 59.840 | 1.487E+09 | 4.553E+08 | 2.395E-03 |
| $4s^2\ 4p^2$ $^3P_1^e$ | $4s\ 4p^3\ ^3D_1^o$ | 158.221 | 1.700E+09 | 2.960E+08 | 6.379E-03 |
| | $4s\ 4p^3\ ^3D_2^o$ | 125.974 | 5.506E+10 | 5.522E+10 | 2.183E-01 |
| | $4s\ 4p^3\ ^3P_0^o$ | 111.510 | 1.339E+11 | 1.250E+11 | 8.320E-02 |
| | $4s\ 4p^3\ ^3P_1^o$ | 108.196 | 2.257E+11 | 1.947E+11 | 3.961E-01 |
| | $4s\ 4p^3\ ^3P_2^o$ | 84.529 | 4.079E+08 | 1.463E+08 | 7.283E-04 |
| | $4s\ 4p^3\ ^3S_1^o$ | 104.683 | 8.171E+10 | 6.065E+10 | 1.342E-01 |
| | $4s\ 4p^3\ ^5S_2^o$ | 187.094 | 9.271E+09 | 8.950E+09 | 8.109E-02 |
| | $4s\ 4p^3\ ^1D_2^o$ | 105.221 | 1.159E+10 | 1.265E+10 | 3.206E-02 |
| | $4s\ 4p^3\ ^1P_1^o$ | 78.731 | 2.258E+10 | 1.115E+10 | 2.098E-02 |
| | $4s^2\ 4p\ 4d\ ^3D_1^o$ | 100.195 | 3.428E+10 | 3.008E+10 | 5.159E-02 |
| | $4s^2\ 4p\ 4d\ ^3D_2^o$ | 97.174 | 1.846E+11 | 1.812E+11 | 4.357E-01 |
| | $4s^2\ 4p\ 4d\ ^3P_0^o$ | 78.917 | 3.356E+11 | 3.644E+11 | 1.044E-01 |
| | $4s^2\ 4p\ 4d\ ^3P_1^o$ | 78.358 | 2.841E+11 | 2.857E+11 | 2.615E-01 |
| | $4s^2\ 4p\ 4d\ ^3P_2^o$ | 76.709 | 2.516E+10 | 2.028E+10 | 3.700E-02 |
| | $4s^2\ 4p\ 4d\ ^3F_2^o$ | 109.630 | 1.843E+09 | 1.550E+09 | 5.536E-03 |
| | $4s^2\ 4p\ 4d\ ^1D_2^o$ | 81.232 | 1.939E+11 | 1.881E+11 | 3.197E-01 |
| | $4s^2\ 4p\ 4d\ ^1P_1^o$ | 71.627 | 8.529E+09 | 7.089E+09 | 6.560E-03 |
| $4s^2\ 4p^2$ $^1D_2^e$ | $4s\ 4p^3\ ^1P_1^o$ | 80.327 | 9.282E+08 | 4.862E+08 | 5.387E-04 |
| | $4s\ 4p^3\ ^1D_2^o$ | 108.092 | 2.585E+11 | 2.539E+11 | 4.528E-01 |
| | $4s\ 4p^3\ ^3D_1^o$ | 164.804 | 1.465E+10 | 8.756E+09 | 3.579E-02 |
| | $4s\ 4p^3\ ^3D_2^o$ | 130.112 | 1.632E+09 | 5.468E+08 | 4.141E-03 |
| | $4s\ 4p^3\ ^3D_3^o$ | 121.351 | 3.836E+10 | 4.802E+10 | 1.186E-01 |
| | $4s\ 4p^3\ ^3P_1^o$ | 111.234 | 1.223E+09 | 3.084E+08 | 1.362E-03 |
| | $4s\ 4p^3\ ^3P_2^o$ | 86.372 | 9.817E+09 | 1.248E+10 | 1.098E-02 |
| | $4s\ 4p^3\ ^5S_2^o$ | 196.369 | 5.182E+09 | 3.061E+09 | 2.996E-02 |
| | $4s\ 4p^3\ ^3S_1^o$ | 107.525 | 2.458E+11 | 1.802E+11 | 2.557E-01 |
| | $4s^2\ 4p\ 4d\ ^3D_1^o$ | 102.796 | 3.885E+09 | 6.184E+09 | 3.693E-03 |
| | $4s^2\ 4p\ 4d\ ^3D_2^o$ | 99.618 | 2.785E+10 | 3.822E+10 | 4.143E-02 |
| | $4s^2\ 4p\ 4d\ ^3D_3^o$ | 79.936 | 4.505E+11 | 4.000E+11 | 6.042E-01 |
| | $4s^2\ 4p\ 4d\ ^3P_1^o$ | 79.940 | 6.291E+10 | 6.734E+10 | 3.616E-02 |
| | $4s^2\ 4p\ 4d\ ^3P_2^o$ | 78.224 | 4.743E+10 | 4.098E+10 | 4.351E-02 |
| | $4s^2\ 4p\ 4d\ ^3F_2^o$ | 112.751 | 1.761E+10 | 2.230E+10 | 3.357E-02 |
| | $4s^2\ 4p\ 4d\ ^3F_3^o$ | 100.052 | 1.104E+11 | 9.274E+10 | 2.320E-01 |
| | $4s^2\ 4p\ 4d\ ^1D_2^o$ | 82.933 | 1.175E+11 | 1.370E+11 | 1.211E-01 |
| | $4s^2\ 4p\ 4d\ ^1P_1^o$ | 72.946 | 1.637E+09 | 3.566E+09 | 7.836E-04 |
| | $4s^2\ 4p\ 4d\ ^1F_3^o$ | 73.411 | 1.226E+11 | 9.260E+10 | 1.386E-01 |
| $4s^2\ 4p^2$ $^3P_2^e$ | $4s\ 4p^3\ ^3D_1^o$ | 316.984 | 8.920E+07 | 9.433E+07 | 8.062E-04 |
| | $4s\ 4p^3\ ^3D_2^o$ | 209.529 | 1.533E+09 | 1.305E+09 | 1.009E-02 |
| | $4s\ 4p^3\ ^3D_3^o$ | 187.707 | 1.613E+10 | 1.743E+10 | 1.193E-01 |
| | $4s\ 4p^3\ ^3P_1^o$ | 164.556 | 4.264E+09 | 1.211E+09 | 1.039E-02 |
| | $4s\ 4p^3\ ^3P_2^o$ | 115.410 | 1.744E+11 | 1.527E+11 | 3.483E-01 |
| | $4s\ 4p^3\ ^3S_1^o$ | 156.566 | 1.094E+10 | 7.622E+09 | 2.412E-02 |
| | $4s\ 4p^3\ ^5S_2^o$ | 458.848 | 1.339E+07 | 2.260E+07 | 4.226E-04 |
| | $4s\ 4p^3\ ^1D_2^o$ | 157.772 | 1.178E+10 | 6.984E+09 | 4.395E-02 |
| | $4s\ 4p^3\ ^1P_1^o$ | 104.866 | 3.061E+11 | 2.359E+11 | 3.028E-01 |
| | $4s^2\ 4p\ 4d\ ^3D_1^o$ | 146.736 | 2.043E+07 | 5.622E+06 | 3.958E-05 |
| | $4s^2\ 4p\ 4d\ ^3D_2^o$ | 140.346 | 1.676E+08 | 2.204E+08 | 4.950E-04 |
| | $4s^2\ 4p\ 4d\ ^3D_3^o$ | 104.200 | 1.807E+10 | 2.085E+10 | 4.118E-02 |
| $4s^2\ 4p^2$ $^3P_2^e$ | $4s^2\ 4p\ 4d\ ^3P_1^o$ | 104.206 | 3.649E+10 | 3.866E+10 | 3.565E-02 |
| | $4s^2\ 4p\ 4d\ ^3P_2^o$ | 101.310 | 1.578E+11 | 1.795E+11 | 2.428E-01 |
| | $4s^2\ 4p\ 4d\ ^3F_2^o$ | 167.898 | 1.015E+08 | 4.549E+08 | 4.289E-04 |
| | $4s^2\ 4p\ 4d\ ^3F_3^o$ | 141.209 | 2.581E+08 | 6.439E+08 | 1.080E-03 |
| | $4s^2\ 4p\ 4d\ ^1F_3^o$ | 93.381 | 4.342E+11 | 3.783E+11 | 7.946E-01 |
| | $4s^2\ 4p\ 4d\ ^1D_2^o$ | 109.351 | 8.799E+05 | 6.378E+08 | 1.577E-06 |
| | $4s^2\ 4p\ 4d\ ^1P_1^o$ | 92.630 | 1.197E+10 | 1.729E+10 | 9.235E-03 |
| $4s^2\ 4p^2$ $^1S_0^e$ | $4s\ 4p^3\ ^1P_1^o$ | 110.815 | 1.153E+11 | 9.789E+10 | 6.368E-01 |
| | $4s\ 4p^3\ ^3S_1^o$ | 170.207 | 1.125E+10 | 1.191E+10 | 1.466E-01 |
| | $4s\ 4p^3\ ^3P_1^o$ | 179.692 | 2.829E+09 | 2.421E+09 | 4.108E-02 |
| | $4s\ 4p^3\ ^3D_1^o$ | 378.377 | 6.670E+06 | 2.447E+08 | 4.295E-04 |
| | $4s^2\ 4p\ 4d\ ^1P_1^o$ | 97.241 | 2.360E+11 | 2.762E+11 | 1.003E+00 |
| | $4s^2\ 4p\ 4d\ ^3P_1^o$ | 110.078 | 2.356E+09 | 2.839E+09 | 1.284E-02 |
| | $4s^2\ 4p\ 4d\ ^3D_1^o$ | 158.653 | 1.549E+07 | 2.772E+06 | 1.754E-04 |
| $4s\ 4p^3$ $^1D_2^o$ | $4p^4\ ^1D_2^e$ | 113.380 | 1.479E+11 | 2.264E+11 | 2.851E-01 |
| | $4p^4\ ^3P_1^e$ | 117.683 | 4.939E+10 | 5.923E+10 | 6.153E-02 |
| | $4p^4\ ^3P_2^e$ | 176.802 | 1.493E+10 | 2.608E+10 | 6.997E-02 |
| $4s\ 4p^3$ $^1P_1^o$ | $4p^4\ ^1D_2^e$ | 177.866 | 2.206E+10 | 3.557E+10 | 1.744E-01 |
| | $4p^4\ ^1S_0^e$ | 336.664 | 2.355E+07 | 1.198E+09 | 1.334E-04 |
| | $4p^4\ ^3P_0^e$ | 115.203 | 3.871E+11 | 5.767E+11 | 2.568E-01 |
| | $4p^4\ ^3P_1^e$ | 188.689 | 6.752E+09 | 1.435E+10 | 3.604E-02 |
| | $4p^4\ ^3P_2^e$ | 406.773 | 6.655E+06 | 5.441E+08 | 2.751E-04 |
| $4s\ 4p^3$ $^3D_1^o$ | $4p^4\ ^1D_2^e$ | 83.310 | 3.371E+07 | 1.080E+07 | 5.847E-05 |
| | $4p^4\ ^1S_0^e$ | 106.934 | 1.600E+11 | 2.592E+11 | 9.142E-02 |
| | $4p^4\ ^3P_0^e$ | 66.394 | 1.765E+08 | 1.766E+08 | 3.888E-05 |
| | $4p^4\ ^3P_1^e$ | 85.609 | 2.372E+10 | 1.682E+10 | 2.606E-02 |
| | $4p^4\ ^3P_2^e$ | 113.127 | 8.339E+10 | 1.083E+11 | 2.666E-01 |
| $4s\ 4p^3$ $^3D_2^o$ | $4p^4\ ^1D_2^e$ | 96.288 | 1.058E+09 | 3.895E+08 | 1.471E-03 |
| | $4p^4\ ^3P_1^e$ | 99.373 | 7.870E+06 | 6.536E+10 | 6.991E-02 |
| | $4p^4\ ^3P_2^e$ | 138.471 | 2.987E+10 | 2.740E+10 | 8.587E-02 |
| $4s\ 4p^3$ $^3D_3^o$ | $4p^4\ ^1D_2^e$ | 101.722 | 5.587E+10 | 6.830E+10 | 6.191E-02 |
| | $4p^4\ ^3P_2^e$ | 149.995 | 4.428E+10 | 6.068E+10 | 1.067E-01 |
| $4s\ 4p^3\ ^3P_0^o$ | $4p^4\ ^3P_1^e$ | 110.699 | 4.317E+10 | 4.488E+10 | 2.379E-01 |
| $4s\ 4p^3$ $^3P_1^o$ | $4p^4\ ^1D_2^e$ | 110.118 | 4.356E+10 | 5.855E+10 | 1.320E-01 |
| | $4p^4\ ^1S_0^e$ | 155.537 | 2.700E+10 | 3.216E+10 | 3.265E-02 |
| | $4p^4\ ^3P_0^e$ | 82.377 | 4.259E+10 | 2.507E+10 | 1.444E-02 |
| | $4p^4\ ^3P_1^e$ | 114.172 | 3.688E+10 | 6.503E+10 | 7.207E-02 |
| | $4p^4\ ^3P_2^e$ | 168.994 | 1.107E+10 | 1.919E+10 | 7.902E-02 |
| $4s\ 4p^3$ $^3P_2^o$ | $4p^4\ ^1D_2^e$ | 154.002 | 4.526E+10 | 7.062E+10 | 1.609E-01 |
| | $4p^4\ ^3P_1^e$ | 162.050 | 2.658E+10 | 3.248E+10 | 6.278E-02 |
| | $4p^4\ ^3P_2^e$ | 300.338 | 4.328E+07 | 9.820E+08 | 5.853E-04 |
| $4s\ 4p^3$ $^3S_1^o$ | $4p^4\ ^1D_2^e$ | 114.011 | 6.637E+09 | 3.732E+10 | 2.156E-02 |
| | $4p^4\ ^1S_0^e$ | 163.420 | 5.905E+10 | 1.127E+11 | 7.881E-02 |
| | $4p^4\ ^3P_0^e$ | 84.537 | 4.174E+08 | 2.550E+09 | 1.491E-04 |
| | $4p^4\ ^3P_1^e$ | 118.362 | 1.341E+11 | 1.921E+11 | 2.816E-01 |
| | $4p^4\ ^3P_2^e$ | 178.340 | 1.261E+10 | 2.722E+10 | 1.002E-01 |
| $4s\ 4p^3$ $^5S_2^o$ | $4p^4\ ^1D_2^e$ | 77.049 | 4.014E+09 | 2.530E+09 | 3.572E-03 |
| | $4p^4\ ^3P_1^e$ | 79.012 | 2.757E+08 | 5.901E+08 | 1.548E-04 |
| | $4p^4\ ^3P_2^e$ | 101.885 | 9.580E+10 | 9.105E+10 | 1.491E-01 |

Table (10): **MCDF** wavelengths (in Å), transition probabilities (in sec$^{-1}$), and oscillator strengths for lines of **Eu XXXII** ion.

| Lower | Upper | MCDF code | | | |
|---|---|---|---|---|---|
| | | λ (Å) | $A_L$ (sec$^{-1}$) | $A_V$ (sec$^{-1}$) | $f_L$ |
| $4s^2\,4p^2$ $^3P_0^e$ | $4s\,4p^3\,^3D_1^o$ | 104.832 | 1.813E+11 | 1.965E+11 | 8.962E-01 |
| | $4s\,4p^3\,^3P_1^o$ | 78.480 | 2.574E+09 | 3.006E+09 | 7.129E-03 |
| | $4s\,4p^3\,^3S_1^o$ | 76.670 | 2.179E+10 | 1.012E+10 | 5.760E-02 |
| | $4s\,4p^3\,^1P_1^o$ | 60.712 | 1.931E+08 | 8.232E+07 | 3.201E-04 |
| | $4s^2\,4p\,4d\,^3D_1^o$ | 75.204 | 5.035E+11 | 5.119E+11 | 1.281E+00 |
| | $4s^2\,4p\,4d\,^3P_1^o$ | 61.175 | 1.472E+08 | 4.956E+07 | 2.478E-04 |
| | $4s^2\,4p\,4d\,^1P_1^o$ | 56.771 | 1.555E+09 | 4.691E+08 | 2.255E-03 |
| $4s^2\,4p^2$ $^3P_1^e$ | $4s\,4p^3\,^3D_1^o$ | 153.285 | 2.044E+09 | 4.206E+08 | 7.199E-03 |
| | $4s\,4p^3\,^3D_2^o$ | 119.471 | 5.964E+10 | 5.990E+10 | 2.127E-01 |
| | $4s\,4p^3\,^3P_0^o$ | 105.768 | 1.496E+11 | 1.394E+11 | 8.362E-02 |
| | $4s\,4p^3\,^3P_1^o$ | 102.809 | 2.535E+11 | 2.192E+11 | 4.017E-01 |
| | $4s\,4p^3\,^3P_2^o$ | 79.635 | 3.867E+08 | 1.269E+08 | 6.127E-04 |
| | $4s\,4p^3\,^3S_1^o$ | 99.724 | 8.784E+10 | 6.570E+10 | 1.310E-01 |
| | $4s\,4p^3\,^5S_2^o$ | 184.983 | 9.762E+09 | 1.004E+10 | 8.347E-02 |
| | $4s\,4p^3\,^1D_2^o$ | 100.078 | 1.346E+10 | 1.459E+10 | 3.368E-02 |
| | $4s\,4p^3\,^1P_1^o$ | 74.317 | 2.325E+10 | 1.138E+10 | 1.925E-02 |
| | $4s^2\,4p\,4d\,^3D_1^o$ | 97.259 | 3.546E+10 | 3.115E+10 | 5.029E-02 |
| | $4s^2\,4p\,4d\,^3D_2^o$ | 95.197 | 1.897E+11 | 1.908E+11 | 4.295E-01 |
| | $4s^2\,4p\,4d\,^3P_0^o$ | 75.540 | 3.637E+11 | 3.944E+11 | 1.037E-01 |
| | $4s^2\,4p\,4d\,^3P_1^o$ | 75.012 | 3.083E+11 | 3.100E+11 | 2.601E-01 |
| | $4s^2\,4p\,4d\,^3P_2^o$ | 73.257 | 2.442E+10 | 1.957E+10 | 3.274E-02 |
| | $4s^2\,4p\,4d\,^3F_2^o$ | 106.435 | 1.957E+09 | 1.650E+09 | 5.540E-03 |
| | $4s^2\,4p\,4d\,^1D_2^o$ | 77.572 | 2.104E+11 | 2.033E+11 | 3.164E-01 |
| | $4s^2\,4p\,4d\,^1P_1^o$ | 68.496 | 8.842E+09 | 7.325E+09 | 6.219E-03 |
| $4s^2\,4p^2$ $^1D_2^e$ | $4s\,4p^3\,^1P_1^o$ | 75.784 | 9.987E+08 | 4.917E+08 | 5.159E-04 |
| | $4s\,4p^3\,^1D_2^o$ | 102.757 | 2.877E+11 | 2.831E+11 | 4.554E-01 |
| | $4s\,4p^3\,^3D_1^o$ | 159.661 | 1.626E+10 | 9.765E+09 | 3.728E-02 |
| | $4s\,4p^3\,^3D_2^o$ | 123.309 | 1.697E+09 | 5.879E+08 | 3.869E-03 |
| | $4s\,4p^3\,^3D_3^o$ | 115.082 | 4.243E+10 | 5.281E+10 | 1.179E-01 |
| | $4s\,4p^3\,^3P_1^o$ | 105.638 | 1.122E+09 | 2.468E+08 | 1.126E-03 |
| | $4s\,4p^3\,^3P_2^o$ | 81.323 | 9.417E+09 | 1.210E+10 | 9.337E-03 |
| | $4s\,4p^3\,^5S_2^o$ | 194.348 | 5.589E+09 | 3.704E+09 | 3.165E-02 |
| | $4s\,4p^3\,^3S_1^o$ | 102.384 | 2.674E+11 | 1.972E+11 | 2.521E-01 |
| | $4s^2\,4p\,4d\,^3D_1^o$ | 99.787 | 4.062E+09 | 6.463E+09 | 3.638E-03 |
| | $4s^2\,4p\,4d\,^3D_2^o$ | 97.617 | 2.671E+10 | 3.766E+10 | 3.815E-02 |
| | $4s^2\,4p\,4d\,^3D_3^o$ | 76.503 | 4.868E+11 | 4.324E+11 | 5.980E-01 |
| | $4s^2\,4p\,4d\,^3P_1^o$ | 76.507 | 6.756E+10 | 7.233E+10 | 3.557E-02 |
| | $4s^2\,4p\,4d\,^3P_2^o$ | 74.682 | 4.642E+10 | 3.958E+10 | 3.882E-02 |
| | $4s^2\,4p\,4d\,^3F_2^o$ | 109.471 | 1.853E+10 | 2.346E+10 | 3.329E-02 |
| | $4s^2\,4p\,4d\,^3F_3^o$ | 96.657 | 1.217E+11 | 1.023E+11 | 2.387E-01 |
| | $4s^2\,4p\,4d\,^1D_2^o$ | 79.172 | 1.322E+11 | 1.530E+11 | 3.815E-02 |
| | $4s^2\,4p\,4d\,^1P_1^o$ | 69.741 | 1.642E+09 | 3.616E+09 | 7.182E-04 |
| | $4s^2\,4p\,4d\,^1F_3^o$ | 70.221 | 1.227E+11 | 9.255E+10 | 1.270E-01 |
| $4s^2\,4p^2$ $^3P_2^e$ | $4s\,4p^3\,^3D_1^o$ | 324.845 | 7.264E+07 | 8.443E+07 | 6.895E-04 |
| | $4s\,4p^3\,^3D_2^o$ | 203.052 | 1.696E+09 | 1.449E+09 | 1.049E-02 |
| | $4s\,4p^3\,^3D_3^o$ | 181.668 | 1.706E+10 | 1.844E+10 | 1.182E-01 |
| | $4s\,4p^3\,^3P_1^o$ | 159.200 | 5.052E+09 | 1.537E+09 | 1.152E-02 |
| | $4s\,4p^3\,^3P_2^o$ | 109.748 | 1.894E+11 | 1.668E+11 | 3.421E-01 |
| | $4s\,4p^3\,^3S_1^o$ | 151.923 | 1.159E+10 | 8.117E+09 | 2.407E-02 |
| | $4s\,4p^3\,^5S_2^o$ | 510.071 | 7.440E+06 | 1.525E+07 | 2.902E-04 |
| | $4s\,4p^3\,^1D_2^o$ | 152.747 | 1.305E+10 | 7.862E+09 | 4.566E-02 |
| | $4s\,4p^3\,^1P_1^o$ | 99.895 | 3.359E+11 | 2.609E+11 | 3.015E-01 |
| | $4s^2\,4p\,4d\,^3D_1^o$ | 146.275 | 1.802E+07 | 4.200E+06 | 3.468E-05 |
| | $4s^2\,4p\,4d\,^3D_2^o$ | 141.660 | 1.447E+08 | 1.929E+08 | 4.354E-04 |
| | $4s^2\,4p\,4d\,^3D_3^o$ | 101.148 | 1.630E+10 | 1.915E+10 | 3.501E-02 |
| $4s^2\,4p^2$ $^3P_2^e$ | $4s^2\,4p\,4d\,^3P_1^o$ | 101.156 | 3.799E+10 | 4.022E+10 | 3.497E-02 |
| | $4s^2\,4p\,4d\,^3P_2^o$ | 97.989 | 1.664E+11 | 1.896E+11 | 2.396E-02 |
| | $4s^2\,4p\,4d\,^3F_2^o$ | 168.068 | 8.485E+07 | 4.168E+08 | 3.593E-04 |
| | $4s^2\,4p\,4d\,^3F_3^o$ | 139.646 | 2.471E+08 | 6.434E+08 | 1.011E-03 |
| | $4s^2\,4p\,4d\,^1F_3^o$ | 90.450 | 4.568E+11 | 3.992E+11 | 7.843E-01 |
| | $4s^2\,4p\,4d\,^1D_2^o$ | 105.867 | 8.867E+07 | 2.734E+08 | 1.490E-04 |
| | $4s^2\,4p\,4d\,^1P_1^o$ | 89.655 | 1.290E+10 | 1.850E+10 | 9.328E-03 |
| $4s^2\,4p^2$ $^1S_0^e$ | $4s\,4p^3\,^1P_1^o$ | 105.405 | 1.263E+11 | 1.077E+11 | 6.310E-01 |
| | $4s\,4p^3\,^3S_1^o$ | 165.046 | 1.186E+10 | 1.264E+10 | 1.453E-01 |
| | $4s\,4p^3\,^3P_1^o$ | 173.669 | 3.065E+09 | 2.601E+09 | 4.157E-02 |
| | $4s\,4p^3\,^3D_1^o$ | 391.383 | 4.735E+06 | 2.260E+08 | 3.262E-04 |
| | $4s^2\,4p\,4d\,^1P_1^o$ | 94.069 | 2.474E+11 | 2.893E+11 | 9.847E-01 |
| | $4s^2\,4p\,4d\,^3P_1^o$ | 106.810 | 2.718E+09 | 3.268E+09 | 1.394E-02 |
| | $4s^2\,4p\,4d\,^3D_1^o$ | 158.401 | 1.279E+07 | 1.702E+06 | 1.443E-04 |
| $4s\,4p^3$ $^1D_2^o$ | $4p^4\,^1D_2^e$ | 107.793 | 1.689E+11 | 2.555E+11 | 2.943E-01 |
| | $4p^4\,^3P_1^e$ | 111.758 | 5.290E+06 | 6.329E+10 | 5.943E-02 |
| | $4p^4\,^3P_2^e$ | 171.387 | 1.558E+10 | 2.740E+10 | 6.859E-02 |
| $4s\,4p^3$ $^1P_1^o$ | $4p^4\,^1D_2^e$ | 172.020 | 2.283E+10 | 3.686E+10 | 1.688E-01 |
| | $4p^4\,^1S_0^e$ | 345.075 | 1.783E+07 | 1.131E+09 | 1.061E-04 |
| | $4p^4\,^3P_0^e$ | 109.312 | 4.348E+11 | 6.396E+11 | 2.597E-01 |
| | $4p^4\,^3P_1^e$ | 182.344 | 7.014E+09 | 1.483E+10 | 3.496E-02 |
| | $4p^4\,^3P_2^e$ | 421.766 | 4.703E+06 | 5.019E+08 | 2.091E-04 |
| $4s\,4p^3$ $^3D_1^o$ | $4p^4\,^1D_2^e$ | 78.460 | 4.096E+07 | 1.214E+07 | 6.301E-05 |
| | $4p^4\,^1S_0^e$ | 101.729 | 1.822E+11 | 2.920E+11 | 9.422E-02 |
| | $4p^4\,^3P_0^e$ | 62.188 | 1.255E+08 | 1.348E+08 | 2.426E-05 |
| | $4p^4\,^3P_1^e$ | 80.539 | 2.437E+10 | 1.679E+10 | 2.370E-02 |
| | $4p^4\,^3P_2^e$ | 107.491 | 9.512E+10 | 1.229E+11 | 2.746E-01 |
| $4s\,4p^3$ $^3D_2^o$ | $4p^4\,^1D_2^e$ | 91.752 | 9.667E+08 | 3.264E+08 | 1.220E-03 |
| | $4p^4\,^3P_1^e$ | 94.609 | 8.795E+10 | 7.350E+10 | 7.081E-02 |
| | $4p^4\,^3P_2^e$ | 134.108 | 3.010E+10 | 2.766E+10 | 8.115E-02 |
| $4s\,4p^3$ $^3D_3^o$ | $4p^4\,^1D_2^e$ | 96.906 | 6.258E+07 | 7.617E+10 | 6.293E-02 |
| | $4p^4\,^3P_2^e$ | 145.413 | 4.612E+06 | 6.314E+10 | 1.044E-01 |
| $4s\,4p^3\,^3P_0^o$ | $4p^4\,^3P_1^e$ | 105.425 | 4.981E+10 | 4.648E+10 | 2.490E-01 |
| $4s\,4p^3$ $^3P_1^o$ | $4p^4\,^1D_2^e$ | 104.796 | 4.964E+10 | 6.647E+10 | 1.362E-02 |
| | $4p^4\,^1S_0^e$ | 150.897 | 2.757E+10 | 3.294E+10 | 3.137E-02 |
| | $4p^4\,^3P_0^e$ | 77.656 | 4.444E+10 | 2.554E+10 | 1.339E-02 |
| | $4p^4\,^3P_1^e$ | 108.539 | 4.376E+10 | 7.568E+10 | 7.728E-02 |
| | $4p^4\,^3P_2^e$ | 163.932 | 1.154E+10 | 2.013E+10 | 7.751E-02 |
| $4s\,4p^3$ $^3P_2^o$ | $4p^4\,^1D_2^e$ | 148.987 | 4.572E+10 | 7.091E+10 | 1.521E-01 |
| | $4p^4\,^3P_1^e$ | 156.669 | 2.862E+10 | 3.498E+10 | 6.319E-02 |
| | $4p^4\,^3P_2^e$ | 305.839 | 3.156E+07 | 8.869E+08 | 4.425E-04 |
| $4s\,4p^3$ $^3S_1^o$ | $4p^4\,^1D_2^e$ | 107.669 | 7.313E+09 | 4.286E+10 | 2.118E-02 |
| | $4p^4\,^1S_0^e$ | 156.927 | 6.229E+10 | 1.143E+11 | 7.666E-02 |
| | $4p^4\,^3P_0^e$ | 79.223 | 4.973E+08 | 2.445E+09 | 1.560E-04 |
| | $4p^4\,^3P_1^e$ | 111.625 | 1.541E+11 | 2.142E+11 | 2.879E-01 |
| | $4p^4\,^3P_2^e$ | 171.074 | 1.311E+10 | 2.753E+10 | 9.587E-02 |
| $4s\,4p^3$ $^5S_2^o$ | $4p^4\,^1D_2^e$ | 72.133 | 4.591E+09 | 2.039E+09 | 3.581E-03 |
| | $4p^4\,^3P_1^e$ | 73.887 | 8.311E+08 | 1.005E+08 | 4.082E-04 |
| | $4p^4\,^3P_2^e$ | 95.960 | 1.098E+11 | 1.219E+11 | 1.516E-01 |

Table 11: Comparison between **MCDF** and **HFR** transition probabilities (in sec$^{-1}$) and oscillator strengths for Ge-like ions.

| Z | Initial | Final | MCDF | | HFR | |
|---|---|---|---|---|---|---|
| | | | $A_L$ (sec$^{-1}$) | $f_L$ | $A_L$ (sec$^{-1}$) | $f_L$ |
| 59 | 4s$^2$ 4p$^2$ $^3$P$_0^e$ | 4s 4p$^3$ $^3$D$_1^o$ | 1.187E+11 | 9.247E-01 | 1.287E+11 | 9.528E-01 |
| | 4s$^2$ 4p$^2$ $^3$P$_1^e$ | 4s 4p$^3$ $^3$S$_1^o$ | 6.672E+10 | 1.470E-01 | 7.340E+10 | 1.603E-01 |
| | 4s$^2$ 4p$^2$ $^3$P$_1^e$ | 4s 4p$^3$ $^3$P$_1^o$ | 1.215E+11 | 2.907E-01 | 1.412E+11 | 3.349E-01 |
| | 4s$^2$ 4p$^2$ $^3$P$_2^e$ | 4s 4p$^3$ $^3$D$_3^o$ | 1.414E+10 | 1.246E-01 | 1.544E+10 | 1.331E-01 |
| | 4s$^2$ 4p$^2$ $^3$P$_2^e$ | 4s 4p$^3$ $^3$P$_2^o$ | 1.376E+11 | 3.724E-01 | 1.358E+11 | 3.639E-01 |
| | 4s$^2$ 4p$^2$ $^3$P$_2^e$ | 4s 4p$^3$ $^1$P$_1^o$ | 2.317E+11 | 3.078E-01 | 2.391E+11 | 3.134E-01 |
| | 4s$^2$ 4p$^2$ $^3$P$_0^e$ | 4s$^2$ 4p 4d $^3$D$_1^o$ | 1.187E+11 | 9.247E-01 | 1.287E+11 | 9.528E-01 |
| | 4s$^2$ 4p$^2$ $^1$D$_2^e$ | 4s$^2$ 4p 4d $^1$F$_3^o$ | 1.197E+11 | 1.786E-01 | 1.013E+11 | 1.510E-01 |
| | 4s$^2$ 4p$^2$ $^1$D$_2^e$ | 4s$^2$ 4p 4d $^3$F$_3^o$ | 8.025E+10 | 2.097E-01 | 9.629E+10 | 2.501E-01 |
| | 4s$^2$ 4p$^2$ $^3$P$_2^e$ | 4s$^2$ 4p 4d $^3$P$_2^o$ | 1.399E+11 | 2.569E-01 | 1.552E+11 | 2.845E-01 |
| | 4s$^2$ 4p$^2$ $^3$P$_2^e$ | 4s$^2$ 4p 4d $^1$F$_3^o$ | 3.689E+11 | 8.266E-01 | 3.859E+11 | 8.650E-01 |
| | 4s 4p$^3$ $^3$P$_0^o$ | 4p$^4$ $^3$P$_1^e$ | 3.151E+10 | 2.367E-01 | 3.580E+10 | 2.564E-01 |
| | 4s 4p$^3$ $^3$P$_1^o$ | 4p$^4$ $^3$P$_2^e$ | 6.086E+09 | 5.276E-02 | 8.334E+09 | 6.900E-02 |
| | 4s 4p$^3$ $^3$P$_2^o$ | 4p$^4$ $^3$P$_2^e$ | 1.014E+08 | 1.308E-03 | 2.512E+08 | 2.979E-03 |
| | 4s 4p$^3$ $^3$D$_1^o$ | 4p$^4$ $^3$P$_1^e$ | 2.698E+10 | 4.320E-02 | 2.017E+10 | 3.118E-02 |
| | 4s 4p$^3$ $^3$D$_2^o$ | 4p$^4$ $^1$D$_2^e$ | 1.338E+09 | 2.501E-03 | 7.168E+08 | 1.327E-03 |
| | 4s 4p$^3$ $^3$S$_1^o$ | 4p$^4$ $^3$P$_2^e$ | 1.207E+10 | 1.189E-01 | 1.493E+10 | 1.399E-01 |
| | 4s 4p$^3$ $^1$P$_1^o$ | 4p$^4$ $^1$D$_2^e$ | 1.974E+10 | 1.928E-01 | 2.272E+10 | 2.128E-01 |
| 60 | 4s$^2$ 4p$^2$ $^3$P$_0^e$ | 4s 4p$^3$ $^3$D$_1^o$ | 1.302E+11 | 9.343E-01 | 1.443E+11 | 9.594E-01 |
| | 4s$^2$ 4p$^2$ $^3$P$_1^e$ | 4s 4p$^3$ $^3$S$_1^o$ | 7.131E+10 | 1.424E-01 | 7.890E+10 | 1.563E-01 |
| | 4s$^2$ 4p$^2$ $^3$P$_1^e$ | 4s 4p$^3$ $^3$P$_1^o$ | 1.362E+11 | 2.932E-01 | 1.587E+11 | 3.380E-01 |
| | 4s$^2$ 4p$^2$ $^3$P$_2^e$ | 4s 4p$^3$ $^3$D$_3^o$ | 1.505E+10 | 1.236E-01 | 1.637E+10 | 1.315E-01 |
| | 4s$^2$ 4p$^2$ $^3$P$_2^e$ | 4s 4p$^3$ $^3$P$_2^o$ | 1.486E+11 | 3.634E-01 | 1.468E+11 | 3.548E-01 |
| | 4s$^2$ 4p$^2$ $^3$P$_2^e$ | 4s 4p$^3$ $^1$P$_1^o$ | 2.542E+11 | 3.059E-01 | 2.619E+11 | 3.105E-01 |
| | 4s$^2$ 4p$^2$ $^3$P$_0^e$ | 4s$^2$ 4p 4d $^3$D$_1^o$ | 4.008E+11 | 1.333E+00 | 4.397E+11 | 1.459E+00 |
| | 4s$^2$ 4p$^2$ $^1$D$_2^e$ | 4s$^2$ 4p 4d $^1$F$_3^o$ | 1.209E+11 | 1.646E-01 | 1.009E+11 | 1.371E-01 |
| | 4s$^2$ 4p$^2$ $^1$D$_2^e$ | 4s$^2$ 4p 4d $^3$F$_3^o$ | 8.976E+10 | 2.177E-01 | 1.066E+11 | 2.571E-01 |
| | 4s$^2$ 4p$^2$ $^3$P$_2^e$ | 4s$^2$ 4p 4d $^3$P$_2^o$ | 1.485E+11 | 2.543E-01 | 1.638E+11 | 2.799E-01 |
| | 4s$^2$ 4p$^2$ $^3$P$_2^e$ | 4s$^2$ 4p 4d $^1$F$_3^o$ | 3.893E+11 | 8.151E-01 | 4.066E+11 | 8.512E-01 |
| | 4s 4p$^3$ $^3$P$_0^o$ | 4p$^4$ $^3$P$_1^e$ | 3.500E+10 | 2.369E-01 | 3.967E+10 | 2.565E-01 |
| | 4s 4p$^3$ $^3$P$_1^o$ | 4p$^4$ $^3$P$_2^e$ | 6.274E+09 | 5.100E-02 | 8.668E+09 | 6.743E-02 |
| | 4s 4p$^3$ $^1$D$_2^o$ | 4p$^4$ $^1$D$_2^e$ | 1.096E+11 | 2.605E-01 | 1.473E+11 | 3.381E-01 |
| | 4s 4p$^3$ $^3$D$_1^o$ | 4p$^4$ $^3$P$_1^e$ | 2.762E+10 | 3.911E-02 | 2.038E+10 | 2.785E-02 |
| | 4s 4p$^3$ $^3$D$_1^o$ | 4p$^4$ $^1$S$_0^e$ | 1.479E+11 | 1.043E-01 | 1.698E+11 | 1.166E-01 |
| | 4s 4p$^3$ $^3$S$_1^o$ | 4p$^4$ $^3$P$_2^e$ | 1.226E+10 | 1.122E-01 | 1.511E+10 | 1.321E-01 |
| | 4s 4p$^3$ $^1$P$_1^o$ | 4p$^4$ $^1$D$_2^e$ | 2.052E+10 | 1.864E-01 | 2.356E+10 | 2.055E-01 |
| 61 | 4s$^2$ 4p$^2$ $^3$P$_0^e$ | 4s 4p$^3$ $^3$D$_1^o$ | 1.444E+11 | 8.778E-01 | 1.616E+11 | 9.683E-01 |
| | 4s$^2$ 4p$^2$ $^3$P$_1^e$ | 4s 4p$^3$ $^3$S$_1^o$ | 7.629E+10 | 1.381E-01 | 8.490E+10 | 1.524E-01 |
| | 4s$^2$ 4p$^2$ $^3$P$_1^e$ | 4s 4p$^3$ $^3$P$_1^o$ | 2.006E+11 | 3.903E-01 | 1.780E+11 | 3.419E-01 |
| | 4s$^2$ 4p$^2$ $^3$P$_2^e$ | 4s 4p$^3$ $^3$D$_3^o$ | 1.597E+10 | 1.226E-01 | 1.733E+10 | 1.303E-01 |
| | 4s$^2$ 4p$^2$ $^3$P$_2^e$ | 4s 4p$^3$ $^3$P$_2^o$ | 1.609E+11 | 3.554E-01 | 1.590E+11 | 3.468E-01 |
| | 4s$^2$ 4p$^2$ $^3$P$_2^e$ | 4s 4p$^3$ $^1$P$_1^o$ | 2.790E+11 | 3.042E-01 | 2.870E+11 | 3.083E-01 |
| | 4s$^2$ 4p$^2$ $^3$P$_0^e$ | 4s$^2$ 4p 4d $^3$D$_1^o$ | 4.274E+11 | 1.297E+00 | 4.760E+11 | 1.442E+00 |
| | 4s$^2$ 4p$^2$ $^1$D$_2^e$ | 4s$^2$ 4p 4d $^1$F$_3^o$ | 1.218E+11 | 1.514E-01 | 1.002E+11 | 1.242E-01 |
| | 4s$^2$ 4p$^2$ $^1$D$_2^e$ | 4s$^2$ 4p 4d $^3$F$_3^o$ | 9.946E+10 | 2.251E-01 | 1.174E+11 | 2.643E-01 |
| | 4s$^2$ 4p$^2$ $^3$P$_2^e$ | 4s$^2$ 4p 4d $^3$P$_2^o$ | 1.572E+11 | 2.512E-01 | 1.723E+11 | 2.754E-01 |
| | 4s$^2$ 4p$^2$ $^3$P$_2^e$ | 4s$^2$ 4p 4d $^1$F$_3^o$ | 4.105E+11 | 8.041E-01 | 4.276E+11 | 8.376E-01 |
| | 4s 4p$^3$ $^3$P$_0^o$ | 4p$^4$ $^3$P$_1^e$ | 3.888E+10 | 2.374E-01 | 4.393E+10 | 2.559E-01 |
| | 4s 4p$^3$ $^3$P$_1^o$ | 4p$^4$ $^3$P$_2^e$ | 1.062E+10 | 8.061E-02 | 9.004E+09 | 6.575E-02 |
| | 4s 4p$^3$ $^1$D$_2^o$ | 4p$^4$ $^1$D$_2^e$ | 1.290E+11 | 2.753E-01 | 1.685E+11 | 3.492E-01 |
| | 4s 4p$^3$ $^3$D$_1^o$ | 4p$^4$ $^3$P$_2^e$ | 7.328E+10 | 2.590E-01 | 8.704E+10 | 3.012E-01 |
| | 4s 4p$^3$ $^3$D$_1^o$ | 4p$^4$ $^1$S$_0^e$ | 1.406E+11 | 8.861E-02 | 1.925E+11 | 1.199E-01 |
| | 4s 4p$^3$ $^3$S$_1^o$ | 4p$^4$ $^3$P$_2^e$ | 1.244E+10 | 1.060E-01 | 1.529E+10 | 1.244E-01 |
| | 4s 4p$^3$ $^1$P$_1^o$ | 4p$^4$ $^1$D$_2^e$ | 2.129E+10 | 1.802E-01 | 2.440E+10 | 1.986E-01 |

Table (11): Continued.

| Z | Initial | Final | MCDF | | HFR | |
|---|---|---|---|---|---|---|
| | | | $A_L$ (sec$^{-1}$) | $f_L$ | $A_L$ (sec$^{-1}$) | $f_L$ |
| **62** | $4s^2\,4p^2\,^3P_0^e$ | $4s\,4p^3\,^3D_1^o$ | 1.624E+11 | 8.878E-01 | 1.808E+11 | 9.750E-01 |
| | $4s^2\,4p^2\,^3P_1^e$ | $4s\,4p^3\,^3S_1^o$ | 8.171E+10 | 1.342E-01 | 9.150E+10 | 1.489E-01 |
| | $4s^2\,4p^2\,^3P_1^e$ | $4s\,4p^3\,^3P_1^o$ | 2.257E+11 | 3.961E-01 | 1.995E+11 | 3.458E-01 |
| | $4s^2\,4p^2\,^3P_2^e$ | $4s\,4p^3\,^3D_3^o$ | 1.613E+10 | 1.193E-01 | 1.829E+10 | 1.288E-01 |
| | $4s^2\,4p^2\,^3P_2^e$ | $4s\,4p^3\,^3P_2^o$ | 1.744E+11 | 3.483E-01 | 1.727E+11 | 3.396E-01 |
| | $4s^2\,4p^2\,^3P_2^e$ | $4s\,4p^3\,^1P_1^o$ | 3.061E+11 | 3.028E-01 | 3.146E+11 | 3.069E-01 |
| | $4s^2\,4p^2\,^3P_0^e$ | $4s^2\,4p\,4d\,^3D_1^o$ | 4.645E+11 | 1.289E+00 | 5.153E+11 | 1.432E+00 |
| | $4s^2\,4p^2\,^1D_2^e$ | $4s^2\,4p\,4d\,^1F_3^o$ | 1.226E+11 | 1.386E-01 | 9.926E+10 | 1.125E-01 |
| | $4s^2\,4p^2\,^1D_2^e$ | $4s^2\,4p\,4d\,^3F_3^o$ | 1.104E+11 | 2.320E-01 | 1.287E+11 | 2.698E-01 |
| | $4s^2\,4p^2\,^3P_2^e$ | $4s^2\,4p\,4d\,^3P_2^o$ | 1.744E+11 | 3.483E-01 | 1.727E+11 | 3.396E-01 |
| | $4s^2\,4p^2\,^3P_2^e$ | $4s^2\,4p\,4d\,^1F_3^o$ | 4.342E+11 | 7.946E-01 | 4.491E+11 | 8.261E-01 |
| | $4s\,4p^3\,^3P_0^o$ | $4p^4\,^3P_1^e$ | 4.317E+10 | 2.379E-01 | 4.870E+10 | 2.564E-01 |
| | $4s\,4p^3\,^3P_1^o$ | $4p^4\,^3P_2^e$ | 1.107E+10 | 7.902E-02 | 9.346E+09 | 6.425E-02 |
| | $4s\,4p^3\,^1D_2^o$ | $4p^4\,^1D_2^e$ | 1.479E+11 | 2.851E-01 | 1.921E+11 | 3.589E-01 |
| | $4s\,4p^3\,^3D_1^o$ | $4p^4\,^3P_2^e$ | 8.339E+10 | 2.666E-01 | 9.874E+10 | 3.089E-01 |
| | $4s\,4p^3\,^3D_1^o$ | $4p^4\,^1S_0^e$ | 1.600E+11 | 9.142E-02 | 2.179E+11 | 1.230E-01 |
| | $4s\,4p^3\,^3S_1^o$ | $4p^4\,^3P_2^e$ | 1.261E+10 | 1.002E-01 | 1.545E+10 | 1.177E-01 |
| | $4s\,4p^3\,^1P_1^o$ | $4p^4\,^1D_2^e$ | 2.206E+10 | 1.744E-01 | 2.522E+10 | 1.923E-01 |
| **63** | $4s^2\,4p^2\,^3P_0^e$ | $4s\,4p^3\,^3D_1^o$ | 1.813E+11 | 8.962E-01 | 8.962E-01 | 8.962E-01 |
| | $4s^2\,4p^2\,^3P_1^e$ | $4s\,4p^3\,^3S_1^o$ | 8.784E+10 | 1.310E-01 | 9.867E+10 | 1.455E-01 |
| | $4s^2\,4p^2\,^3P_1^e$ | $4s\,4p^3\,^3P_1^o$ | 2.535E+11 | 4.017E-01 | 2.234E+11 | 3.490E-01 |
| | $4s^2\,4p^2\,^3P_2^e$ | $4s\,4p^3\,^3D_3^o$ | 1.706E+10 | 1.182E-01 | 1.926E+10 | 1.277E-01 |
| | $4s^2\,4p^2\,^3P_2^e$ | $4s\,4p^3\,^3P_2^o$ | 1.894E+11 | 3.421E-01 | 1.878E+11 | 3.342E-01 |
| | $4s^2\,4p^2\,^3P_2^e$ | $4s\,4p^3\,^1P_1^o$ | 3.359E+11 | 3.015E-01 | 3.450E+11 | 3.048E-01 |
| | $4s^2\,4p^2\,^3P_0^e$ | $4s^2\,4p\,4d\,^3D_1^o$ | 5.035E+11 | 1.281E+00 | 5.583E+11 | 1.419E+00 |
| | $4s^2\,4p^2\,^1D_2^e$ | $4s^2\,4p\,4d\,^1F_3^o$ | 1.227E+11 | 1.270E-01 | 9.823E+10 | 1.019E-01 |
| | $4s^2\,4p^2\,^1D_2^e$ | $4s^2\,4p\,4d\,^3F_3^o$ | 1.217E+11 | 2.387E-01 | 1.405E+11 | 2.754E-01 |
| | $4s^2\,4p^2\,^3P_2^e$ | $4s^2\,4p\,4d\,^3P_2^o$ | 1.664E+11 | 2.396E-01 | 1.894E+11 | 2.655E-01 |
| | $4s^2\,4p^2\,^3P_2^e$ | $4s^2\,4p\,4d\,^1F_3^o$ | 4.568E+11 | 7.843E-01 | 4.711E+11 | 8.129E-01 |
| | $4s\,4p^3\,^3P_0^o$ | $4p^4\,^3P_1^e$ | 4.981E+10 | 2.490E-01 | 5.393E+10 | 2.564E-01 |
| | $4s\,4p^3\,^3P_1^o$ | $4p^4\,^3P_2^e$ | 1.154E+10 | 7.751E-02 | 9.690E+09 | 6.293E-02 |
| | $4s\,4p^3\,^1D_2^o$ | $4p^4\,^1D_2^e$ | 1.689E+11 | 2.943E-01 | 2.180E+11 | 3.682E-01 |
| | $4s\,4p^3\,^3D_1^o$ | $4p^4\,^3P_2^e$ | 9.512E+10 | 2.746E-01 | 1.118E+11 | 3.161E-01 |
| | $4s\,4p^3\,^3D_1^o$ | $4p^4\,^1S_0^e$ | 1.822E+11 | 9.422E-02 | 2.463E+11 | 1.259E-01 |
| | $4s\,4p^3\,^3S_1^o$ | $4p^4\,^3P_2^e$ | 1.311E+10 | 9.587E-02 | 1.561E+10 | 1.114E-01 |
| | $4s\,4p^3\,^1P_1^o$ | $4p^4\,^1D_2^e$ | 2.283E+10 | 1.688E-01 | 2.602E+10 | 1.862E-01 |

Table 12: Collision Strengths of allowed transitions for **Pr XXVIII** ion.

| Lower | Upper | Incident Energy (eV) | | | | | | | | |
|---|---|---|---|---|---|---|---|---|---|---|
| | | 200 | 500 | 1000 | 1500 | 2000 | 3000 | 4000 | 5000 | 10000 |
| $4s^2\ 4p^2$ $^3P_0^e$ | $4s\ 4p^3\ ^3D_1^o$ | 9.386E-01 | 1.488E+00 | 1.870E+00 | 2.088E+00 | 2.241E+00 | 2.455E+00 | 2.606E+00 | 2.722E+00 | 3.080E+00 |
| | $4s\ 4p^3\ ^3P_1^o$ | 1.740E-03 | 3.219E-03 | 4.222E-03 | 4.791E-03 | 5.191E-03 | 5.749E-03 | 6.142E-03 | 6.445E-03 | 7.376E-03 |
| | $4s\ 4p^3\ ^3S_1^o$ | 4.262E-02 | 8.192E-02 | 1.085E-01 | 1.236E-01 | 1.342E-01 | 1.491E-01 | 1.595E-01 | 1.675E-01 | 1.923E-01 |
| | $4s^2\ 4p\ 4d\ ^3D_1^o$ | 5.231E-01 | 1.104E+00 | 1.495E+00 | 1.717E+00 | 1.873E+00 | 2.091E+00 | 2.244E+00 | 2.362E+00 | 2.725E+00 |
| | $4s^2\ 4p\ 4d\ ^3P_1^o$ | 1.116E-04 | 3.155E-04 | 4.488E-04 | 5.245E-04 | 5.775E-04 | 6.516E-04 | 7.038E-04 | 7.440E-04 | 8.676E-04 |
| $4s^2\ 4p^2$ $^3P_1^e$ | $4s\ 4p^3\ ^3D_1^o$ | 1.994E-02 | 2.805E-02 | 3.378E-02 | 3.707E-02 | 3.938E-02 | 4.262E-02 | 4.490E-02 | 4.666E-02 | 5.208E-02 |
| | $4s\ 4p^3\ ^3D_2^o$ | 9.749E-01 | 1.457E+00 | 1.795E+00 | 1.989E+00 | 2.125E+00 | 2.315E+00 | 2.449E+00 | 2.552E+00 | 2.870E+00 |
| | $4s\ 4p^3\ ^3P_0^o$ | 2.575E-01 | 4.066E-01 | 5.104E-01 | 5.696E-01 | 6.111E-01 | 6.693E-01 | 7.102E-01 | 7.418E-01 | 8.390E-01 |
| | $4s\ 4p^3\ ^3P_1^o$ | 8.426E-01 | 1.351E+00 | 1.703E+00 | 1.904E+00 | 2.046E+00 | 2.243E+00 | 2.382E+00 | 2.489E+00 | 2.819E+00 |
| | $4s\ 4p^3\ ^3P_2^o$ | 2.076E-03 | 3.845E-03 | 5.046E-03 | 5.729E-03 | 6.207E-03 | 6.876E-03 | 7.347E-03 | 7.711E-03 | 8.827E-03 |
| | $4s\ 4p^3\ ^3S_1^o$ | 3.887E-01 | 6.365E-01 | 8.078E-01 | 9.054E-01 | 9.740E-01 | 1.070E+00 | 1.137E+00 | 1.189E+00 | 1.349E+00 |
| | $4s\ 4p^3\ ^5S_2^o$ | 5.597E-01 | 7.555E-01 | 8.958E-01 | 9.765E-01 | 1.033E+00 | 1.113E+00 | 1.169E+00 | 1.212E+00 | 1.345E+00 |
| | $4s\ 4p^3\ ^1D_2^o$ | 7.205E-02 | 1.176E-01 | 1.492E-01 | 1.672E-01 | 1.798E-01 | 1.975E-01 | 2.100E-01 | 2.196E-01 | 2.491E-01 |
| | $4s^2\ 4p\ 4d\ ^1D_2^o$ | 4.270E-01 | 8.774E-01 | 1.180E+00 | 1.352E+00 | 1.472E+00 | 1.641E+00 | 1.759E+00 | 1.851E+00 | 2.132E+00 |
| | $4s^2\ 4p\ 4d\ ^3D_1^o$ | 1.122E-01 | 1.980E-01 | 2.570E-01 | 2.906E-01 | 3.142E-01 | 3.472E-01 | 3.705E-01 | 3.884E-01 | 4.435E-01 |
| | $4s^2\ 4p\ 4d\ ^3D_2^o$ | 7.488E-01 | 1.406E+00 | 1.862E+00 | 2.123E+00 | 2.307E+00 | 2.563E+00 | 2.744E+00 | 2.884E+00 | 3.313E+00 |
| | $4s^2\ 4p\ 4d\ ^3P_0^o$ | 1.265E-01 | 2.671E-01 | 3.616E-01 | 4.152E-01 | 4.529E-01 | 5.054E-01 | 5.425E-01 | 5.710E-01 | 6.588E-01 |
| | $4s^2\ 4p\ 4d\ ^3P_1^o$ | 3.078E-01 | 6.529E-01 | 8.847E-01 | 1.016E+00 | 1.109E+00 | 1.238E+00 | 1.329E+00 | 1.399E+00 | 1.614E+00 |
| | $4s^2\ 4p\ 4d\ ^3P_2^o$ | 4.413E-02 | 1.123E-01 | 1.595E-01 | 1.866E-01 | 2.057E-01 | 2.324E-01 | 2.512E-01 | 2.657E-01 | 3.104E-01 |
| $4s^2\ 4p^2$ $^1D_2^e$ | $4s\ 4p^3\ ^1P_1^o$ | 1.101E-06 | 1.953E-06 | 1.986E-06 | 2.004E-06 | 2.017E-06 | 2.035E-06 | 2.048E-06 | 2.058E-06 | 2.090E-06 |
| | $4s\ 4p^3\ ^1D_2^o$ | 2.302E+00 | 3.688E+00 | 4.649E+00 | 5.197E+00 | 5.582E+00 | 6.120E+00 | 6.499E+00 | 6.792E+00 | 7.691E+00 |
| | $4s\ 4p^3\ ^3D_1^o$ | 2.981E-01 | 4.156E-01 | 4.992E-01 | 5.471E-01 | 5.808E-01 | 6.280E-01 | 6.613E-01 | 6.870E-01 | 7.660E-01 |
| | $4s\ 4p^3\ ^3D_2^o$ | 3.175E-02 | 4.665E-02 | 5.712E-02 | 6.312E-02 | 6.733E-02 | 7.322E-02 | 7.737E-02 | 8.057E-02 | 9.042E-02 |
| | $4s\ 4p^3\ ^3D_3^o$ | 8.394E-01 | 1.277E+00 | 1.583E+00 | 1.758E+00 | 1.881E+00 | 2.053E+00 | 2.174E+00 | 2.268E+00 | 2.555E+00 |
| | $4s\ 4p^3\ ^3P_1^o$ | 8.853E-02 | 1.398E-01 | 1.755E-01 | 1.959E-01 | 2.102E-01 | 2.302E-01 | 2.443E-01 | 2.551E-01 | 2.885E-01 |
| | $4s\ 4p^3\ ^3P_2^o$ | 7.927E-02 | 1.441E-01 | 1.883E-01 | 2.134E-01 | 2.310E-01 | 2.556E-01 | 2.730E-01 | 2.864E-01 | 3.275E-01 |
| | $4s^2\ 4p\ 4d\ ^1P_1^o$ | 5.932E-03 | 1.258E-02 | 1.369E-02 | 1.439E-02 | 1.491E-02 | 1.565E-02 | 1.618E-02 | 1.660E-02 | 1.789E-02 |
| | $4s^2\ 4p\ 4d\ ^1D_2^o$ | 2.270E-01 | 4.747E-01 | 6.440E-01 | 7.405E-01 | 8.083E-01 | 9.030E-01 | 9.697E-01 | 1.021E+00 | 1.179E+00 |
| | $4s^2\ 4p\ 4d\ ^1F_3^o$ | 1.943E-01 | 5.355E-01 | 7.730E-01 | 9.092E-01 | 1.005E+00 | 1.139E+00 | 1.234E+00 | 1.307E+00 | 1.532E+00 |
| | $4s^2\ 4p\ 4d\ ^3D_1^o$ | 1.152E-02 | 2.536E-02 | 3.199E-02 | 3.590E-02 | 3.868E-02 | 4.260E-02 | 4.537E-02 | 4.752E-02 | 5.415E-02 |
| | $4s^2\ 4p\ 4d\ ^3D_2^o$ | 1.140E-01 | 2.312E-01 | 3.127E-01 | 3.595E-01 | 3.925E-01 | 4.387E-01 | 4.713E-01 | 4.964E-01 | 5.739E-01 |
| | $4s^2\ 4p\ 4d\ ^3D_3^o$ | 1.358E+00 | 2.751E+00 | 3.678E+00 | 4.204E+00 | 4.573E+00 | 5.087E+00 | 5.450E+00 | 5.729E+00 | 6.587E+00 |
| | $4s^2\ 4p\ 4d\ ^3P_1^o$ | 9.110E-02 | 1.784E-01 | 2.356E-01 | 2.679E-01 | 2.906E-01 | 3.223E-01 | 3.445E-01 | 3.617E-01 | 4.144E-01 |
| | $4s^2\ 4p\ 4d\ ^3P_2^o$ | 1.808E-01 | 3.316E-01 | 4.218E-01 | 4.723E-01 | 5.076E-01 | 5.567E-01 | 5.912E-01 | 6.178E-01 | 6.993E-01 |
| | $4s^2\ 4p\ 4d\ ^3F_2^o$ | 1.061E-01 | 2.028E-01 | 2.693E-01 | 3.076E-01 | 3.346E-01 | 3.724E-01 | 3.991E-01 | 4.197E-01 | 4.832E-01 |
| | $4s^2\ 4p\ 4d\ ^3F_3^o$ | 5.979E-01 | 1.129E+00 | 1.500E+00 | 1.712E+00 | 1.862E+00 | 2.071E+00 | 2.218E+00 | 2.332E+00 | 2.682E+00 |
| $4s^2\ 4p^2$ $^3P_2^e$ | $4s\ 4p^3\ ^3D_1^o$ | 3.731E-02 | 4.728E-02 | 5.455E-02 | 5.874E-02 | 6.169E-02 | 6.584E-02 | 6.877E-02 | 7.103E-02 | 7.798E-02 |
| | $4s\ 4p^3\ ^3D_2^o$ | 1.364E-01 | 1.805E-01 | 2.123E-01 | 2.306E-01 | 2.434E-01 | 2.615E-01 | 2.742E-01 | 2.840E-01 | 3.142E-01 |
| | $4s\ 4p^3\ ^3D_3^o$ | 1.572E+00 | 2.134E+00 | 2.536E+00 | 2.767E+00 | 2.929E+00 | 3.157E+00 | 3.318E+00 | 3.442E+00 | 3.823E+00 |
| | $4s\ 4p^3\ ^3P_1^o$ | 2.098E-02 | 2.910E-02 | 3.487E-02 | 3.817E-02 | 4.050E-02 | 4.375E-02 | 4.605E-02 | 4.782E-02 | 5.327E-02 |
| | $4s\ 4p^3\ ^3P_2^o$ | 2.066E+00 | 3.213E+00 | 4.012E+00 | 4.468E+00 | 4.788E+00 | 5.237E+00 | 5.552E+00 | 5.796E+00 | 6.545E+00 |
| | $4s\ 4p^3\ ^3S_1^o$ | 2.218E-01 | 3.143E-01 | 3.797E-01 | 4.173E-01 | 4.436E-01 | 4.806E-01 | 5.066E-01 | 5.267E-01 | 5.885E-01 |
| | $4s\ 4p^3\ ^5S_2^o$ | 2.768E-02 | 3.374E-02 | 3.820E-02 | 4.078E-02 | 4.260E-02 | 4.516E-02 | 4.697E-02 | 4.837E-02 | 5.267E-02 |
| | $4s\ 4p^3\ ^1D_2^o$ | 3.512E-01 | 4.955E-01 | 5.978E-01 | 6.564E-01 | 6.976E-01 | 7.553E-01 | 7.960E-01 | 8.274E-01 | 9.240E-01 |
| | $4s^2\ 4p\ 4d\ ^1D_2^o$ | 2.577E-02 | 4.020E-02 | 4.276E-02 | 4.431E-02 | 4.543E-02 | 4.702E-02 | 4.815E-02 | 4.903E-02 | 5.175E-02 |
| | $4s^2\ 4p\ 4d\ ^3D_1^o$ | 3.065E-04 | 5.163E-04 | 6.603E-04 | 7.432E-04 | 8.016E-04 | 8.834E-04 | 9.412E-04 | 9.857E-04 | 1.123E-03 |
| | $4s^2\ 4p\ 4d\ ^3D_2^o$ | 3.795E-03 | 6.893E-03 | 8.897E-03 | 1.005E-02 | 1.087E-02 | 1.202E-02 | 1.283E-02 | 1.345E-02 | 1.537E-02 |
| | $4s^2\ 4p\ 4d\ ^3D_3^o$ | 3.719E-01 | 5.647E-01 | 6.852E-01 | 7.532E-01 | 8.007E-01 | 8.670E-01 | 9.136E-01 | 9.495E-01 | 1.060E+00 |
| | $4s^2\ 4p\ 4d\ ^3P_1^o$ | 1.751E-01 | 2.801E-01 | 3.499E-01 | 3.895E-01 | 4.172E-01 | 4.560E-01 | 4.832E-01 | 5.043E-01 | 5.689E-01 |
| | $4s^2\ 4p\ 4d\ ^3P_2^o$ | 8.656E-01 | 1.526E+00 | 1.982E+00 | 2.242E+00 | 2.424E+00 | 2.679E+00 | 2.858E+00 | 2.997E+00 | 3.423E+00 |
| $4s^2\ 4p^2$ $^1S_0^e$ | $4s\ 4p^3\ ^1P_1^o$ | 6.689E-01 | 1.060E+00 | 1.332E+00 | 1.488E+00 | 1.597E+00 | 1.749E+00 | 1.856E+00 | 1.939E+00 | 2.194E+00 |
| | $4s\ 4p^3\ ^3S_1^o$ | 3.226E-01 | 4.473E-01 | 5.361E-01 | 5.871E-01 | 6.230E-01 | 6.732E-01 | 7.086E-01 | 7.359E-01 | 8.200E-01 |
| | $4s^2\ 4p\ 4d\ ^1P_1^o$ | 6.775E-01 | 1.214E+00 | 1.583E+00 | 1.793E+00 | 1.940E+00 | 2.146E+00 | 2.292E+00 | 2.404E+00 | 2.748E+00 |

Table 13: Collision Strengths of allowed transitions for **Nd XXIX** ion.

| Lower | Upper | Incident Energy (eV) | | | | | | | | |
|---|---|---|---|---|---|---|---|---|---|---|
| | | 200 | 500 | 1000 | 1500 | 2000 | 3000 | 4000 | 5000 | 10000 |
| $4s^2\,4p^2$ $^3P_0^e$ | $4s\,4p^3\,^3D_1^o$ | 8.631E-01 | 1.395E+00 | 1.762E+00 | 1.971E+00 | 2.117E+00 | 2.322E+00 | 2.467E+00 | 2.578E+00 | 2.921E+00 |
| | $4s\,4p^3\,^3P_1^o$ | 1.288E-03 | 2.493E-03 | 3.299E-03 | 3.756E-03 | 4.076E-03 | 4.523E-03 | 4.838E-03 | 5.081E-03 | 5.826E-03 |
| | $4s\,4p^3\,^3S_1^o$ | 3.346E-02 | 6.745E-02 | 9.016E-02 | 1.030E-01 | 1.121E-01 | 1.247E-01 | 1.336E-01 | 1.404E-01 | 1.615E-01 |
| | $4s^2\,4p\,4d\,^3D_1^o$ | 4.653E-01 | 1.025E+00 | 1.396E+00 | 1.607E+00 | 1.754E+00 | 1.960E+00 | 2.105E+00 | 2.217E+00 | 2.561E+00 |
| | $4s^2\,4p\,4d\,^3P_1^o$ | 7.739E-05 | 2.491E-04 | 3.585E-04 | 4.205E-04 | 4.639E-04 | 5.245E-04 | 5.672E-04 | 6.001E-04 | 7.012E-04 |
| $4s^2\,4p^2$ $^3P_1^e$ | $4s\,4p^3\,^3D_1^o$ | 2.337E-02 | 3.306E-02 | 3.989E-02 | 4.380E-02 | 4.655E-02 | 5.040E-02 | 5.311E-02 | 5.521E-02 | 6.164E-02 |
| | $4s\,4p^3\,^3D_2^o$ | 8.646E-01 | 1.311E+00 | 1.622E+00 | 1.799E+00 | 1.924E+00 | 2.098E+00 | 2.221E+00 | 2.316E+00 | 2.608E+00 |
| | $4s\,4p^3\,^3P_0^o$ | 2.353E-01 | 3.783E-01 | 4.770E-01 | 5.333E-01 | 5.728E-01 | 6.280E-01 | 6.669E-01 | 6.969E-01 | 7.891E-01 |
| | $4s\,4p^3\,^3P_1^o$ | 7.742E-01 | 1.266E+00 | 1.604E+00 | 1.797E+00 | 1.932E+00 | 2.121E+00 | 2.254E+00 | 2.357E+00 | 2.673E+00 |
| | $4s\,4p^3\,^3P_2^o$ | 1.526E-03 | 2.949E-03 | 3.902E-03 | 4.444E-03 | 4.823E-03 | 5.353E-03 | 5.726E-03 | 6.013E-03 | 6.897E-03 |
| | $4s\,4p^3\,^3S_1^o$ | 3.450E-01 | 5.761E-01 | 7.345E-01 | 8.247E-01 | 8.880E-01 | 9.764E-01 | 1.039E+00 | 1.087E+00 | 1.234E+00 |
| | $4s\,4p^3\,^5S_2^o$ | 5.656E-01 | 7.667E-01 | 9.102E-01 | 9.927E-01 | 1.051E+00 | 1.132E+00 | 1.189E+00 | 1.234E+00 | 1.370E+00 |
| | $4s\,4p^3\,^1D_2^o$ | 6.904E-02 | 1.152E-01 | 1.469E-01 | 1.650E-01 | 1.777E-01 | 1.954E-01 | 2.079E-01 | 2.175E-01 | 2.471E-01 |
| | $4s^2\,4p\,4d\,^1D_2^o$ | 3.784E-01 | 8.093E-01 | 1.095E+00 | 1.257E+00 | 1.370E+00 | 1.529E+00 | 1.640E+00 | 1.726E+00 | 1.991E+00 |
| | $4s^2\,4p\,4d\,^3D_1^o$ | 1.039E-01 | 1.855E-01 | 2.413E-01 | 2.731E-01 | 2.953E-01 | 3.265E-01 | 3.484E-01 | 3.653E-01 | 4.173E-01 |
| | $4s^2\,4p\,4d\,^3D_2^o$ | 7.018E-01 | 1.338E+00 | 1.778E+00 | 2.030E+00 | 2.206E+00 | 2.453E+00 | 2.626E+00 | 2.761E+00 | 3.173E+00 |
| | $4s^2\,4p\,4d\,^3P_0^o$ | 1.126E-01 | 2.480E-01 | 3.377E-01 | 3.886E-01 | 4.243E-01 | 4.741E-01 | 5.092E-01 | 5.362E-01 | 6.193E-01 |
| | $4s^2\,4p\,4d\,^3P_1^o$ | 2.742E-01 | 6.070E-01 | 8.276E-01 | 9.526E-01 | 1.040E+00 | 1.163E+00 | 1.249E+00 | 1.315E+00 | 1.519E+00 |
| | $4s^2\,4p\,4d\,^3P_2^o$ | 3.346E-02 | 9.127E-02 | 1.311E-01 | 1.539E-01 | 1.699E-01 | 1.924E-01 | 2.082E-01 | 2.204E-01 | 2.580E-01 |
| $4s^2\,4p^2$ $^1D_2^e$ | $4s\,4p^3\,^1P_1^o$ | 5.584E-07 | 1.664E-06 | 1.954E-06 | 2.125E-06 | 2.245E-06 | 2.416E-06 | 2.537E-06 | 2.631E-06 | 2.922E-06 |
| | $4s\,4p^3\,^1D_2^o$ | 2.105E+00 | 3.442E+00 | 4.362E+00 | 4.887E+00 | 5.255E+00 | 5.769E+00 | 6.131E+00 | 6.410E+00 | 7.269E+00 |
| | $4s\,4p^3\,^3D_1^o$ | 3.038E-01 | 4.257E-01 | 5.121E-01 | 5.615E-01 | 5.963E-01 | 6.450E-01 | 6.794E-01 | 7.059E-01 | 7.874E-01 |
| | $4s\,4p^3\,^3D_2^o$ | 2.769E-02 | 4.122E-02 | 5.068E-02 | 5.609E-02 | 5.988E-02 | 6.519E-02 | 6.894E-02 | 7.182E-02 | 8.070E-02 |
| | $4s\,4p^3\,^3D_3^o$ | 7.570E-01 | 1.170E+00 | 1.457E+00 | 1.621E+00 | 1.736E+00 | 1.897E+00 | 2.010E+00 | 2.097E+00 | 2.366E+00 |
| | $4s\,4p^3\,^3P_1^o$ | 8.150E-02 | 1.311E-01 | 1.654E-01 | 1.849E-01 | 1.987E-01 | 2.178E-01 | 2.313E-01 | 2.417E-01 | 2.738E-01 |
| | $4s\,4p^3\,^3P_2^o$ | 5.931E-02 | 1.123E-01 | 1.480E-01 | 1.682E-01 | 1.824E-01 | 2.023E-01 | 2.162E-01 | 2.270E-01 | 2.601E-01 |
| | $4s^2\,4p\,4d\,^1P_1^o$ | 4.973E-03 | 1.190E-02 | 1.289E-02 | 1.349E-02 | 1.395E-02 | 1.460E-02 | 1.506E-02 | 1.543E-02 | 1.656E-02 |
| | $4s^2\,4p\,4d\,^1D_2^o$ | 2.128E-01 | 4.615E-01 | 6.298E-01 | 7.255E-01 | 7.926E-01 | 8.865E-01 | 9.525E-01 | 1.004E+00 | 1.160E+00 |
| | $4s^2\,4p\,4d\,^1F_3^o$ | 1.537E-01 | 4.570E-01 | 6.666E-01 | 7.867E-01 | 8.712E-01 | 9.896E-01 | 1.073E+00 | 1.137E+00 | 1.336E+00 |
| | $4s^2\,4p\,4d\,^3D_1^o$ | 1.022E-02 | 2.409E-02 | 3.042E-02 | 3.414E-02 | 3.678E-02 | 4.051E-02 | 4.315E-02 | 4.519E-02 | 5.150E-02 |
| | $4s^2\,4p\,4d\,^3D_2^o$ | 1.021E-01 | 2.114E-01 | 2.869E-01 | 3.303E-01 | 3.608E-01 | 4.036E-01 | 4.337E-01 | 4.570E-01 | 5.286E-01 |
| | $4s^2\,4p\,4d\,^3D_3^o$ | 1.206E+00 | 2.546E+00 | 3.424E+00 | 3.922E+00 | 4.270E+00 | 4.756E+00 | 5.099E+00 | 5.362E+00 | 6.173E+00 |
| | $4s^2\,4p\,4d\,^3P_1^o$ | 8.009E-02 | 1.633E-01 | 2.170E-01 | 2.473E-01 | 2.685E-01 | 2.981E-01 | 3.190E-01 | 3.350E-01 | 3.844E-01 |
| | $4s^2\,4p\,4d\,^3P_2^o$ | 1.485E-01 | 2.830E-01 | 3.602E-01 | 4.032E-01 | 4.333E-01 | 4.751E-01 | 5.044E-01 | 5.270E-01 | 5.963E-01 |
| | $4s^2\,4p\,4d\,^3F_2^o$ | 1.003E-01 | 1.942E-01 | 2.585E-01 | 2.954E-01 | 3.214E-01 | 3.578E-01 | 3.835E-01 | 4.034E-01 | 4.644E-01 |
| | $4s^2\,4p\,4d\,^3F_3^o$ | 5.853E-01 | 1.121E+00 | 1.494E+00 | 1.706E+00 | 1.856E+00 | 2.066E+00 | 2.213E+00 | 2.327E+00 | 2.677E+00 |
| $4s^2\,4p^2$ $^3P_2^e$ | $4s\,4p^3\,^3D_1^o$ | 3.292E-02 | 4.159E-02 | 4.789E-02 | 5.153E-02 | 5.410E-02 | 5.770E-02 | 6.024E-02 | 6.220E-02 | 6.824E-02 |
| | $4s\,4p^3\,^3D_2^o$ | 1.381E-01 | 1.832E-01 | 2.156E-01 | 2.343E-01 | 2.474E-01 | 2.658E-01 | 2.787E-01 | 2.887E-01 | 3.195E-01 |
| | $4s\,4p^3\,^3D_3^o$ | 1.494E+00 | 2.035E+00 | 2.421E+00 | 2.642E+00 | 2.798E+00 | 3.016E+00 | 3.170E+00 | 3.289E+00 | 3.655E+00 |
| | $4s\,4p^3\,^3P_1^o$ | 2.508E-02 | 3.498E-02 | 4.198E-02 | 4.599E-02 | 4.881E-02 | 5.276E-02 | 5.554E-02 | 5.769E-02 | 6.429E-02 |
| | $4s\,4p^3\,^3P_2^o$ | 1.851E+00 | 2.924E+00 | 3.668E+00 | 4.091E+00 | 4.389E+00 | 4.805E+00 | 5.098E+00 | 5.324E+00 | 6.019E+00 |
| | $4s\,4p^3\,^3S_1^o$ | 2.139E-01 | 3.043E-01 | 3.680E-01 | 4.045E-01 | 4.302E-01 | 4.661E-01 | 4.914E-01 | 5.110E-01 | 5.710E-01 |
| | $4s\,4p^3\,^5S_2^o$ | 2.432E-02 | 2.953E-02 | 3.337E-02 | 3.559E-02 | 3.716E-02 | 3.936E-02 | 4.092E-02 | 4.212E-02 | 4.582E-02 |
| | $4s\,4p^3\,^1D_2^o$ | 3.511E-01 | 4.982E-01 | 6.021E-01 | 6.615E-01 | 7.034E-01 | 7.619E-01 | 8.032E-01 | 8.350E-01 | 9.330E-01 |
| | $4s^2\,4p\,4d\,^1D_2^o$ | 2.597E-02 | 3.956E-02 | 4.057E-02 | 4.118E-02 | 4.163E-02 | 4.227E-02 | 4.272E-02 | 4.308E-02 | 4.419E-02 |
| | $4s^2\,4p\,4d\,^3D_1^o$ | 2.678E-04 | 4.501E-04 | 5.744E-04 | 6.459E-04 | 6.963E-04 | 7.668E-04 | 8.165E-04 | 8.550E-04 | 9.732E-04 |
| | $4s^2\,4p\,4d\,^3D_2^o$ | 3.072E-03 | 5.626E-03 | 7.251E-03 | 8.189E-03 | 8.850E-03 | 9.778E-03 | 1.043E-02 | 1.094E-02 | 1.250E-02 |
| | $4s^2\,4p\,4d\,^3D_3^o$ | 3.184E-01 | 4.869E-01 | 5.890E-01 | 6.465E-01 | 6.867E-01 | 7.426E-01 | 7.820E-01 | 8.123E-01 | 9.054E-01 |
| | $4s^2\,4p\,4d\,^3P_1^o$ | 1.645E-01 | 2.658E-01 | 3.324E-01 | 3.702E-01 | 3.966E-01 | 4.335E-01 | 4.594E-01 | 4.794E-01 | 5.409E-01 |
| | $4s^2\,4p\,4d\,^3P_2^o$ | 8.042E-01 | 1.441E+00 | 1.877E+00 | 2.126E+00 | 2.300E+00 | 2.544E+00 | 2.716E+00 | 2.848E+00 | 3.255E+00 |
| $4s^2\,4p^2$ $^1S_0^e$ | $4s\,4p^3\,^1P_1^o$ | 6.048E-01 | 9.758E-01 | 1.232E+00 | 1.378E+00 | 1.480E+00 | 1.623E+00 | 1.724E+00 | 1.802E+00 | 2.041E+00 |
| | $4s\,4p^3\,^3S_1^o$ | 3.077E-01 | 4.281E-01 | 5.136E-01 | 5.626E-01 | 5.970E-01 | 6.453E-01 | 6.793E-01 | 7.056E-01 | 7.864E-01 |
| | $4s^2\,4p\,4d\,^1P_1^o$ | 6.254E-01 | 1.138E+00 | 1.487E+00 | 1.686E+00 | 1.826E+00 | 2.021E+00 | 2.158E+00 | 2.264E+00 | 2.590E+00 |

Table 14: Collision Strengths of allowed transitions for **Pm XXX** ion.

| Lower | Upper | Incident Energy (eV) | | | | | | | | |
|---|---|---|---|---|---|---|---|---|---|---|
| | | 200 | 500 | 1000 | 1500 | 2000 | 3000 | 4000 | 5000 | 10000 |
| $4s^2\,4p^2$ $^3P_0^e$ | $4s\,4p^3\,^3D_1^o$ | 7.393E-01 | 1.219E+00 | 1.548E+00 | 1.734E+00 | 1.865E+00 | 2.049E+00 | 2.178E+00 | 2.277E+00 | 2.583E+00 |
| | $4s\,4p^3\,^3P_1^o$ | 3.640E-03 | 7.489E-03 | 1.002E-02 | 1.146E-02 | 1.246E-02 | 1.386E-02 | 1.485E-02 | 1.561E-02 | 1.795E-02 |
| | $4s\,4p^3\,^3S_1^o$ | 2.548E-02 | 5.478E-02 | 7.404E-02 | 8.495E-02 | 9.259E-02 | 1.033E-01 | 1.108E-01 | 1.166E-01 | 1.343E-01 |
| | $4s^2\,4p\,4d\,^3D_1^o$ | 4.014E-01 | 9.317E-01 | 1.278E+00 | 1.474E+00 | 1.611E+00 | 1.802E+00 | 1.937E+00 | 2.041E+00 | 2.360E+00 |
| | $4s^2\,4p\,4d\,^3P_1^o$ | 2.558E-05 | 9.768E-05 | 1.408E-04 | 1.651E-04 | 1.820E-04 | 2.057E-04 | 2.224E-04 | 2.352E-04 | 2.746E-04 |
| $4s^2\,4p^2$ $^3P_1^e$ | $4s\,4p^3\,^3D_1^o$ | 2.947E-02 | 4.182E-02 | 5.050E-02 | 5.547E-02 | 5.896E-02 | 6.385E-02 | 6.729E-02 | 6.995E-02 | 7.812E-02 |
| | $4s\,4p^3\,^3D_2^o$ | 7.669E-01 | 1.181E+00 | 1.468E+00 | 1.631E+00 | 1.746E+00 | 1.906E+00 | 2.019E+00 | 2.107E+00 | 2.375E+00 |
| | $4s\,4p^3\,^3P_0^o$ | 2.147E-01 | 3.521E-01 | 4.462E-01 | 4.997E-01 | 5.373E-01 | 5.898E-01 | 6.268E-01 | 6.553E-01 | 7.430E-01 |
| | $4s\,4p^3\,^3P_1^o$ | 9.408E-01 | 1.570E+00 | 2.000E+00 | 2.244E+00 | 2.415E+00 | 2.655E+00 | 2.823E+00 | 2.953E+00 | 3.353E+00 |
| | $4s\,4p^3\,^3P_2^o$ | 1.110E-03 | 2.258E-03 | 3.015E-03 | 3.444E-03 | 3.745E-03 | 4.164E-03 | 4.460E-03 | 4.687E-03 | 5.386E-03 |
| | $4s\,4p^3\,^3S_1^o$ | 3.059E-01 | 5.220E-01 | 6.688E-01 | 7.523E-01 | 8.108E-01 | 8.926E-01 | 9.502E-01 | 9.946E-01 | 1.131E+00 |
| | $4s\,4p^3\,^5S_2^o$ | 5.605E-01 | 7.632E-01 | 9.074E-01 | 9.901E-01 | 1.048E+00 | 1.130E+00 | 1.187E+00 | 1.232E+00 | 1.368E+00 |
| | $4s\,4p^3\,^1D_2^o$ | 6.777E-02 | 1.155E-01 | 1.480E-01 | 1.665E-01 | 1.794E-01 | 1.975E-01 | 2.103E-01 | 2.201E-01 | 2.503E-01 |
| | $4s^2\,4p\,4d\,^1D_2^o$ | 3.334E-01 | 7.474E-01 | 1.018E+00 | 1.170E+00 | 1.278E+00 | 1.427E+00 | 1.532E+00 | 1.613E+00 | 1.862E+00 |
| | $4s^2\,4p\,4d\,^3D_1^o$ | 9.689E-02 | 1.749E-01 | 2.278E-01 | 2.579E-01 | 2.790E-01 | 3.084E-01 | 3.291E-01 | 3.451E-01 | 3.943E-01 |
| | $4s^2\,4p\,4d\,^3D_2^o$ | 6.607E-01 | 1.277E+00 | 1.701E+00 | 1.942E+00 | 2.112E+00 | 2.348E+00 | 2.515E+00 | 2.644E+00 | 3.040E+00 |
| | $4s^2\,4p\,4d\,^3P_0^o$ | 9.979E-02 | 2.294E-01 | 3.140E-01 | 3.619E-01 | 3.955E-01 | 4.423E-01 | 4.752E-01 | 5.007E-01 | 5.787E-01 |
| | $4s^2\,4p\,4d\,^3P_1^o$ | 2.425E-01 | 5.650E-01 | 7.751E-01 | 8.941E-01 | 9.774E-01 | 1.094E+00 | 1.176E+00 | 1.239E+00 | 1.432E+00 |
| | $4s^2\,4p\,4d\,^3P_2^o$ | 2.497E-02 | 7.406E-02 | 1.076E-01 | 1.268E-01 | 1.403E-01 | 1.592E-01 | 1.725E-01 | 1.828E-01 | 2.144E-01 |
| $4s^2\,4p^2$ $^1D_2^e$ | $4s\,4p^3\,^1P_1^o$ | 4.362E-07 | 1.967E-06 | 2.627E-06 | 3.009E-06 | 3.280E-06 | 3.660E-06 | 3.930E-06 | 4.138E-06 | 4.782E-06 |
| | $4s\,4p^3\,^1D_2^o$ | 1.937E+00 | 3.231E+00 | 4.114E+00 | 4.616E+00 | 4.969E+00 | 5.461E+00 | 5.808E+00 | 6.075E+00 | 6.897E+00 |
| | $4s\,4p^3\,^3D_1^o$ | 3.332E-01 | 4.680E-01 | 5.631E-01 | 6.176E-01 | 6.559E-01 | 7.095E-01 | 7.472E-01 | 7.764E-01 | 8.661E-01 |
| | $4s\,4p^3\,^3D_2^o$ | 2.400E-02 | 3.626E-02 | 4.477E-02 | 4.963E-02 | 5.304E-02 | 5.781E-02 | 6.117E-02 | 6.376E-02 | 7.172E-02 |
| | $4s\,4p^3\,^3D_3^o$ | 6.829E-01 | 1.074E+00 | 1.343E+00 | 1.497E+00 | 1.605E+00 | 1.756E+00 | 1.862E+00 | 1.944E+00 | 2.196E+00 |
| | $4s\,4p^3\,^3P_1^o$ | 2.567E-03 | 4.304E-03 | 5.504E-03 | 6.187E-03 | 6.667E-03 | 7.338E-03 | 7.810E-03 | 8.175E-03 | 9.295E-03 |
| | $4s\,4p^3\,^3P_2^o$ | 4.412E-02 | 8.767E-02 | 1.166E-01 | 1.330E-01 | 1.445E-01 | 1.605E-01 | 1.718E-01 | 1.805E-01 | 2.072E-01 |
| | $4s^2\,4p\,4d\,^1P_1^o$ | 4.088E-03 | 1.127E-02 | 1.215E-02 | 1.268E-02 | 1.308E-02 | 1.365E-02 | 1.405E-02 | 1.437E-02 | 1.537E-02 |
| | $4s^2\,4p\,4d\,^1D_2^o$ | 1.973E-01 | 4.465E-01 | 6.130E-01 | 7.075E-01 | 7.738E-01 | 8.663E-01 | 9.315E-01 | 9.817E-01 | 1.136E+00 |
| | $4s^2\,4p\,4d\,^1F_3^o$ | 1.192E-01 | 3.886E-01 | 5.732E-01 | 6.788E-01 | 7.532E-01 | 8.572E-01 | 9.306E-01 | 9.873E-01 | 1.161E+00 |
| | $4s^2\,4p\,4d\,^3D_1^o$ | 9.070E-03 | 2.291E-02 | 2.896E-02 | 3.249E-02 | 3.500E-02 | 3.854E-02 | 4.105E-02 | 4.299E-02 | 4.897E-02 |
| | $4s^2\,4p\,4d\,^3D_2^o$ | 9.169E-02 | 1.940E-01 | 2.643E-01 | 3.045E-01 | 3.329E-01 | 3.726E-01 | 4.005E-01 | 4.222E-01 | 4.886E-01 |
| | $4s^2\,4p\,4d\,^3D_3^o$ | 1.071E+00 | 2.354E+00 | 3.183E+00 | 3.651E+00 | 3.979E+00 | 4.436E+00 | 4.758E+00 | 5.006E+00 | 5.768E+00 |
| | $4s^2\,4p\,4d\,^3P_1^o$ | 7.005E-02 | 1.498E-01 | 2.003E-01 | 2.288E-01 | 2.487E-01 | 2.765E-01 | 2.960E-01 | 3.111E-01 | 3.574E-01 |
| | $4s^2\,4p\,4d\,^3P_2^o$ | 1.209E-01 | 2.419E-01 | 3.080E-01 | 3.447E-01 | 3.703E-01 | 4.059E-01 | 4.308E-01 | 4.501E-01 | 5.090E-01 |
| | $4s^2\,4p\,4d\,^3F_2^o$ | 9.533E-02 | 1.867E-01 | 2.488E-01 | 2.844E-01 | 3.095E-01 | 3.447E-01 | 3.695E-01 | 3.886E-01 | 4.475E-01 |
| | $4s^2\,4p\,4d\,^3F_3^o$ | 5.729E-01 | 1.113E+00 | 1.486E+00 | 1.699E+00 | 1.849E+00 | 2.058E+00 | 2.206E+00 | 2.319E+00 | 2.669E+00 |
| $4s^2\,4p^2$ $^3P_2^e$ | $4s\,4p^3\,^3D_1^o$ | 2.478E-02 | 3.111E-02 | 3.572E-02 | 3.838E-02 | 4.025E-02 | 4.288E-02 | 4.474E-02 | 4.617E-02 | 5.059E-02 |
| | $4s\,4p^3\,^3D_2^o$ | 1.393E-01 | 1.852E-01 | 2.181E-01 | 2.371E-01 | 2.504E-01 | 2.690E-01 | 2.822E-01 | 2.923E-01 | 3.236E-01 |
| | $4s\,4p^3\,^3D_3^o$ | 1.420E+00 | 1.941E+00 | 2.311E+00 | 2.524E+00 | 2.673E+00 | 2.883E+00 | 3.030E+00 | 3.144E+00 | 3.495E+00 |
| | $4s\,4p^3\,^3P_1^o$ | 8.778E-02 | 1.232E-01 | 1.481E-01 | 1.624E-01 | 1.725E-01 | 1.866E-01 | 1.965E-01 | 2.041E-01 | 2.276E-01 |
| | $4s\,4p^3\,^3P_2^o$ | 1.658E+00 | 2.667E+00 | 3.360E+00 | 3.754E+00 | 4.031E+00 | 4.418E+00 | 4.691E+00 | 4.901E+00 | 5.547E+00 |
| | $4s\,4p^3\,^3S_1^o$ | 2.059E-01 | 2.941E-01 | 3.561E-01 | 3.915E-01 | 4.164E-01 | 4.513E-01 | 4.759E-01 | 4.948E-01 | 5.531E-01 |
| | $4s\,4p^3\,^5S_2^o$ | 2.200E-02 | 2.662E-02 | 3.002E-02 | 3.199E-02 | 3.338E-02 | 3.533E-02 | 3.671E-02 | 3.777E-02 | 4.105E-02 |
| | $4s\,4p^3\,^1D_2^o$ | 3.641E-01 | 5.187E-01 | 6.274E-01 | 6.896E-01 | 7.334E-01 | 7.946E-01 | 8.377E-01 | 8.710E-01 | 9.733E-01 |
| | $4s^2\,4p\,4d\,^1D_2^o$ | 2.707E-02 | 4.090E-02 | 4.118E-02 | 4.130E-02 | 4.140E-02 | 4.155E-02 | 4.165E-02 | 4.173E-02 | 4.199E-02 |
| | $4s^2\,4p\,4d\,^3D_1^o$ | 2.317E-04 | 3.874E-04 | 4.926E-04 | 5.531E-04 | 5.956E-04 | 6.553E-04 | 6.973E-04 | 7.298E-04 | 8.297E-04 |
| | $4s^2\,4p\,4d\,^3D_2^o$ | 2.810E-03 | 5.127E-03 | 6.595E-03 | 7.441E-03 | 8.037E-03 | 8.873E-03 | 9.463E-03 | 9.918E-03 | 1.132E-02 |
| | $4s^2\,4p\,4d\,^3D_3^o$ | 2.716E-01 | 4.173E-01 | 5.028E-01 | 5.507E-01 | 5.842E-01 | 6.309E-01 | 6.637E-01 | 6.889E-01 | 7.665E-01 |
| | $4s^2\,4p\,4d\,^3P_1^o$ | 1.544E-01 | 2.524E-01 | 3.160E-01 | 3.519E-01 | 3.771E-01 | 4.122E-01 | 4.369E-01 | 4.560E-01 | 5.144E-01 |
| | $4s^2\,4p\,4d\,^3P_2^o$ | 7.458E-01 | 1.360E+00 | 1.777E+00 | 2.015E+00 | 2.181E+00 | 2.414E+00 | 2.578E+00 | 2.704E+00 | 3.093E+00 |
| $4s^2\,4p^2$ $^1S_0^e$ | $4s\,4p^3\,^1P_1^o$ | 5.462E-01 | 8.989E-01 | 1.140E+00 | 1.277E+00 | 1.374E+00 | 1.508E+00 | 1.603E+00 | 1.676E+00 | 1.901E+00 |
| | $4s\,4p^3\,^3S_1^o$ | 2.935E-01 | 4.098E-01 | 4.921E-01 | 5.392E-01 | 5.724E-01 | 6.188E-01 | 6.515E-01 | 6.767E-01 | 7.544E-01 |
| | $4s^2\,4p\,4d\,^1P_1^o$ | 5.774E-01 | 1.068E+00 | 1.400E+00 | 1.589E+00 | 1.721E+00 | 1.906E+00 | 2.036E+00 | 2.136E+00 | 2.445E+00 |

Table 15: Collision Strengths of allowed transitions for **Sm XXXI** ion.

| Lower | Upper | Incident Energy (eV) | | | | | | | | |
|---|---|---|---|---|---|---|---|---|---|---|
| | | 300 | 500 | 1000 | 1500 | 2000 | 3000 | 4000 | 5000 | 10000 |
| $4s^2\ 4p^2$ $^3P_0^e$ | $4s\ 4p^3\ ^3D_1^o$ | 8.957E-01 | 1.143E+00 | 1.459E+00 | 1.638E+00 | 1.764E+00 | 1.940E+00 | 2.064E+00 | 2.159E+00 | 2.453E+00 |
| | $4s\ 4p^3\ ^3P_1^o$ | 4.365E-03 | 6.069E-03 | 8.209E-03 | 9.418E-03 | 1.026E-02 | 1.145E-02 | 1.228E-02 | 1.292E-02 | 1.488E-02 |
| | $4s\ 4p^3\ ^3S_1^o$ | 3.236E-02 | 4.567E-02 | 6.239E-02 | 7.184E-02 | 7.845E-02 | 8.768E-02 | 9.417E-02 | 9.917E-02 | 1.145E-01 |
| | $4s^2\ 4p\ 4d\ ^3D_1^o$ | 6.038E-01 | 8.671E-01 | 1.197E+00 | 1.384E+00 | 1.515E+00 | 1.697E+00 | 1.825E+00 | 1.924E+00 | 2.228E+00 |
| | $4s^2\ 4p\ 4d\ ^3P_1^o$ | 6.532E-05 | 1.048E-04 | 1.532E-04 | 1.804E-04 | 1.994E-04 | 2.259E-04 | 2.446E-04 | 2.589E-04 | 3.030E-04 |
| $4s^2\ 4p^2$ $^3P_1^e$ | $4s\ 4p^3\ ^3D_1^o$ | 3.860E-02 | 4.603E-02 | 5.567E-02 | 6.119E-02 | 6.506E-02 | 7.048E-02 | 7.430E-02 | 7.724E-02 | 8.630E-02 |
| | $4s\ 4p^3\ ^3D_2^o$ | 8.599E-01 | 1.066E+00 | 1.331E+00 | 1.482E+00 | 1.588E+00 | 1.736E+00 | 1.840E+00 | 1.920E+00 | 2.168E+00 |
| | $4s\ 4p^3\ ^3P_0^o$ | 2.577E-01 | 3.279E-01 | 4.177E-01 | 4.688E-01 | 5.046E-01 | 5.545E-01 | 5.897E-01 | 6.169E-01 | 7.003E-01 |
| | $4s\ 4p^3\ ^3P_1^o$ | 1.155E+00 | 1.480E+00 | 1.896E+00 | 2.132E+00 | 2.297E+00 | 2.528E+00 | 2.691E+00 | 2.816E+00 | 3.201E+00 |
| | $4s\ 4p^3\ ^3P_2^o$ | 1.248E-03 | 1.727E-03 | 2.330E-03 | 2.670E-03 | 2.908E-03 | 3.241E-03 | 3.475E-03 | 3.655E-03 | 4.209E-03 |
| | $4s\ 4p^3\ ^3S_1^o$ | 3.665E-01 | 4.735E-01 | 6.099E-01 | 6.873E-01 | 7.415E-01 | 8.173E-01 | 8.706E-01 | 9.117E-01 | 1.038E+00 |
| | $4s\ 4p^3\ ^5S_2^o$ | 6.453E-01 | 7.556E-01 | 8.996E-01 | 9.822E-01 | 1.040E+00 | 1.122E+00 | 1.179E+00 | 1.223E+00 | 1.359E+00 |
| | $4s\ 4p^3\ ^1D_2^o$ | 8.774E-02 | 1.134E-01 | 1.461E-01 | 1.647E-01 | 1.777E-01 | 1.959E-01 | 2.087E-01 | 2.186E-01 | 2.489E-01 |
| | $4s^2\ 4p\ 4d\ ^1D_2^o$ | 4.957E-01 | 6.982E-01 | 9.524E-01 | 1.096E+00 | 1.197E+00 | 1.337E+00 | 1.435E+00 | 1.511E+00 | 1.745E+00 |
| | $4s^2\ 4p\ 4d\ ^3D_1^o$ | 1.250E-01 | 1.644E-01 | 2.145E-01 | 2.429E-01 | 2.629E-01 | 2.907E-01 | 3.103E-01 | 3.254E-01 | 3.719E-01 |
| | $4s^2\ 4p\ 4d\ ^3D_2^o$ | 8.909E-01 | 1.210E+00 | 1.619E+00 | 1.851E+00 | 2.014E+00 | 2.242E+00 | 2.402E+00 | 2.526E+00 | 2.906E+00 |
| | $4s^2\ 4p\ 4d\ ^3P_0^o$ | 1.488E-01 | 2.131E-01 | 2.938E-01 | 3.394E-01 | 3.713E-01 | 4.158E-01 | 4.471E-01 | 4.712E-01 | 5.454E-01 |
| | $4s^2\ 4p\ 4d\ ^3P_1^o$ | 3.661E-01 | 5.261E-01 | 7.267E-01 | 8.400E-01 | 9.194E-01 | 1.030E+00 | 1.108E+00 | 1.168E+00 | 1.352E+00 |
| | $4s^2\ 4p\ 4d\ ^3P_2^o$ | 3.673E-02 | 5.807E-02 | 8.517E-02 | 1.006E-01 | 1.115E-01 | 1.267E-01 | 1.375E-01 | 1.458E-01 | 1.712E-01 |
| $4s^2\ 4p^2$ $^1D_2^e$ | $4s\ 4p^3\ ^1P_1^o$ | 1.341E-03 | 1.904E-03 | 2.610E-03 | 3.009E-03 | 3.288E-03 | 3.677E-03 | 3.951E-03 | 4.162E-03 | 4.811E-03 |
| | $4s\ 4p^3\ ^1D_2^o$ | 2.198E+00 | 2.818E+00 | 3.608E+00 | 4.057E+00 | 4.372E+00 | 4.812E+00 | 5.121E+00 | 5.360E+00 | 6.094E+00 |
| | $4s\ 4p^3\ ^3D_1^o$ | 3.847E-01 | 4.568E-01 | 5.506E-01 | 6.042E-01 | 6.419E-01 | 6.947E-01 | 7.319E-01 | 7.606E-01 | 8.488E-01 |
| | $4s\ 4p^3\ ^3D_2^o$ | 2.918E-02 | 3.590E-02 | 4.454E-02 | 4.947E-02 | 5.292E-02 | 5.775E-02 | 6.115E-02 | 6.378E-02 | 7.185E-02 |
| | $4s\ 4p^3\ ^3D_3^o$ | 7.234E-01 | 9.036E-01 | 1.135E+00 | 1.267E+00 | 1.359E+00 | 1.488E+00 | 1.579E+00 | 1.649E+00 | 1.865E+00 |
| | $4s\ 4p^3\ ^3P_1^o$ | 6.723E-03 | 8.627E-03 | 1.107E-02 | 1.245E-02 | 1.343E-02 | 1.479E-02 | 1.574E-02 | 1.648E-02 | 1.875E-02 |
| | $4s\ 4p^3\ ^3P_2^o$ | 3.267E-02 | 4.492E-02 | 6.036E-02 | 6.910E-02 | 7.521E-02 | 8.375E-02 | 8.975E-02 | 9.438E-02 | 1.086E-01 |
| | $4s^2\ 4p\ 4d\ ^1P_1^o$ | 8.264E-03 | 1.068E-02 | 1.144E-02 | 1.188E-02 | 1.220E-02 | 1.268E-02 | 1.302E-02 | 1.328E-02 | 1.411E-02 |
| | $4s^2\ 4p\ 4d\ ^1D_2^o$ | 3.053E-01 | 4.340E-01 | 5.969E-01 | 6.891E-01 | 7.538E-01 | 8.440E-01 | 9.075E-01 | 9.565E-01 | 1.107E+00 |
| | $4s^2\ 4p\ 4d\ ^1F_3^o$ | 1.996E-01 | 3.266E-01 | 4.882E-01 | 5.805E-01 | 6.455E-01 | 7.364E-01 | 8.005E-01 | 8.500E-01 | 1.002E+00 |
| | $4s^2\ 4p\ 4d\ ^3D_1^o$ | 1.559E-02 | 2.162E-02 | 2.736E-02 | 3.069E-02 | 3.306E-02 | 3.640E-02 | 3.877E-02 | 4.060E-02 | 4.624E-02 |
| | $4s^2\ 4p\ 4d\ ^3D_2^o$ | 1.258E-01 | 1.773E-01 | 2.428E-01 | 2.803E-01 | 3.066E-01 | 3.435E-01 | 3.695E-01 | 3.896E-01 | 4.513E-01 |
| | $4s^2\ 4p\ 4d\ ^3D_3^o$ | 1.554E+00 | 2.187E+00 | 2.978E+00 | 3.424E+00 | 3.736E+00 | 4.171E+00 | 4.477E+00 | 4.713E+00 | 5.437E+00 |
| | $4s^2\ 4p\ 4d\ ^3P_1^o$ | 9.877E-02 | 1.371E-01 | 1.847E-01 | 2.115E-01 | 2.302E-01 | 2.562E-01 | 2.746E-01 | 2.887E-01 | 3.321E-01 |
| | $4s^2\ 4p\ 4d\ ^3P_2^o$ | 1.622E-01 | 2.127E-01 | 2.705E-01 | 3.025E-01 | 3.248E-01 | 3.557E-01 | 3.774E-01 | 3.942E-01 | 4.454E-01 |
| | $4s^2\ 4p\ 4d\ ^3F_2^o$ | 1.306E-01 | 1.780E-01 | 2.378E-01 | 2.721E-01 | 2.962E-01 | 3.300E-01 | 3.538E-01 | 3.722E-01 | 4.288E-01 |
| | $4s^2\ 4p\ 4d\ ^3F_3^o$ | 8.038E-01 | 1.094E+00 | 1.466E+00 | 1.678E+00 | 1.826E+00 | 2.034E+00 | 2.181E+00 | 2.294E+00 | 2.642E+00 |
| $4s^2\ 4p^2$ $^3P_2^e$ | $4s\ 4p^3\ ^3D_1^o$ | 2.435E-02 | 2.735E-02 | 3.134E-02 | 3.364E-02 | 3.526E-02 | 3.753E-02 | 3.914E-02 | 4.038E-02 | 4.419E-02 |
| | $4s\ 4p^3\ ^3D_2^o$ | 1.611E-01 | 1.865E-01 | 2.198E-01 | 2.390E-01 | 2.525E-01 | 2.713E-01 | 2.846E-01 | 2.949E-01 | 3.265E-01 |
| | $4s\ 4p^3\ ^3D_3^o$ | 1.588E+00 | 1.859E+00 | 2.213E+00 | 2.417E+00 | 2.559E+00 | 2.759E+00 | 2.900E+00 | 3.009E+00 | 3.344E+00 |
| | $4s\ 4p^3\ ^3P_1^o$ | 1.121E-01 | 1.330E-01 | 1.602E-01 | 1.757E-01 | 1.867E-01 | 2.020E-01 | 2.127E-01 | 2.210E-01 | 2.466E-01 |
| | $4s\ 4p^3\ ^3P_2^o$ | 1.930E+00 | 2.436E+00 | 3.083E+00 | 3.452E+00 | 3.710E+00 | 4.071E+00 | 4.325E+00 | 4.521E+00 | 5.123E+00 |
| | $4s\ 4p^3\ ^3S_1^o$ | 2.375E-01 | 2.838E-01 | 3.440E-01 | 3.784E-01 | 4.025E-01 | 4.363E-01 | 4.601E-01 | 4.785E-01 | 5.350E-01 |
| | $4s\ 4p^3\ ^5S_2^o$ | 2.173E-02 | 2.398E-02 | 2.698E-02 | 2.871E-02 | 2.994E-02 | 3.166E-02 | 3.288E-02 | 3.382E-02 | 3.671E-02 |
| | $4s\ 4p^3\ ^1D_2^o$ | 4.377E-01 | 5.227E-01 | 6.331E-01 | 6.962E-01 | 7.406E-01 | 8.027E-01 | 8.464E-01 | 8.801E-01 | 9.839E-01 |
| | $4s^2\ 4p\ 4d\ ^1D_2^o$ | 4.126E-02 | 4.521E-02 | 4.541E-02 | 4.542E-02 | 4.543E-02 | 4.544E-02 | 4.544E-02 | 4.544E-02 | 4.545E-02 |
| | $4s^2\ 4p\ 4d\ ^3D_1^o$ | 2.691E-04 | 3.419E-04 | 4.337E-04 | 4.863E-04 | 5.233E-04 | 5.752E-04 | 6.118E-04 | 6.401E-04 | 7.269E-04 |
| | $4s^2\ 4p\ 4d\ ^3D_2^o$ | 2.911E-03 | 3.817E-03 | 4.907E-03 | 5.533E-03 | 5.975E-03 | 6.594E-03 | 7.030E-03 | 7.368E-03 | 8.406E-03 |
| | $4s^2\ 4p\ 4d\ ^3D_3^o$ | 2.967E-01 | 3.574E-01 | 4.289E-01 | 4.688E-01 | 4.968E-01 | 5.356E-01 | 5.629E-01 | 5.839E-01 | 6.484E-01 |
| | $4s^2\ 4p\ 4d\ ^3P_1^o$ | 1.912E-01 | 2.398E-01 | 3.005E-01 | 3.348E-01 | 3.587E-01 | 3.922E-01 | 4.157E-01 | 4.339E-01 | 4.896E-01 |
| | $4s^2\ 4p\ 4d\ ^3P_2^o$ | 9.803E-01 | 1.291E+00 | 1.688E+00 | 1.914E+00 | 2.072E+00 | 2.293E+00 | 2.448E+00 | 2.568E+00 | 2.937E+00 |
| $4s^2\ 4p^2$ $^1S_0^e$ | $4s\ 4p^3\ ^1P_1^o$ | 6.505E-01 | 8.288E-01 | 1.057E+00 | 1.186E+00 | 1.277E+00 | 1.403E+00 | 1.493E+00 | 1.561E+00 | 1.773E+00 |
| | $4s\ 4p^3\ ^3S_1^o$ | 3.316E-01 | 3.924E-01 | 4.717E-01 | 5.170E-01 | 5.489E-01 | 5.935E-01 | 6.250E-01 | 6.493E-01 | 7.239E-01 |
| | $4s^2\ 4p\ 4d\ ^1P_1^o$ | 7.553E-01 | 1.003E+00 | 1.319E+00 | 1.498E+00 | 1.624E+00 | 1.799E+00 | 1.923E+00 | 2.018E+00 | 2.310E+00 |

Table 16: Collision Strengths of allowed transitions for **Eu XXXII** ion.

| Lower | Upper | Incident Energy (eV) | | | | | | | | |
|---|---|---|---|---|---|---|---|---|---|---|
| | | 300 | 500 | 1000 | 1500 | 2000 | 3000 | 4000 | 5000 | 10000 |
| $4s^2\ 4p^2$ $^3P_0^e$ | $4s\ 4p^3\ ^3D_1^o$ | 8.344E-01 | 1.074E+00 | 1.378E+00 | 1.551E+00 | 1.672E+00 | 1.840E+00 | 1.959E+00 | 2.051E+00 | 2.333E+00 |
| | $4s\ 4p^3\ ^3P_1^o$ | 3.513E-03 | 4.984E-03 | 6.819E-03 | 7.853E-03 | 8.576E-03 | 9.584E-03 | 1.029E-02 | 1.084E-02 | 1.252E-02 |
| | $4s\ 4p^3\ ^3S_1^o$ | 2.631E-02 | 3.791E-02 | 5.238E-02 | 6.054E-02 | 6.625E-02 | 7.421E-02 | 7.980E-02 | 8.412E-02 | 9.737E-02 |
| | $4s^2\ 4p\ 4d\ ^3D_1^o$ | 5.542E-01 | 8.073E-01 | 1.123E+00 | 1.301E+00 | 1.425E+00 | 1.599E+00 | 1.721E+00 | 1.815E+00 | 2.104E+00 |
| | $4s^2\ 4p\ 4d\ ^3P_1^o$ | 6.249E-05 | 1.039E-04 | 1.542E-04 | 1.823E-04 | 2.020E-04 | 2.294E-04 | 2.487E-04 | 2.636E-04 | 3.092E-04 |
| $4s^2\ 4p^2$ $^3P_1^e$ | $4s\ 4p^3\ ^3D_1^o$ | 4.189E-02 | 5.004E-02 | 6.060E-02 | 6.664E-02 | 7.087E-02 | 7.680E-02 | 8.098E-02 | 8.420E-02 | 9.411E-02 |
| | $4s\ 4p^3\ ^3D_2^o$ | 7.723E-01 | 9.640E-01 | 1.210E+00 | 1.349E+00 | 1.447E+00 | 1.584E+00 | 1.680E+00 | 1.755E+00 | 1.983E+00 |
| | $4s\ 4p^3\ ^3P_0^o$ | 2.379E-01 | 3.054E-01 | 3.913E-01 | 4.400E-01 | 4.742E-01 | 5.219E-01 | 5.555E-01 | 5.813E-01 | 6.609E-01 |
| | $4s\ 4p^3\ ^3P_1^o$ | 1.079E+00 | 1.395E+00 | 1.797E+00 | 2.025E+00 | 2.184E+00 | 2.407E+00 | 2.564E+00 | 2.685E+00 | 3.056E+00 |
| | $4s\ 4p^3\ ^3P_2^o$ | 9.370E-04 | 1.321E-03 | 1.801E-03 | 2.071E-03 | 2.260E-03 | 2.524E-03 | 2.710E-03 | 2.853E-03 | 3.292E-03 |
| | $4s\ 4p^3\ ^3S_1^o$ | 3.306E-01 | 4.309E-01 | 5.581E-01 | 6.301E-01 | 6.806E-01 | 7.511E-01 | 8.007E-01 | 8.389E-01 | 9.563E-01 |
| | $4s\ 4p^3\ ^5S_2^o$ | 6.617E-01 | 7.741E-01 | 9.208E-01 | 1.005E+00 | 1.064E+00 | 1.147E+00 | 1.205E+00 | 1.250E+00 | 1.389E+00 |
| | $4s\ 4p^3\ ^1D_2^o$ | 8.499E-02 | 1.108E-01 | 1.436E-01 | 1.622E-01 | 1.753E-01 | 1.934E-01 | 2.062E-01 | 2.161E-01 | 2.464E-01 |
| | $4s^2\ 4p\ 4d\ ^1D_2^o$ | 4.512E-01 | 6.447E-01 | 8.860E-01 | 1.022E+00 | 1.117E+00 | 1.250E+00 | 1.343E+00 | 1.415E+00 | 1.636E+00 |
| | $4s^2\ 4p\ 4d\ ^3D_1^o$ | 1.172E-01 | 1.546E-01 | 2.022E-01 | 2.292E-01 | 2.480E-01 | 2.744E-01 | 2.930E-01 | 3.073E-01 | 3.512E-01 |
| | $4s^2\ 4p\ 4d\ ^3D_2^o$ | 8.603E-01 | 1.171E+00 | 1.566E+00 | 1.791E+00 | 1.948E+00 | 2.168E+00 | 2.323E+00 | 2.443E+00 | 2.810E+00 |
| | $4s^2\ 4p\ 4d\ ^3P_0^o$ | 1.366E-01 | 1.983E-01 | 2.754E-01 | 3.188E-01 | 3.491E-01 | 3.915E-01 | 4.213E-01 | 4.442E-01 | 5.147E-01 |
| | $4s^2\ 4p\ 4d\ ^3P_1^o$ | 3.363E-01 | 4.902E-01 | 6.820E-01 | 7.901E-01 | 8.657E-01 | 9.712E-01 | 1.045E+00 | 1.102E+00 | 1.278E+00 |
| | $4s^2\ 4p\ 4d\ ^3P_2^o$ | 2.898E-02 | 4.715E-02 | 7.011E-02 | 8.320E-02 | 9.239E-02 | 1.053E-01 | 1.143E-01 | 1.213E-01 | 1.429E-01 |
| $4s^2\ 4p^2$ $^1D_2^e$ | $4s\ 4p^3\ ^1P_1^o$ | 1.143E-03 | 1.657E-03 | 2.298E-03 | 2.659E-03 | 2.912E-03 | 3.264E-03 | 3.511E-03 | 3.702E-03 | 4.289E-03 |
| | $4s\ 4p^3\ ^1D_2^o$ | 2.037E+00 | 2.634E+00 | 3.393E+00 | 3.823E+00 | 4.124E+00 | 4.545E+00 | 4.841E+00 | 5.069E+00 | 5.771E+00 |
| | $4s\ 4p^3\ ^3D_1^o$ | 3.856E-01 | 4.587E-01 | 5.534E-01 | 6.076E-01 | 6.457E-01 | 6.990E-01 | 7.365E-01 | 7.655E-01 | 8.545E-01 |
| | $4s\ 4p^3\ ^3D_2^o$ | 2.513E-02 | 3.112E-02 | 3.880E-02 | 4.317E-02 | 4.623E-02 | 5.051E-02 | 5.353E-02 | 5.585E-02 | 6.300E-02 |
| | $4s\ 4p^3\ ^3D_3^o$ | 6.626E-01 | 8.339E-01 | 1.053E+00 | 1.177E+00 | 1.265E+00 | 1.386E+00 | 1.472E+00 | 1.538E+00 | 1.742E+00 |
| | $4s\ 4p^3\ ^3P_1^o$ | 5.098E-03 | 6.604E-03 | 8.526E-03 | 9.618E-03 | 1.038E-02 | 1.145E-02 | 1.220E-02 | 1.278E-02 | 1.457E-02 |
| | $4s\ 4p^3\ ^3P_2^o$ | 2.474E-02 | 3.466E-02 | 4.708E-02 | 5.409E-02 | 5.899E-02 | 6.583E-02 | 7.065E-02 | 7.436E-02 | 8.574E-02 |
| | $4s^2\ 4p\ 4d\ ^1P_1^o$ | 7.531E-03 | 1.014E-02 | 1.086E-02 | 1.123E-02 | 1.151E-02 | 1.193E-02 | 1.222E-02 | 1.246E-02 | 1.318E-02 |
| | $4s^2\ 4p\ 4d\ ^1D_2^o$ | 2.887E-01 | 4.160E-01 | 5.762E-01 | 6.667E-01 | 7.301E-01 | 8.186E-01 | 8.809E-01 | 9.289E-01 | 1.076E+00 |
| | $4s^2\ 4p\ 4d\ ^1F_3^o$ | 1.636E-01 | 2.758E-01 | 4.177E-01 | 4.988E-01 | 5.557E-01 | 6.355E-01 | 6.917E-01 | 7.350E-01 | 8.684E-01 |
| | $4s^2\ 4p\ 4d\ ^3D_1^o$ | 1.448E-02 | 2.060E-02 | 2.614E-02 | 2.933E-02 | 3.159E-02 | 3.479E-02 | 3.705E-02 | 3.880E-02 | 4.420E-02 |
| | $4s^2\ 4p\ 4d\ ^3D_2^o$ | 1.135E-01 | 1.604E-01 | 2.197E-01 | 2.535E-01 | 2.773E-01 | 3.106E-01 | 3.341E-01 | 3.522E-01 | 4.079E-01 |
| | $4s^2\ 4p\ 4d\ ^3D_3^o$ | 1.420E+00 | 2.026E+00 | 2.778E+00 | 3.202E+00 | 3.497E+00 | 3.910E+00 | 4.200E+00 | 4.424E+00 | 5.110E+00 |
| | $4s^2\ 4p\ 4d\ ^3P_1^o$ | 8.962E-02 | 1.262E-01 | 1.711E-01 | 1.964E-01 | 2.140E-01 | 2.386E-01 | 2.559E-01 | 2.692E-01 | 3.100E-01 |
| | $4s^2\ 4p\ 4d\ ^3P_2^o$ | 1.380E-01 | 1.827E-01 | 2.325E-01 | 2.599E-01 | 2.789E-01 | 3.053E-01 | 3.238E-01 | 3.381E-01 | 3.817E-01 |
| | $4s^2\ 4p\ 4d\ ^3F_2^o$ | 1.245E-01 | 1.706E-01 | 2.283E-01 | 2.614E-01 | 2.846E-01 | 3.172E-01 | 3.401E-01 | 3.578E-01 | 4.123E-01 |
| | $4s^2\ 4p\ 4d\ ^3F_3^o$ | 7.878E-01 | 1.078E+00 | 1.449E+00 | 1.660E+00 | 1.808E+00 | 2.014E+00 | 2.160E+00 | 2.273E+00 | 2.618E+00 |
| $4s^2\ 4p^2$ $^3P_2^e$ | $4s\ 4p^3\ ^3D_1^o$ | 2.183E-02 | 2.446E-02 | 2.795E-02 | 2.996E-02 | 3.139E-02 | 3.338E-02 | 3.479E-02 | 3.587E-02 | 3.922E-02 |
| | $4s\ 4p^3\ ^3D_2^o$ | 1.615E-01 | 1.872E-01 | 2.208E-01 | 2.401E-01 | 2.536E-01 | 2.727E-01 | 2.861E-01 | 2.964E-01 | 3.282E-01 |
| | $4s\ 4p^3\ ^3D_3^o$ | 1.514E+00 | 1.775E+00 | 2.115E+00 | 2.310E+00 | 2.447E+00 | 2.639E+00 | 2.774E+00 | 2.879E+00 | 3.200E+00 |
| | $4s\ 4p^3\ ^3P_1^o$ | 1.193E-01 | 1.418E-01 | 1.710E-01 | 1.877E-01 | 1.994E-01 | 2.158E-01 | 2.274E-01 | 2.363E-01 | 2.638E-01 |
| | $4s\ 4p^3\ ^3P_2^o$ | 1.752E+00 | 2.228E+00 | 2.835E+00 | 3.179E+00 | 3.421E+00 | 3.758E+00 | 3.995E+00 | 4.179E+00 | 4.741E+00 |
| | $4s\ 4p^3\ ^3S_1^o$ | 2.287E-01 | 2.738E-01 | 3.321E-01 | 3.654E-01 | 3.889E-01 | 4.216E-01 | 4.447E-01 | 4.625E-01 | 5.172E-01 |
| | $4s\ 4p^3\ ^5S_2^o$ | 1.745E-02 | 1.916E-02 | 2.145E-02 | 2.277E-02 | 2.371E-02 | 2.502E-02 | 2.595E-02 | 2.667E-02 | 2.887E-02 |
| | $4s\ 4p^3\ ^1D_2^o$ | 4.371E-01 | 5.230E-01 | 6.342E-01 | 6.978E-01 | 7.424E-01 | 8.049E-01 | 8.489E-01 | 8.828E-01 | 9.872E-01 |
| | $4s^2\ 4p\ 4d\ ^1D_2^o$ | 4.409E-02 | 4.890E-02 | 4.953E-02 | 4.972E-02 | 4.984E-02 | 5.001E-02 | 5.012E-02 | 5.020E-02 | 5.045E-02 |
| | $4s^2\ 4p\ 4d\ ^3D_1^o$ | 2.395E-04 | 3.036E-04 | 3.840E-04 | 4.300E-04 | 4.623E-04 | 5.077E-04 | 5.396E-04 | 5.643E-04 | 6.402E-04 |
| | $4s^2\ 4p\ 4d\ ^3D_2^o$ | 2.663E-03 | 3.476E-03 | 4.448E-03 | 5.005E-03 | 5.397E-03 | 5.947E-03 | 6.335E-03 | 6.635E-03 | 7.557E-03 |
| | $4s^2\ 4p\ 4d\ ^3D_3^o$ | 2.533E-01 | 3.051E-01 | 3.645E-01 | 3.976E-01 | 4.207E-01 | 4.528E-01 | 4.754E-01 | 4.928E-01 | 5.460E-01 |
| | $4s^2\ 4p\ 4d\ ^3P_1^o$ | 1.812E-01 | 2.279E-01 | 2.859E-01 | 3.186E-01 | 3.415E-01 | 3.734E-01 | 3.958E-01 | 4.131E-01 | 4.661E-01 |
| | $4s^2\ 4p\ 4d\ ^3P_2^o$ | 9.193E-01 | 1.218E+00 | 1.598E+00 | 1.813E+00 | 1.964E+00 | 2.175E+00 | 2.324E+00 | 2.438E+00 | 2.790E+00 |
| $4s^2\ 4p^2$ $^1S_0^e$ | $4s\ 4p^3\ ^1P_1^o$ | 5.954E-01 | 7.647E-01 | 9.801E-01 | 1.102E+00 | 1.188E+00 | 1.308E+00 | 1.392E+00 | 1.457E+00 | 1.656E+00 |
| | $4s\ 4p^3\ ^3S_1^o$ | 3.169E-01 | 3.756E-01 | 4.519E-01 | 4.955E-01 | 5.262E-01 | 5.691E-01 | 5.993E-01 | 6.227E-01 | 6.944E-01 |
| | $4s^2\ 4p\ 4d\ ^1P_1^o$ | 7.070E-01 | 9.443E-01 | 1.245E+00 | 1.415E+00 | 1.535E+00 | 1.701E+00 | 1.818E+00 | 1.909E+00 | 2.187E+00 |

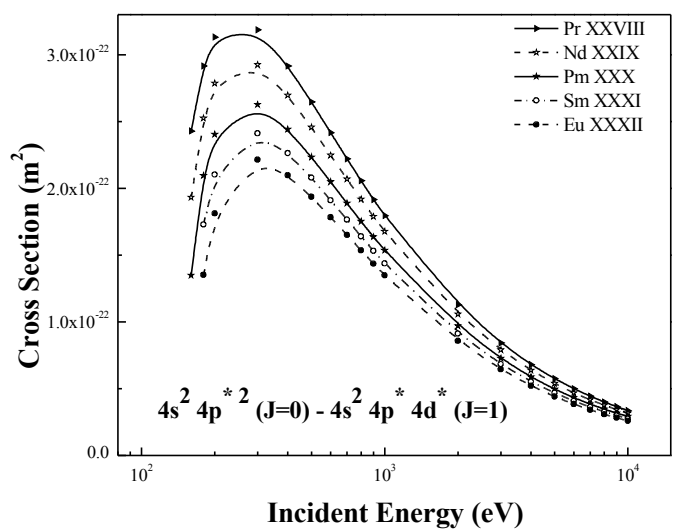

Fig. 1:
The behavior of cross section (m²) of
$4s^2 4p^{*2} (J=0) - 4s^2 4p^* 4d^* (J=1)$
for Ge-like ions

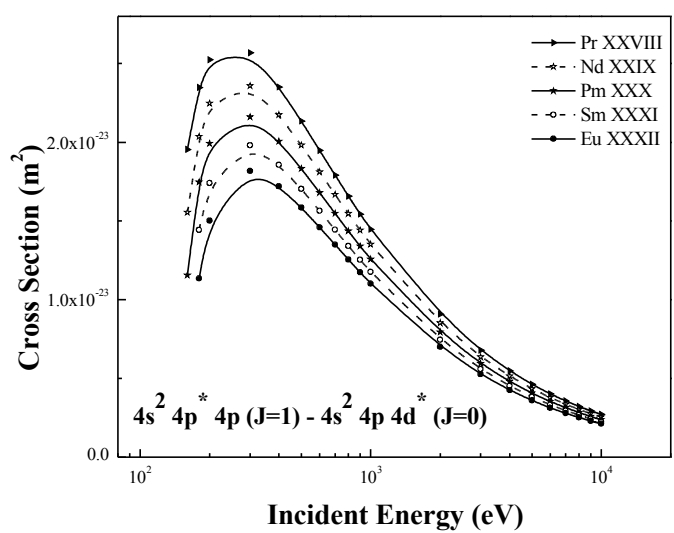

Fig. 2:
The behavior of cross section (m²) of
$4s^2 4p^* 4p (J=1) - 4s^2 4p 4d^* (J=0)$
for Ge-like ions.

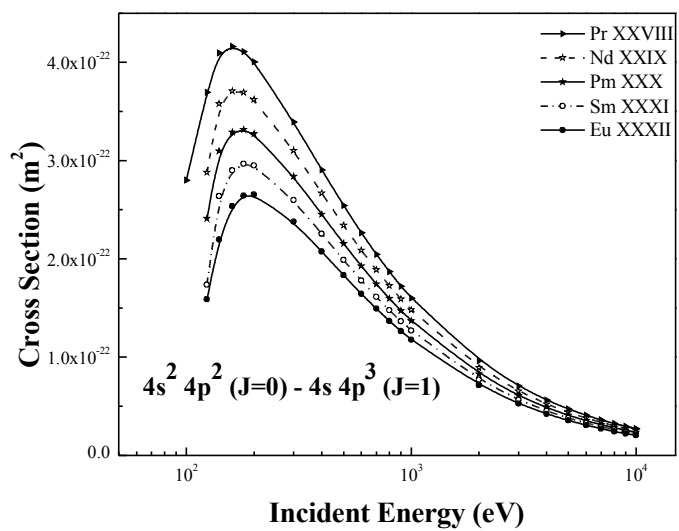

Fig. 3:
The behavior of cross section (m²) of
$4s^2 4p^2 (J=0) - 4s 4p^3 (J=1)$
for Ge-like ions.

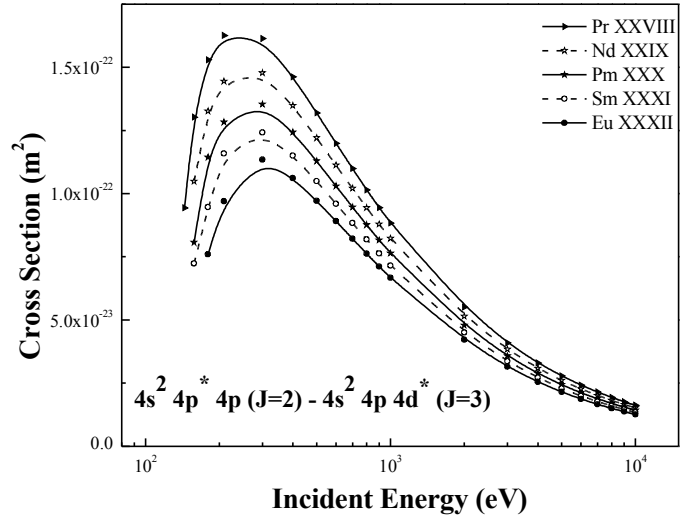

Fig. 4:
The behavior of cross section (m²) of
$4s^2 4p^* 4p\ (J=2) - 4s^2 4p\ 4d^*\ (J=3)$
for Ge-like ions.

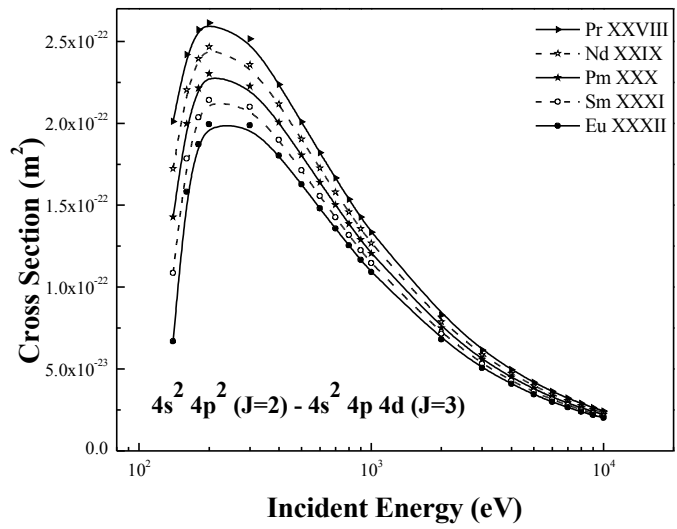

Fig. 5:
The behavior of cross section (m²) of
$4s^2 4p^2\ (J=2) - 4s^2 4p\ 4d\ (J=3)$
for Ge-like ions.